\documentclass[12pt]{article}
\usepackage{axodraw}
\def\CellGroup{\bgroup}
\def\endCellGroup{\egroup}

\newcommand{\be}{\begin{equation}}
\newcommand{\ee}{\end{equation}}
\newcommand{\beq}{\begin{equation}}
\newcommand{\eeq}{\end{equation}}
\newcommand{\bea}{\begin{eqnarray}}
\newcommand{\eea}{\end{eqnarray}}

\def\Re{\mathop{\mbox{Re}}}

\def\trace{{\mathrm{tr}}}

\newcommand{\spqr}{SPQ$_{\textrm{cd}}$R\ }

\newcommand{\nn}{\nonumber}

\def\barr{\begin{array}}
\def\earr{\end{array}}
\def\la{\langle}
\def\ra{\rangle}

\def\l{\left}
\def\r{\right}

\def\ln#1{\mbox{log}{\l( #1 \r)}}
\def\op{{\mathcal{O}}}
\def\ep{\epsilon}

\def\ChPT{$\chi$PT~}

\def\qu{{\vec{q}_1}}
\def\qd{{\vec{q}_2}}

\def\eq#1{{eq.~(\ref{#1})}}
\def\eqs#1#2{{ eqs.~(\ref{#1})} and {(\ref{#2})}}
\def\Eq#1{{Eq.~(\ref{#1})}}
\def\tab#1{{tab.~\ref{#1}}}

\def\sec#1{{\,sec.~\ref{#1}}}
\def\Sec#1{{\,Sec.~\ref{#1}}}
\def\secs#1#2{{\,sec.~\ref{#1}} and {\ref{#2}}}
\def\fig#1{{ fig.~\ref{#1}}}

\newcommand{\PL}[3]{{\it Phys.\ Lett.\ }         {\bf #1}, {#3}
{(#2)}}

\newcommand{\PRL}[3]{{\it Phys.\ Rev.\ Lett.\ }  {\bf #1}, {#3}
{(#2)}}

\newcommand{\PRD}[3]{{\it Phys.\ Rev.\ }     {\bf D#1}, {#3}
{(#2)}}
\newcommand{\NP}[3]{{\it Nucl.\ Phys.\ }         {\bf #1}, {#3}
{(#2)}}
\newcommand{\ZP}[3]{{\it Z.\ Phys.\ }            {\bf #1}, {#3}
{(#2)}}
\newcommand{\JHEP}[3]{{\it JHEP\ }               {\bf #1}, {#3}
{(#2)}}

\newcommand{\CMP}[3]{{\it Commun.\  Math.\  Phys.\  }  {\bf #1}, {#3}
{(#2)}}

\begin{document}
%\begin{fmffile}{villafmf}
%%%%%%%%%%% Titlepage

\begin{titlepage}
\begin{flushright}
CERN-TH/2002-130\\ %
ROMA--1337/02\\  %
SHEP 02--13\\ %
SISSA 47/02 EP\\ %
\end{flushright}
\vskip 0.5cm
\begin{center}
{\Large \bf \boldmath{${K^+\to\pi^+\pi^0}$} Decays on Finite
Volumes and at Next-to-Leading Order in the Chiral Expansion}

\vskip1cm {\large\bf C.-J.D.~Lin$^a$, G.~Martinelli$^b$,
E.~Pallante$^c$,\\ C.T.~Sachrajda$^{a,d}$, G.~Villadoro$^b$}\\

\vspace{.5cm} {\normalsize {\sl $^a$ Dept. of Physics and
Astronomy, Univ. of Southampton,\\ Southampton, SO17 1BJ, UK \\
\vspace{.2cm} $^b$ Dip. di Fisica, Univ. di Roma ``La Sapienza"
and INFN,\\ Sezione di Roma, P.le A. Moro 2, I-00185 Rome, Italy\\ %
\vspace{.2cm} $^c$ SISSA, Via Beirut 2-4, 34013, Trieste, Italy
\\
\vspace{.2cm} $^d$ Theory Division, CERN, CH-1211 Geneva 23,
Switzerland}}\\ \vspace{.2cm} \vspace{.2cm} \vskip1.0cm

{\large\bf Abstract:\\[10pt]} \parbox[t]{\textwidth}{{We present
the ingredients necessary for the determination of physical
$K\to\pi\pi$ decay amplitudes for $\Delta I=3/2$ transitions, from
lattice simulations at unphysical kinematics and the use of chiral
perturbation theory at next-to-leading order. In particular we
derive the expressions for the matrix elements
$_{I=2}\la\,\pi\pi|\op_W|K\ra$, where $\op_W$ is one of the
operators appearing in the $\Delta S=1$ weak Hamiltonian, in terms
of low-energy constants at next-to-leading order in the chiral
expansion. The one-loop chiral corrections are evaluated for arbitrary
masses and momenta, both in full QCD and in the quenched
approximation. We also investigate the finite-volume
effects in this procedure.}}

\end{center}

\end{titlepage}

\setcounter{footnote}{0}
\setcounter{equation}{0}

\section{Introduction}\label{sec:intro}

The need for a quantitative control of strong-interaction effects
in $K\to\pi\pi$ decays is underlined by the recent measurement of
a non-zero value of $\varepsilon^\prime/\varepsilon$~\cite{epeexp}
(demonstrating the existence of direct CP-violation) and the
long-standing $\Delta I=1/2$ rule. A precise evaluation of
non-perturbative QCD effects in nonleptonic kaon decays by lattice
simulations, although possible in principle, is a truly
extraordinary challenge, with a number of major theoretical and
computational obstacles to be overcome. Ultimately, of course, we
would wish to perform the simulations in full QCD, with physical
quark masses on a lattice which is both large enough and
sufficiently fine-grained for finite-volume effects and lattice
artefacts to be negligible (or at least under control). We are
some way from being able to do this, and therefore need rely on
approximations. In this paper we explain our strategy, first
presented in ref.~\cite{spqr}, for improving one of the key
approximations, the use of Chiral Perturbation Theory ($\chi$PT).
Specifically we propose to perform the calculations at
next-to-leading order (NLO) in the chiral expansion. In this paper
we present the necessary formulae for $\Delta I=3/2$ decays and
calculate the chiral logarithms for both full QCD and for the
quenched theory. The chiral logarithms have been evaluated for
arbitrary meson masses and momenta.

In our calculations we incorporate the recent developments in our
understanding of finite-volume effects in nonleptonic kaon
decays~\cite{LL,dlinetal,cairns}. At NLO in $\chi$PT, for $\Delta
I=3/2$ transitions, the corrections which vanish only as powers of
the volume can be eliminated using the techniques presented in
these papers, even in the quenched approximation (note however, that the
generalization of the techniques in refs.~\cite{LL,dlinetal,cairns}
to the case when the two-pions have non-zero total momentum
has yet to be performed). This point is
explained in some detail in \sec{sec:chptfv} below. This is not
the case however, when applying quenched \ChPT to $\Delta I=1/2$
transitions. For these decays, the lack of unitarity in the
quenched theory leads to a number of subtleties and major
difficulties which need to be overcome in order to understand the
volume dependence and to extract the
amplitudes~\cite{BG,GP,dt12quenched}. We expect similar
difficulties for $\Delta I=3/2$ transitions beyond one-loop order
in chiral perturbation theory in quenched QCD.

In this paper we do not discuss the \textit{ultra-violet} problem,
i.e. the construction of finite matrix elements of renormalized
operators constructed from the bare lattice ones. For $\Delta
I=1/2$ transitions this involves the subtraction of terms which
diverge as inverse powers of the lattice spacing (and for fermion
actions which explicitly break chiral symmetry, such as the Wilson
action, there is also mixing with other operators of dimension 6).
The subtraction of power divergences is unavoidable, and different
lattice formulations of QCD, including ones which preserve chiral
symmetry, are being used in attempts to perform these subtractions
effectively and to determine the matrix elements as precisely as
possible. The strategy discussed in this paper for $\Delta I=3/2$
decays can be applied to any of these formulations of lattice
fermions.

The non-perturbative QCD effects in $\Delta I=3/2$ $K\to\pi\pi$
decays are contained in the matrix elements of the following three
operators which appear in the $\Delta S=1$ effective Hamiltonian:
\begin{eqnarray}
\op_4 & = & (\bar s_\alpha d_\alpha)_L\,(\bar u_\beta u_\beta -
\bar d_\beta d_\beta)_L + (\bar s_\alpha u_\alpha)_L(\bar u_\beta
d_\beta)_L\,;\label{eq:o4def}\\ %
\op_7 & = & \frac32(\bar s_\alpha d_\alpha)_L\,\sum_{q=u,d,s,c}
e_q(\bar q_\beta q_\beta)_R\label{eq:o7def}\,;\\ %
\op_8 & = & \frac32(\bar s_\alpha d_\beta)_L\,\sum_{q=u,d,s,c}
e_q(\bar q_\beta q_\alpha)_R\,,\label{eq:o8def}
\end{eqnarray}
where $\alpha,\beta$ are colour indices and $e_q$ is the charge of
quark $q$. The subscripts $L,R$ signify ``left" and ``right" so
that $(\bar q_1 q_2)_{L,R}=\bar q_1\gamma_\mu(1\mp\gamma_5)q_2$.
$\op_4$ transforms as the (27,1) representation of the
$SU(3)_L\times SU(3)_R$ chiral symmetry and the electroweak
penguin (EWP) operators $\op_{7,8}$ transform as the (8,8)
representation. $\op_4$ is a $\Delta I=3/2$ operator, whereas
$\op_{7,8}$ have both $\Delta I=1/2$ and 3/2 components. The
definition of the EWP operators $\op_{7,8}$ coincides with that
used in refs.~\cite{ciuchini,buras} (where they were denoted by
$Q_{7,8}$). Our definition of $\op_4$ is that introduced in
ref.~\cite{shifman}. In terms of the operators defined in
refs.~\cite{ciuchini,buras} $\op_4$ is equal to the $\Delta I=3/2$
component of $Q_1+Q_2$. In this paper we study the $K\to\pi\pi$
matrix elements of these operators, with the two pions in an $I=2$
state, so that only the $\Delta I=3/2$ components of $\op_{7,8}$
contribute.

In this project we are evaluating $K\to\pi\pi$ matrix elements
directly. All but one of the previous lattice studies evaluated
$K\to\pi$ matrix elements of the form $\la\,\pi|\op_W|K\ra$ and
used lowest order \ChPT to relate them to matrix elements
$\la\,\pi\pi|\op_W|K\ra$, following the original proposal in
ref.~\cite{BPRD} (see also refs.~\cite{bmgg,cmp}). ($\op_W$
represents any of the $\Delta S=1$ operators which appear in the
effective weak Hamiltonian.) The exception is the quenched study
of the CP-conserving $K\to\pi\pi$ $\Delta I=3/2$ decays in
ref.~\cite{aoki}, where the two pions were at rest. The previous
studies are therefore not sensitive to final-state interactions.

In ref.~\cite{spqr} it was shown that it is possible to obtain all
the low-energy constants necessary to determine the physical
$\Delta I=3/2$ decay amplitudes at NLO in the chiral expansion by
performing simulations with a simple range of momenta for the
mesons~\footnote{The generalization of the formalism to $\Delta
I=1/2$ decays is currently under investigation}. Specifically the
kaon and one of the final-state pions are at rest and the second
pion has an arbitrary momentum (the energy of this second pion
will be referred to as $E_\pi$ in the following). In order to
ensure that the final state has isospin equal to 2, we have to
symmetrise the final state over the two pions. Since $K\to\pi\pi$
matrix elements with this set of momenta are being computed by the
\spqr collaboration~\cite{spqr2}, we refer to it as SPQR
kinematics. Of course, it is also possible to determine the
low-energy constants which appear at NLO in the chiral expansion
from simulations using other kinematical
configurations~\footnote{In ref.~\cite{laiho} it is suggested that
an alternative set of kaon matrix elements be used to determine
the low energy constants. We comment on this proposal in
\sec{sec:concs}.}. For this reason in the following we will
present the results with arbitrary kaon and pion masses and
momenta.

Among the previous lattice studies are two recent determinations
of $K\to\pi\pi$ matrix elements from computations of $K\to\pi$
matrix elements with quenched domain-wall fermions and using \ChPT
at lowest order~\cite{CPPACS,RBC}. The very interesting results
for the octet enhancement, Re${\cal A}_0/$Re${\cal A}_2$=
9\,--\,12~\cite{CPPACS} and 24\,--\,27~\cite{RBC} (compared to the
experimental value of 22.2) and for
$\varepsilon^\prime/\varepsilon$=
\{($-$7)\,--\,($-$2)\}$10^{-4}$~\cite{CPPACS} and
\{($-$8)\,--\,($-$4)\}$10^{-4}$~\cite{RBC} (compared to the
experimental value of $+(17.2\pm 1.8)\,10^{-4}$), will serve as
important benchmarks for future calculations. (${\cal A}_I$ is the
$K\to\pi\pi$ amplitude for decays into two-pion states with
isospin $I$.) These papers also contain detailed discussions of
the systematic errors and the difficulties in estimating and
controlling them. The importance of a reliable error analysis is
underlined by a comparison of the results of the two
collaborations and with the experimental values. For example, the
cancellations which occur between different contributions to
$\varepsilon^\prime/\varepsilon$ (in particular between the matrix
elements of operators conventionally called $\op_6$ and $\op_8$)
imply that each component must be determined with good precision.
In this paper we tackle one of the main sources of systematic
error, the use of \ChPT at lowest order.

The remainder of the paper is organized as follows. In the next
section we explain our strategy in detail. We express the matrix
elements in terms of the low-energy constants and demonstrate that
it is possible to determine the physical matrix elements at NLO in
the chiral expansion from numerical simulations which are
currently feasible. We present a brief introduction to quenched
and unquenched \ChPT in \sec{sec:chptintro}. \Sec{sec:chptfv} is
devoted to a study of finite volume effects in \ChPT. The NLO
results are presented in \sec{sec:logs} for \textit{physical}
kinematics, i.e. with the momenta of the mesons satisfying
$p_{K^+}=p_{\pi^+}+p_{\pi^0}$, but with the masses being
arbitrary. In this section we also present the results for the
special cases $m_K=2m_\pi$ and $m_K=m_\pi$, in both cases with
each of the pions at rest, and compare our results with those of
previous papers where this is possible. The expressions for the
matrix elements at NLO in the chiral expansion for the SPQR
kinematics are presented in appendices~\ref{subsec:spqrfull} and
\ref{subsec:spqrquenched} for full QCD and the quenched
approximation respectively. The results for general kinematics,
i.e. with arbitrary masses and momenta, are very lengthy and we
present them on the web site~\cite{quenchedfull}, from which
interested users can import them into their programs. For full QCD
we also present the results for general kinematics in
appendix~\ref{subsec:generalfull}. \Sec{sec:concs} contains our
conclusions. There is one other appendix which contains integrals
used in the study of finite-volume effects.

\section{The Strategy}
\label{sec:strategy} In this section we present our strategy for
the determination of $K\to\pi\pi$ matrix elements using chiral
perturbation theory at next-to-leading order. We envisage
computing the matrix elements in numerical simulations for a
variety of (unphysical) masses and momenta. The generic form for
the behaviour of the matrix element of a weak operator $\op_W$
with masses and momenta is
\begin{equation}
\la\,\pi^+\pi^0\,|{\cal O}_W\,|\,K^+\,\ra =
A\,\rho\bigg(1+\textrm{chiral logs}\bigg)+\sum_i \lambda_iB_i,
\label{eq:chiralogeneric}\end{equation} where $A$ and the $B_i$
are kinematic factors and $\rho$ and the $\lambda_i$ are
low-energy constants. (In the applications below $\rho$
corresponds to $\alpha^{(27,1)}$ or $\gamma^{(8,8)}$ and the
$\lambda_i$'s correspond to the $\beta_i$'s or the $\delta_i$'s.)
$A$ is $O(p^2)$ when ${\cal O}=\op_4$ and $O(p^0)$ when
$\op=\op_{7,8}$ and the $B_i$ are $O(p^4)$ and $O(p^2)$ when
$\op=\op_4$ and $\op=\op_{7,8}$ respectively. In both cases there
is only one low-energy constant at lowest order of \ChPT and
several at next-to-leading order. For the
EWP operators $\op_{7,8}$ the low energy constants $\rho$ and
$\lambda_i$ include a factor of the electromagnetic coupling
$e^2$. The expression ``chiral logs" on the right-hand side of
\eq{eq:chiralogeneric} is shorthand for the full one-loop
contribution, which includes both terms containing logarithms and
also others without logarithms.

Of course we wish to determine the matrix element for a physical
kaon decaying into two physical pions. Our strategy for doing this
at NLO in the chiral expansion is as follows~\cite{spqr}:
\begin{enumerate}
\item Calculate the chiral logarithms at one-loop order in \ChPT
for general masses and momenta. In this paper we perform this
calculation for the $\Delta I$ = 3/2 operators $\op_{4,7,8}$ in
both full QCD and in the quenched approximation. Once the logarithms have
been computed, we have the formula for the behaviour of each
matrix element in terms of the low-energy constants $\rho$ and
$\lambda_i$.
\item Determine the matrix elements in \eq{eq:chiralogeneric}
using lattice simulations for a variety of masses and momenta. The
kinematic range must be sufficient to enable the low-energy
constants to be obtained from a fit to the behaviour of the matrix
elements with the masses and momenta.\label{en:fit}
\item Use the low-energy constants determined in item \ref{en:fit} and
the expression in \eq{eq:chiralogeneric} with $m_K$ and $m_\pi$
equal to their physical values and zero energy-momentum transfer
at the operator (i.e. with $p_{K^+}=p_{\pi^+}+p_{\pi^0}$) to
determine the physical matrix elements. In extracting the physical
amplitudes we assume that the higher order terms in the chiral
expansion can be safely neglected in the region of the
extrapolation.
\end{enumerate}

For compactness of notation, in the following we represent the
low-energy constants by the same symbols in both full QCD and in
the quenched theory. It must be stressed however, that the
low-energy constants are not expected to be the same in the two
cases. Indeed their dependence on the renormalization scale is in
general different.

In the remainder of this section we give the expressions for the
matrix elements of $\op_{4,7,8}$ in terms of the low energy
constants.

\subsection{Counterterms for $\mathbf{{\cal O}_4}$ up to $\mathbf{O(p^4)}$}
\label{subsec:chiralo4gen}

The general expression for the matrix element of ${\cal O}_4$,
which transforms as a (27,1) under the $SU(3)\times SU(3)$ chiral
symmetry, for arbitrary meson masses and momenta is
\begin{eqnarray}
 \la\,\pi^+\pi^0\,|{\cal O}_4\,|\,K^+\,\ra& = &
 -\frac{\sqrt{2}}{f_K\,f_\pi^2}\,\Bigg\{\alpha^{(27,1)}\,\Bigl(\,8p_{\pi^{0}}\cdot p_{K}
     +6p_{\pi^{0}}\cdot p_{\pi^{+}} - 2p_{\pi^{+}}\cdot p_{K}+\textrm{chiral logs}
     \Bigr)\nonumber\\
  &&+\beta_{2}\,\Bigl(\,-24m_{\pi}^{2}m_{K}^{2} + 24m_{\pi}^{4}\Bigr)\nonumber\\
  &&\hspace{-1.25in}+\beta_{4}\Big(\,16m_{\pi}^{2}\,(p_{\pi^{0}}\cdot p_{K}) +
  16m_{K}^{2}\,(p_{\pi^{0}}\cdot p_{K}) +
  24m_{\pi}^{2}\,(p_{\pi^{0}}\cdot p_{\pi^{+}})\nonumber\\
  &&-4m_{\pi}^{2}\,(p_{\pi^{+}}\cdot p_{K}) -
  4m_{K}^{2}\,(p_{\pi^{+}}\cdot p_{K})\Big)\nonumber\\
  &&\hspace{-0.8in}+\beta_{5}\,\Bigl(-16m_{\pi}^{2}\,(p_{\pi^{0}}\cdot p_{K})
  -12m_{K}^{2}\,(p_{\pi^{0}}\cdot p_{\pi^{+}})
  +4m_{\pi}^{2}\,(p_{\pi^{+}}\cdot p_{K})\Big)\nonumber\\
  &&\hspace{-1.25in}+\beta_{7}\Big(16m_{\pi}^{2}\,(p_{\pi^{0}}\cdot p_{K})
               +32m_{K}^{2}\,(p_{\pi^{0}}\cdot p_{K})
               +12m_{\pi}^{2}\,(p_{\pi^{0}}\cdot p_{\pi^{+}})\nonumber\\
   & &\hspace{-0.5in}+24m_{K}^{2}\,(p_{\pi^{0}}\cdot p_{\pi^{+}})
         -4m_{\pi}^{2}\,(p_{\pi^{+}}\cdot p_{K})
         -8m_{K}^{2}\,(p_{\pi^{+}}\cdot
         p_{K})\Big)\nonumber\\
&&\hspace{-1.5in}+\beta_{22}\Big(24(p_{K} \cdot p_{\pi^{0}})\,
                   (p_{K} \cdot p_{\pi^{+}})
         -8(p_{K} \cdot p_{\pi^{0}})\,
       (p_{\pi^{0}} \cdot p_{\pi^{+}})
         + 32(p_{K} \cdot p_{\pi^{+}})\,
      (p_{\pi^{0}} \cdot p_{\pi^{+}})\Big)
\nonumber\\
   &&\hspace{-1in}+\beta_{24}\Big(128(p_{K} \cdot p_{\pi^{0}})^{2}
         -96(p_{\pi^{0}} \cdot p_{\pi^{+}})^{2}
         -32(p_{K} \cdot p_{\pi^{+}})^{2}\Big)\,\Bigg\}\ .
\label{eq:chiralo4gen}\end{eqnarray} $f_{\pi}$ and $f_K$ are the
physical pion and kaon decay constants. At NLO in the chiral
expansion they are given in terms of the parameter $f$ of the
chiral Lagrangian (see sec.~\ref{sec:chptintro} below) and the
Gasser-Leutwyler coefficients in ref.~\cite{ga-le} for full QCD
and in ref.~\cite{bg} for quenched QCD. $\alpha^{(27,1)}$ is the
leading order ($O(p^2)$) low-energy constant whereas the
$\beta_{i}$ ($i=2,4,5,7,22,24)$ are the $O(p^4)$ counterterms,
where the subscript corresponds to the numbering of the operators
in ref.~\cite{KMW}. These operators are listed explicitly in
\eq{o271nl} of \sec{sec:chptintro}. Thus for a set of simulations,
with a range of quark masses and momenta in the chiral regime, the
mass and momentum dependence is given by \eq{eq:chiralo4gen}. We
envisage determining the low-energy constants $\alpha^{(27,1)}$
and $\beta_i$ in lattice simulations (or at least a sufficient
subset of these constants). We have calculated the chiral
logarithms for arbitrary meson masses and momenta, for full QCD
and in the quenched approximation and we present the results on
the web site~\cite{quenchedfull}. For full QCD we also present the
results in appendix~\ref{subsec:generalfull}.

It is possible to determine the physical matrix elements of
$\op_4$ by performing simulations with the SPQR kinematics. This
will become clear in the following subsections where we rewrite
\eq{eq:chiralo4gen} for physical kinematics
(subsection~\ref{subsec:chiralo4phys}) and for SPQR kinematics
(subsection~\ref{subsec:chiralo4spqr}). The corresponding one-loop
expressions with the chiral logarithms included are given in:
\begin{itemize}
\item eqs.~(\ref{eAmPhys}) -- (\ref{eAmPhyspiu}) for the
physical kinematics in full QCD;
\item eqs.~(\ref{eqn:AmpQ}) -- (\ref{eq:Cdef}) for the
physical kinematics in quenched QCD;
\item eqs.~(\ref{eq:spqrfull271}) -- (\ref{eq:icd5def}) for the
SPQR kinematics in full QCD;
\item eqs.~(\ref{eq:spqrquenched271}) -- (\ref{eq:icdqgen}) for the
SPQR kinematics in quenched QCD.
\end{itemize}

\subsubsection{The Matrix Element of $\mathbf{{\cal O}_4}$ for
Physical Kinematics}\label{subsec:chiralo4phys}

For the physical matrix element, when there is no momentum
insertion at the weak operator ${\cal O}_4$, \eq{eq:chiralo4gen}
reduces to
\begin{eqnarray}
 \la\,\pi^+\pi^0\,|{\cal O}_4\,|\,K^+\,\ra_{\textrm{\small{phys}}}
 & = &-\frac{6\sqrt{2}}
 {f_K\,f_\pi^2}\Bigg\{\alpha^{(27,1)} \bigg(\,m^{2}_K- m^{2}_{\pi}+\textrm{chiral logs}
 \bigg)\nonumber\\
 &&\hspace{-1.3in}+ (\beta_{4}-\beta_{5}+4\beta_{7}+ 2 \beta_{22})
     m^{4}_{K}+ (4\beta_{2}-4\beta_{4}-2\beta_{7}-16\beta_{24})
     m^{4}_{\pi}\nonumber\\
 &&\hspace{-0.8in} +(-4\beta_{2}+3\beta_{4}+\beta_{5}-2\beta_{7}-2\beta_{22}
      +16\beta_{24}) m^{2}_{K}m^{2}_{\pi}\Bigg\}\ .
\label{eq:chiralo4phys}\end{eqnarray} Of course it must be
remembered that the chiral logarithms depend on the kinematics,
however, as they are calculable we do not exhibit this dependence
explicitly here. In principle, by varying the masses and momenta
of the pions it is possible to determine $\alpha^{(27,1)}$ and the
$\beta_i$'s but in practice this is not the optimal strategy. For
each set of pion masses and momenta one would have to tune the
mass of the strange quark in order to ensure energy conservation,
and in addition one would have to extract the matrix elements of
non-leading (excited) energy levels. In the following subsection
we demonstrate that by performing the simulations with the SPQR
kinematics it is possible to determine the combinations of the
low-energy constants which appear in \eq{eq:chiralo4phys}.

\subsubsection{The Matrix Element of $\mathbf{{\cal O}_4}$ for
the SPQR Kinematics}\label{subsec:chiralo4spqr}

For the SPQR kinematics the matrix element in \eq{eq:chiralo4gen}
reduces to
\begin{eqnarray}
 \la\,\pi^+\pi^0\,|{\cal O}_4\,|\,K^+\,\ra_{\textrm{\scriptsize{SPQR}}}
 & = &-\frac{6\sqrt{2}}
 {f_K\,f_\pi^2}\Bigg\{\alpha^{(27,1)}\Big(E_{\pi}m_{\pi}+
  \frac12m_K\,(E_{\pi} + m_{\pi})
  +\textrm{chiral logs}\Big)\nonumber\\
  &&\hspace{-1.5in}+ 4 \beta_{2} m^{4}_{\pi}
  +(4\beta_{4}+2\beta_{7})E_{\pi}m^{3}_{\pi}
  +(\beta_{4}-\beta_{5}+\beta_{7})m^{3}_{\pi}m_{K}\nonumber\\
  &&\hspace{-2in}+ (\beta_{4}-\beta_{5}+\beta_{7}+2\beta_{22})
              E_{\pi}m^{2}_{\pi}m_{K}
  + (-4\beta_{2}+8\beta_{24})m^{2}_{\pi}m^{2}_{K}
+ (\beta_{4}+2\beta_{7})E_{\pi}m^{3}_{K}\nonumber\\
 && \hspace{-1in}+ (-2\beta_{5}+4\beta_{7}+4\beta_{22})
           E_{\pi}m_{\pi}m^{2}_{K}
  + (\beta_{4}+2\beta_{7})m_{\pi}m^{3}_{K}\nonumber\\
  &&\hspace{-1in} +(-16\beta_{24})E^{2}_{\pi}m^{2}_{\pi}
  + 2\beta_{22}E^{2}_{\pi} m_{\pi} m_{K}
  + 8\beta_{24}E^{2}_{\pi}m^{2}_{K}\,
     \Bigg\}\,.
\end{eqnarray}
It can readily be seen that by fitting the lattice results for the
matrix elements as a function of the masses and $E_\pi$, the
combinations of low-energy constants needed to determine the
physical matrix elements in \eq{eq:chiralo4phys} can be obtained.

\subsection{Counterterms for $\mathbf{{\cal O}_{7,8}}$ up to $\mathbf{O(p^2)}$}
\label{subsec:chiralo78gen}

We now present the corresponding formulae for the $\Delta I=3/2$
EWP  operators ${\cal O}_{7,8}$. In this case the matrix elements
start at $O(p^0)$ and at NLO we have to consider all possible
terms of $O(p^2)$. The general expression for these matrix
elements up to this order is
\begin{eqnarray}
 \la\,\pi^+\pi^0\,|{\cal O}_{7,8}\,|\,K^+\,\ra& = &
 \frac{2\sqrt{2}}{f_K\,f_\pi^2}\,\Bigg\{\gamma^{(8,8)}\,(1 + \textrm{
 chiral logs })  +\nn\\
&&\hspace{-1.3in}\delta_1\left(\frac{1}{3} (p_{\pi^{0}}\cdot
p_{K})
                    - (p_{\pi^{0}}\cdot p_{\pi^{+}})
                   + \frac{2}{3} (p_{\pi^{+}}\cdot p_{K})\right)
-\delta_2\left( (p_{\pi^{0}}\cdot p_{K})
                  + (p_{\pi^{0}}\cdot p_{\pi^{+}}) \right )
                  \label{eq:chiralo78gen}\\
&&\hspace{-1.4in}-\delta_3 \left (\frac{4}{3}  (p_{\pi^{0}}\cdot
p_{K})
                    +\frac{2}{3} (p_{\pi^{+}}\cdot p_{K})\right )
\nonumber + (\delta_4+\delta_5) \left (
2m^{2}_{K}+4m^{2}_{\pi}\right ) + \delta_6 \left(
4m^{2}_{K}+2m^{2}_{\pi}\right ) \,\Bigg\}\,.
\end{eqnarray} The $O(p^2)$ counterterms
$\delta_i$, $i=1$ -- 6, correspond to the operators with the same
numbers in ref.~\cite{CG}. We give these operators explicitly in
\eq{o88nl} of \sec{sec:chptintro}. A complete list of the
$O(p^2)$ independent counterterms can be found in ref.~\cite{einp}.

Since $\op_7$ and $\op_8$ both
transform under the same representation, the generic form of the
chiral expansion is the same but the values of the
$\gamma^{(8,8)}$ and the $\delta_i$ on the right-hand side of
\eq{eq:chiralo78gen} depends on the operator. We leave this
dependence implicit.

We present the results for the chiral logarithms for arbitrary
meson masses and momenta, together with those for $\op_4$, on the
web site~\cite{quenchedfull}. For full QCD we also present the
results in appendix~\ref{subsec:generalfull}.

We rewrite \eq{eq:chiralo4gen} for physical and SPQR kinematics in
\secs{subsec:chiralo78phys}{subsec:chiralo78spqr} respectively.
The one-loop expressions with the chiral logarithms included are
given in:
\begin{itemize}
\item eqs.~(\ref{eAmPhys78}) and (\ref{eq:jsdef}) for the
physical kinematics in full QCD;
\item eqs.~(\ref{eAmpQ78}) and (\ref{eq:jqsdef}) for the
physical kinematics in quenched QCD;
\item eqs.~(\ref{eq:spqrfull88}) -- (\ref{eq:jcdfulldef}) for the
SPQR kinematics in full QCD;
\item eqs.~(\ref{eq:spqrquenched88}) -- (\ref{eq:jcdqgen}) for the
SPQR kinematics in quenched QCD.
\end{itemize}

\subsubsection{The Matrix Element of $\mathbf{{\cal O}_{7,8}}$ for
Physical Kinematics}\label{subsec:chiralo78phys}

For the physical kinematics \eq{eq:chiralo78gen} reduces to
\begin{eqnarray}
 \la\,\pi^+\pi^0\,|{\cal
 O}_{7,8}\,|\,K^+\,\ra_{\textrm{\small{phys}}}
 & = &\frac{2\sqrt{2}}{f_K\,f_\pi^2}\,\Bigg\{\gamma^{(8,8)}\,(1 + \textrm{
 chiral logs }) +\label{eq:chiralo78phys}\\
 &&\hspace{-2in}\Big(-\delta_2-\delta_3+2(\delta_4+\delta_5)+4\delta_6\Big)\,m_K^2
+\Big(\delta_1+\delta_2+4(\delta_4+\delta_5)+2\delta_6\Big)m_\pi^2\Bigg\}.\nn
\end{eqnarray}

\subsubsection{The Matrix Element of $\mathbf{{\cal O}_{7,8}}$ for
the SPQR Kinematics}\label{subsec:chiralo78spqr}

Following the same strategy as for ${\cal O}_4$, we envisage
determining the $\gamma^{(8,8)}$'s and $\delta_i$'s by performing
the simulations with the SPQR kinematics. In this case
\eq{eq:chiralo78gen} reduces to
\begin{eqnarray}
 \la\,\pi^+\pi^0\,|{\cal
 O}_{7,8}\,|\,K^+\,\ra_{\textrm{\scriptsize{SPQR}}}
 & = &\frac{2\sqrt{2}}{f_K\,f_\pi^2}\,\Bigg\{\gamma^{(8,8)}\,(1 + \textrm{
 chiral logs }) +
 2\Big(\delta_4+\delta_5+2\delta_6\Big)\,m_K^2
 +\nonumber\\
 &&\hspace{-2in}\frac12\Big(\delta_1-\delta_2-2\delta_3\Big)(m_\pi+E_\pi)m_K
 +2\Big(2(\delta_4+\delta_5)+\delta_6\Big)m_\pi^2
  -\Big(\delta_1+\delta_2\Big)m_\pi E_\pi \Bigg\}.
\label{eq:chiralo78spqr}\end{eqnarray} It is straightforward to
verify that if one determines $\gamma^{(8,8)}$ and the
combinations of the $\delta_i$'s which appear in
\eq{eq:chiralo78spqr} by fitting lattice data for SPQR kinematics,
then one has all the necessary ingredients to determine the
physical matrix elements using \eq{eq:chiralo78phys}.

\section{Quenched and Unquenched \ChPT for $K\to\pi\pi$ Decays}
\label{sec:chptintro}

In this section we outline the main ingredients of the one-loop
\ChPT calculations. The generic form of the (27,1) component of
the weak Lagrangian for $K\to\pi\pi$ up to $O(p^4)$ is~\cite{KMW}
\begin{equation}
{\cal L}_{\textrm{eff}}^{(27,1)}=
-\alpha^{(27,1)}\,\op^{(27,1)}_{2}-\sum_{i=2,4,5,7,22,24}\beta_{i}
\,\op^{(27,1)}_{4,i}+\mbox{h.c.}+O(p^6)
\label{leff4}\end{equation} where $\op^{(27,1)}_{2}$ is the
leading order (\,$O(p^{2})$\,) operator
\begin{equation}
\op^{(27,1)}_{2}=  T^{ij}_{kl} (L_{\mu})^{k}_{\mbox{ }i}
     (L^{\mu})^{l}_{\mbox{ }j},
\label{o271l}\end{equation}
and  $\op^{(27,1)}_{4,i}$ are the operators which appear at NLO
($O(p^4)$)
\begin{eqnarray}
 \op^{(27,1)}_{4,2} &=& - T^{ij}_{kl}
   (P)^{k}_{\mbox{ }i} (P)^{l}_{\mbox{ }j} ,\nonumber\\
 \op^{(27,1)}_{4,4} &=&  T^{ij}_{kl}
    (L_{\mu})^{k}_{\mbox{ }i}
   \{L^{\mu},\, S\}^{l}_{\mbox{ }j} ,\nonumber\\
 \op^{(27,1)}_{4,5} &=& i\, T^{ij}_{kl}
    (L_{\mu})^{k}_{\mbox{ }i}
    \left[ P, L^{\mu}\right]^{l}_{\mbox{ }j} ,\nonumber\\
 \op^{(27,1)}_{4,7} &=& T^{ij}_{kl}
   (L_{\mu})^{k}_{\mbox{ }i} (L^{\mu})^{l}_{\mbox{ }j}
  \ \trace\left[ S \right ] ,\nonumber\\
 \op^{(27,1)}_{4,22} &=&  i\,T^{ij}_{kl}
   (L_{\mu})^{k}_{\mbox{ }i}
  \left[ L_{\nu}, W^{\mu\nu}\right]^{l}_{\mbox{ }j} ,\nonumber\\
 \op^{(27,1)}_{4,24} &=&   T^{ij}_{kl}
   (W_{\mu\nu})^{k}_{\mbox{ }i} (W^{\mu\nu})^{l}_{\mbox{ }j} .
\label{o271nl}\end{eqnarray}
The tensor $T^{ij}_{kl}$ is defined as
\bea
 & &T^{13}_{12}=T^{31}_{12}=T^{13}_{21}=T^{31}_{21}=\frac{1}{2} ,\nonumber\\
 & &T^{23}_{22}=T^{32}_{22}=-\frac{1}{2} ,
\eea
with all other components zero.  $L_{\mu}$ and $W_{\mu\nu}$ are
defined by
\bea
 L_{\mu} &=& i\,\Sigma^{\dagger}\,\partial_{\mu}\Sigma ,\nonumber\\
 W_{\mu\nu} &=& 2\,\left (
   \partial_{\mu}L_{\nu}+\partial_{\nu}L_{\mu}\right )\ ,
\eea and $S$ and $P$ by
 \begin{equation}
 S = \Sigma^{\dagger}\,\chi +
 \chi^{\dagger}\,\Sigma\hspace{0.2in}\textrm{and}\hspace{.2in}
 P = i\,\left (\Sigma^{\dagger}\,\chi -
 \chi^{\dagger}\,\Sigma\right )
\end{equation}
with
\begin{equation}
 \chi = \frac{-2\la 0|\bar{u}u+\bar{d}d|0\ra}{f^{2}}\,{\cal M}\,.
 \label{eq:chidef}\end{equation}
$\Sigma$ is the conventional exponential representation of the
pseudo-Goldstone boson field
\begin{equation}
 \Sigma = {\mathrm{exp}}\left ( \frac{2i\Phi}{f}\right )\ \
 \textrm{\ \ with\ \ }\ \  \Phi =  \left ( \begin{array}{ccc}
            \frac{\pi^{0}}{\sqrt{2}}+\frac{\eta}{\sqrt{6}} &
           \pi^{+} &  K^{+} \\
           \pi^{-} & -\frac{\pi^{0}}{\sqrt{2}}+\frac{\eta}{\sqrt{6}} &
           K^{0} \\
           K^{-} & \bar{K}^{0} & -\sqrt{\frac{2}{3}}\eta
          \end{array} \right ) ,
\label{Sigmadef}\end{equation} and ${\cal M}$ is the quark
mass-matrix ${\cal M}=\textrm{diag}(m_u,m_d,m_s)$.

For the (8,8) component of the $\Delta S=1$ weak Lagrangian the
generic form is~\cite{CG,einp}:
\begin{equation} {\cal L}_{\textrm{eff}}^{(8,8)}=
 -\gamma^{(8,8)}\,\op^{(8,8)}_{0}
 -\sum_{i=1}^{6}\delta_{i}\,\op^{(8,8)}_{2,i} + O(p^4),
\end{equation}
where $\op^{(8,8)}_{0}$ is the leading order (\,$O(p^{0})$\,)
operator
\beq
 \op^{(8,8)}_{0} = \trace\left[ \lambda_{6}\,\Sigma^{\dagger}\, Q\,
\Sigma\right] , \label{o88l}\eeq
and $\op^{(8,8)}_{2,i}$ are the NLO ($O(p^{2})$) operators
\bea
 \op^{(8,8)}_{2,1} &=& \trace\left[ \lambda_{6}\,L_{\mu}
    \Sigma^{\dagger}\,Q\,\Sigma\, L^{\mu}\right] ,\nonumber\\
 \op^{(8,8)}_{2,2} &=& \trace\left[\lambda_{6}\,
     L_{\mu} \right]\,
      \trace\left[  \Sigma^{\dagger}\,Q\,\Sigma\,
     L^{\mu} \right],\nonumber\\
 \op^{(8,8)}_{2,3} &=& \trace\left[ \lambda_{6}\,\{
\Sigma^{\dagger}\,Q\,\Sigma\, , L_\mu\, L^\mu\}\right] ,\nonumber\\
 \op^{(8,8)}_{2,4} &=& \trace\left[ \lambda_{6}\,
\{\Sigma^{\dagger}\,Q\,\Sigma\, , S\}\right] ,\nonumber\\
 \op^{(8,8)}_{2,5} &=& i\,\trace\left[ \lambda_{6}\,
     \left [ \Sigma^{\dagger}\,Q\,\Sigma , P\right ]\right] ,\nonumber\\
 \op^{(8,8)}_{2,6} &=& \trace\left[ \lambda_{6}\,\Sigma^{\dagger}\, Q\,
   \Sigma\right]\ \trace\left[ S\right]\, .\label{o88nl}
\eea
$Q$ and $\lambda_6$ are the charge and  Gell-Mann  matrices
respectively:
\begin{equation}
 Q = \left ( \begin{array}{ccc}
        \frac{2}{3} & 0 & 0\\
        0 & -\frac{1}{3} & 0\\
        0 & 0 & -\frac{1}{3}
       \end{array} \right ) \textrm{\ \ \ and\ \ \ }
\lambda_{6} = \left ( \begin{array}{ccc}
        0 & 0 & 0\\
        0 & 0 & 1\\
        0 & 1 & 0
       \end{array} \right ).
\label{eq:matricesdef}\end{equation}

For the strong interactions we only need the leading order chiral
Lagrangian
\begin{equation}
{\cal
L}_{\textrm{\scriptsize{strong}}}=\frac{f^2}{8}\trace\left[(\partial_\mu
\Sigma^\dagger)\,(\partial^\mu\Sigma) + \Sigma^\dagger\chi +
\chi^\dagger\Sigma\,\right]\ . \label{eq:lstrong}\end{equation}
The NLO terms in ${\cal L}_{\textrm{\scriptsize{strong}}}$, which
are proportional to the Gasser-Leutwyler $L_i$ coefficients,
contribute to the wave function renormalization. These effects are
reabsorbed in the renormalization of the decay constant leading to
the replacement of the factor $1/f^3$ by $1/(f_Kf_\pi^2)$ in all
the $K\to\pi\pi$ matrix elements. We therefore do not present the
NLO terms in the strong chiral Lagrangian.

\subsection{Quenched \ChPT}
\label{subsec:qchpt}

For the calculation in quenched \ChPT we use the strong Lagrangian
introduced in ref.~\cite{bg}
\begin{equation}
{\cal
L}_{\textrm{\scriptsize{strong}}}=\frac{f^2}{8}\textrm{str}\,\left[(\partial_\mu\Sigma^\dagger)\,
(\partial^\mu\Sigma) + \Sigma^\dagger\chi +
\chi^\dagger\Sigma\,\right] -
m_0^2\Phi_0^2+\alpha\,(\partial_\mu\Phi_0)(\partial^\mu\Phi_0)\ .
\label{eq:lstrongq}\end{equation} In \eq{eq:lstrongq} the trace
over the chiral group indices in \eq{eq:lstrong} has been replaced
by the supertrace over the indices of the graded group
$SU(3|3)_L\times SU(3|3)_R$. The fields $\Sigma$ and $\chi$ are
now graded extensions of $\Sigma$ and $\chi$ defined in
\eqs{Sigmadef}{eq:chidef}.
Since in the quenched approximation the singlet field remains degenerate
with the pion and has to be treated as dynamical, the graded extension of
the matrix $\Phi$ is not supertraceless. We define the singlet field
$\Phi_0=$str$[\Phi]/\sqrt{6}$ proportional to the graded extension of the
$\eta^\prime$.

In the weak operators each field in eqs.~(\ref{o271l}),
(\ref{o271nl}), (\ref{o88l}) and (\ref{o88nl}) is replaced by its
graded extension, and every trace by the corresponding supertrace.
Although we use the following extension for the charge matrix in
the quenched approximation
\begin{equation}
Q = \textrm{diag}(2/3,-1/3,-1/3,2/3,-1/3,-1/3),
\end{equation}
the results for $\Delta I=3/2$ transitions do not depend on the
choice of extension in the SU(2) (isospin) limit. This is because
the EWP operators are always inserted on external quark lines, and
hence the matrix elements are independent of the charges assigned
to the ghost particles. For the same reason, for the present
calculation we can simply extend $\lambda_6$ to a six-by-six block
diagonal matrix:
\begin{equation}
\lambda_6\to \l(\begin{array}{cc}\lambda_6 & 0 \\ 0 & 0
\end{array}\r).
\end{equation}

\section{\ChPT \, on a Finite Volume}
\label{sec:chptfv}

In this section  we study, at one-loop order in \ChPT, the
Euclidean correlation functions relevant for the finite-volume
lattice determination of $K \to \pi\pi$ decay amplitudes. As
argued in the section~\ref{sec:strategy}, such a study is a
necessary step to extract, by extrapolation, the physical
amplitudes from correlation functions which at present can only be
computed  at unphysical values of masses and momenta~\cite{spqr}.
Although the explicit discussion is presented for full QCD, it
also applies to quenched (and partially quenched) QCD for $\Delta
I=3/2$ decays (for $\Delta I=1/2$ decays in quenched QCD this is
not the case~\cite{dt12quenched}).

A significant step towards enabling the determination of the
physical matrix element of ${\cal O}_4$ from matrix elements
computed in lattice simulations at unphysical kinematics, has been
the evaluation, at one-loop order in perturbation theory in finite
volumes, of $K\to\pi\pi$ amplitudes with $m_K=m_\pi$ and
$m_K=2m_\pi$ and with all the mesons at
rest\,\cite{GL,GL2,Glpart}. (For a discussion of the advantages of
using these kinematic points in the evaluation of $\Delta I=1/2$
decay amplitudes at leading order in the chiral expansion see
refs.~\cite{BDHS,dawson}.) The one-loop generating functional in
the presence of weak interactions, also in the quenched
approximation, was considered in ref.\,\cite{EP}.

In discussing the behaviour of the correlation functions and
matrix elements with the volume, we envisage varying the volume at
fixed physics, i.e. with the masses, energies and momenta constant
in physical units. From our explicit computations we see that all
the one-loop corrections can be classified into five categories:
\begin{enumerate}
\item The infinite-volume chiral corrections to the
matrix elements of the weak operators. These diverge in \ChPT, and
are renormalized by operators of higher dimension~\cite{KMW}, see
sec.~\ref{sec:logs};
\item The finite-volume  shift to the two-pion energy,
$\Delta W=W -2 E_{\pi}$, which is  independent of the weak
operators, and only depends on the isospin of the two-pion state.
We illustrate this point by showing explicitly that our
calculations give the same shift $\Delta W$ for the operators
$\op_{4}$, $\op^{3/2}_{7}$ and $\op^{3/2}_{8}$ (where the
superscript 3/2 indicates the $\Delta I =3/2$ component of the
operators);
\item The corrections which shift the energy argument ($E$)  of
the amplitude ${\cal A}(E)$ from its infinite volume value to the
finite volume one, i.e.  ${\cal A}(2 E_{\pi}) \to {\cal A}(W)$. We
show that in earlier studies the shift of the argument of ${\cal
A}(E)$ was wrongly interpreted as a finite-volume correction to
the amplitude~\cite{GL,GL2};
\item \label{item:sources} The finite volume corrections to the
sink (source) which is used to annihilate (create) the two pions.
These corrections obviously depend on the sink/source and can be
eliminated non-perturbatively by studying the source-sink
correlation functions. In the following we consider two classes of
correlation functions. In the first the two pions are annihilated
at the same time (as used, for example, in ref.~\cite{LL}) and in
the second one pion is annihilated much later than the other (as
used, for example, in refs.~\cite{dlinetal,cairns,MT}).
\item Finally we have the Lellouch-L\"uscher (LL) finite-volume
corrections to the matrix elements themselves.
\end{enumerate}

A very important result is that the finite-volume correction to
the amplitude, in one-loop \ChPT and in the center of mass frame,
is the universal LL factor with
both types of sources mentioned in item~\ref{item:sources} above.
This will be explained in \sec{skpplphi4} below.
The generalization of the LL factor to the case where the two pions have
a non-zero total three-momentum is being studied (see
sec.\,\ref{subsubsec:moving}).

The starting point for our studies is the Euclidean correlation
function
\begin{equation}
{\cal C}_{\qu \qd}(t_K,t_1,t_2)=\la 0 |\pi_\qu(t_1) \pi_\qd(t_2)
\op_W(0) K^\dagger_{\vec 0}(t_K) |0 \ra  \, ,
\label{eq:first}\end{equation}  where
\begin{equation}
\pi_{\vec{q}}(t)=\int d^3x\ \pi(\vec{x},t)\,
e^{i\vec{q}\vec{x}}\end{equation} is the Fourier transform of the
one-pion interpolating field (with a similar definition for the
kaon) and $t_K<0<t_2\le t_1$. $\op_W$ is one of the $\Delta I=3/2$
weak operators, and from the correlation functions in
\eq{eq:first} we wish to determine the matrix element
$\la\pi\pi|\op_W|K\ra$. In the following subsections we explain,
at one-loop order in \ChPT, what the finite-volume effects in the
correlation function are, and how one recovers the LL factor in
the matrix elements.

In order to simplify the discussion of the different corrections
occurring at one loop, while maintaining all the essential
ingredients, we begin in \sec{skpplphi4} with a study of the
correlation functions in a $\lambda \Phi^4$ theory (for the strong
interactions) and later, in \secs{subsec:kppchptfv}{sec:kppFV}
generalise the discussion to \ChPT.

\subsection{$K^+\to\pi^+\pi^0$ Decays on a Finite Volume
in a $\lambda \Phi^4$ Model}
\label{skpplphi4}

In this section we illustrate the structure of the relations
between correlation functions in finite and infinite volumes, by
considering $K \rightarrow \pi \pi$ decays in a simple model in
which the interactions between the pions are described by a
$\lambda \Phi^4$ theory. The bare $K\to\pi\pi$ vertex is taken to
be a constant ($\lambda_{K\pi\pi}$). This model contains all the
essential ingredients necessary to illustrate the key features of
finite-volume effects. In the following section we generalise this
discussion to \ChPT.

At tree-level, the correlation function in \eq{eq:first} is given
by
\begin{eqnarray}
{\cal C}^{(0)}_{\qu\qd}(t_K,t_1,t_2) &\equiv & {\cal
A}^{(0)}(E_{T}) \times {\cal G}^{(0)}(t_K,t_1,t_2,E_{T}) \nn
\\&=& \lambda_{K\pi\pi}
\times \frac{1}{2m_K \, 2E_\qu \, 2E_\qd}
e^{-m_K|t_K|} \, e^{-E_\qu (t_1-t_2)}\, e^{-E_{T} \, t_2} \, ,
\end{eqnarray}
where $E_{\vec{q}_i}^2=|\vec{q}_i|^2+m_\pi^2$ and the total
two-pion energy is given by $E_{T}=E_\qu +E_\qd$. ${\cal
A}^{(0)}(E_T)\ (=\lambda_{K\pi\pi})$ is the tree-level component
of the $K\to\pi\pi$ decay amplitude.

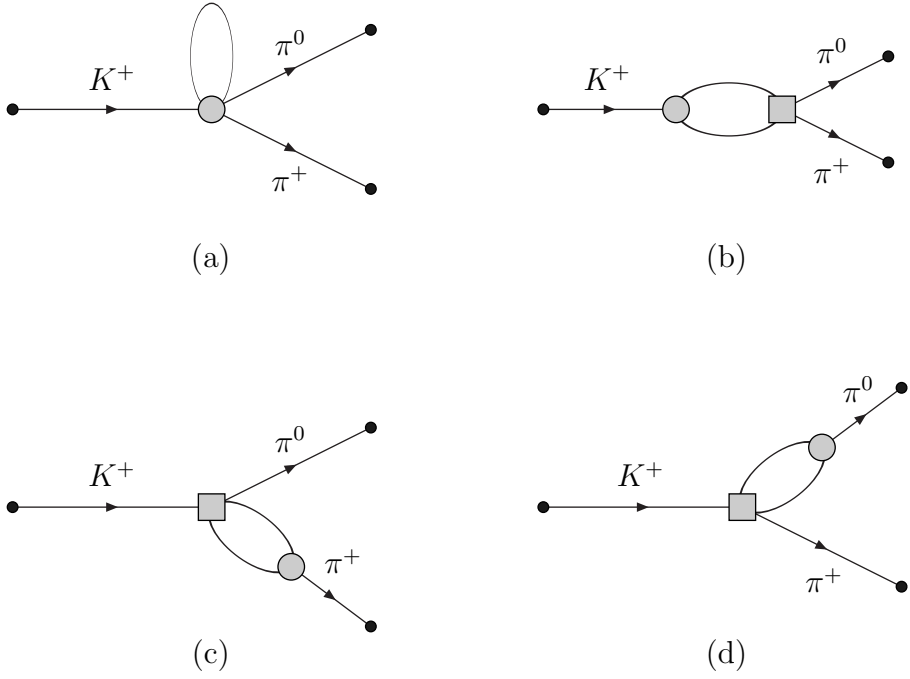
\begin{figure}[t]
\begin{center}
\begin{picture}(400,250)(0,-65)
\ArrowLine(0,150)(75,150)\ArrowLine(75,150)(135,180)
\ArrowLine(75,150)(135,120) \Oval(75,170)(20,8)(0)
\GCirc(75,150){5}{0.8}
\GCirc(0,150){2}{0.1}\GCirc(135,180){2}{0.1}
\GCirc(135,120){2}{0.1}
\Text(37.5,158)[b]{$K^+$}\Text(105,171)[b]{$\pi^0$}
\Text(105,129)[t]{$\pi^+$} \Text(75,100)[t]{(a)}
\ArrowLine(200,150)(250,150) \ArrowLine(290,150)(330,170)
\ArrowLine(290,150)(330,130) \Oval(270,150)(10,20)(0)
\GCirc(200,150){2}{0.1}\GCirc(330,170){2}{0.1}
\GCirc(330,130){2}{0.1} \Text(225,158)[b]{$K^+$}
\Text(310,168)[b]{$\pi^0$}\Text(310,132)[t]{$\pi^+$}
\GCirc(250,150){5}{0.8} \GBoxc(290,150)(10,10){0.8}
\Text(270,100)[t]{(b)}
\ArrowLine(0,0)(75,0)\ArrowLine(75,0)(135,30)
\ArrowLine(105,-22.5)(135,-45) \Oval(90,-11.25)(9,18.75)(323.13)
\GCirc(0,0){2}{0.1}\GCirc(135,30){2}{0.1}\GCirc(135,-45){2}{0.1}
\GBoxc(75,0)(10,10){0.8}\GCirc(105,-22.5){5}{0.8}
\Text(37.5,8)[b]{$K^+$}\Text(105,21)[b]{$\pi^0$}
\Text(125,-25)[b]{$\pi^+$} \Text(75,-50)[t]{(c)}
\ArrowLine(200,0)(275,0)\ArrowLine(275,0)(335,-30)
\ArrowLine(305,22.5)(335,45) \Oval(290,11.25)(9,18.75)(36.87)
\GCirc(200,0){2}{0.1}\GCirc(335,-30){2}{0.1}\GCirc(335,45){2}{0.1}
\GBoxc(275,0)(10,10){0.8}\GCirc(305,22.5){5}{0.8}
\Text(237.5,8)[b]{$K^+$}\Text(320,40)[b]{$\pi^0$}
\Text(307,-23)[t]{$\pi^+$} \Text(270,-50)[t]{(d)}
\end{picture}
\caption{One-Loop Chiral Perturbation Theory Diagrams for the
$K^+\to\pi^0\pi^+$ decay. The grey circle (square) represents weak
(strong) vertices. In addition one must also calculate the
contributions from the diagrams which renormalize the wave
functions of the external mesons. \label{kppgraph}}
\end{center}
\end{figure}

At one-loop order, we insert an interaction vertex (${\cal O}_s=-
\lambda\left|\pi^0(x_s)\right|^2\left|\pi^+(x_s)\right|^2/4$,
where $\pi$ represents the pion field) into
the correlation function, and its contribution is given by the
expression
\begin{equation}
{\cal C}^{(1)}_{\qu \qd}(t_K,t_1,t_2)=\la 0 |\,T[\,\pi_\qu(t_1)
\pi_\qd(t_2) \op_W(0) \, \Bigg[\int{d^4x_s \op_s(x_s)}\Bigg]\,
K_{\vec 0}^\dagger(t_K)\,]\,|0 \ra \label{ClFV}
\end{equation}
corresponding to the Feynman diagram in \fig{kppgraph}(b). In this
case the weak operator is simply given by the triple scalar
vertex,
$\op_W(0)=\lambda_{K\pi\pi}\,K(0)\,\pi^\dagger(0)\,\pi^\dagger(0)$.

In \ChPT, of the four diagrams shown in \fig{kppgraph} (and the
graphs which renormalize the external meson fields) it is only
diagram (b) which contains finite-volume effects which decrease as
inverse powers of the volume. For the remaining diagrams the
finite volume effects decrease exponentially and will be
neglected. It is for this reason that the evaluation of this graph
in the $\lambda \Phi^4$ model illustrates the key properties of
finite volume effects in \ChPT. Note however, that the evaluation
of the graph in \fig{kppgraph}(b) is different in this model and
\ChPT\ (for example both the weak and strong vertices in \ChPT\
depend in general on the momenta).

The contribution from the diagram of fig.~\ref{kppgraph}\,(b) to
the correlation function is:
\begin{eqnarray}
&&\hspace{-0.4in}-\lambda_{K\pi\pi}\,\lambda\frac{e^{-m_K|t_K|}}{2
m_K}\frac{1}{L^3}\int dt_s \frac{e^{-E_{\qu}|t_1-t_s|}}{2 E_{\qu}}
\frac{e^{-E_{\qd}|t_2-t_s|}}{2 E_{\qd}} \sum_{\vec k,\vec l} \int
\frac{dk_4}{2 \pi} \frac{dl_4}{2 \pi}
\frac{e^{-i(k_4+l_4)t_s}}{(k_4^2+w_1^2)(l_4^2+w_2^2)}\delta^3_
{\vec{k}+\vec{l},\,\vec q}\nn \\ &=&-\lambda_{K\pi\pi}\,\lambda
\times \frac{e^{-m_K|t_K|}}{2 m_K} \frac{1}{L^3} \sum_{\vec k,\vec
l}\delta^3_ {\vec{k}+\vec{l},\,\vec q} \int dt_s
\frac{e^{-E_{\qu}|t_1-t_s|}}{2 E_{\qu}}
\frac{e^{-E_{\qd}|t_2-t_s|}}{2 E_{\qd}}
 \frac{e^{-w_1|t_s|}}{2w_1}\frac{e^{-w_2|t_s|}}{2w_2}  \nn\\
&=& -\lambda_{K\pi\pi}\,\lambda \times
\frac{e^{-m_K|t_K|}}{2m_K}\frac{1}{L^3}\frac{1}{16E_\qu
E_\qd}\sum_{\vec k,\vec l}\delta^3_ {\vec{k}+\vec{l},\,\vec q}\
\frac{1}{w_1 w_2}\{T_1+T_2+T_3+T_4 \}\,, \label{eintts}
\end{eqnarray}
where
\begin{equation}
w_1^2=m_1^2+|\vec{k}|^2\,,\hspace{0.25in}
w_2^2=m_2^2+|\vec{l}|^2
\end{equation}
and $\vec{q} = \qu + \qd$. $T_1$\,--\,$T_4$ are the contributions
to the integrations over $t_s$ from the intervals $(-\infty,0)$,
$(0,t_2)$, $(t_2,t_1)$ and $(t_1,\infty)$ respectively and are
given by
\begin{eqnarray}
T_1&=&\frac{e^{-E_\qu t_1-E_\qd t_2}}{E_\qu+E_\qd+w_1+w_2}\,.
\nonumber\rule[-0.15in]{0in}{0.5in}\\
T_2&=&\frac{e^{-E_\qu(t_1-t_2)-(w_1+w_2)t_2}-e^{-E_\qd t_2-E_\qu
t_1}}{E_\qu+E_\qd-(w_1+w_2)} \rule[-0.15in]{0in}{0.5in}\nonumber\\
T_3&=&\frac{e^{-E_\qd
(t_1-t_2)-(w_1+w_2)t_1}-e^{-E_\qu(t_1-t_2)-(w_1+w_2)t_2}}{E_\qu-E_\qd-w_1-w_2}
\rule[-0.15in]{0in}{0.5in}\\ T_4&=&\frac{e^{-E_\qd
(t_1-t_2)-(w_1+w_2)t_1}}{E_\qu+E_\qd+w_1+w_2} \nonumber
\rule[-0.15in]{0in}{0.5in}\ .
\end{eqnarray}

For simplicity we choose $|\qu|=|\qd|$ and take $m_1=m_2$, so that
$\vec q=0$, $w_1=w_2\equiv w$ and $E_\qu=E_\qd\equiv E$. The
generalization to the SPQR kinematics is straightforward in
perturbation theory, but as described in
sec.\,\ref{subsubsec:moving}, the generalization of the LL-factor
to the case where the two pions have a non-zero total momentum has not
been performed yet. Factorizing the tree level correlation
function, we obtain \bea & & {\cal C}^{(1)}_{\qu \qd}(t_K,t_1,t_2)
= -
 {\cal G}^{(0)}(t_K,t_1,t_2,2 E)   \times
\lambda \times \label{eq:csimple}\\ & & \frac{1}{L^3}
\sum_{\vec{k}} \frac{ {\cal A}^{(0)}(2 w)} {(2w)^2} \l
\{\frac{1}{2(E+w)} +\frac{e^{2(E-w)t_2}-1}{2(E-w)}+
\frac{e^{2Et_2-2wt_1}-e^{2(E-w)t_2}}{-2w}+
\frac{e^{2Et_2-2wt_1}}{2(E+w)} \r  \}  \, .
\nn\rule[-0.15in]{0in}{0.5in} \eea Note that in this specific
example the weak amplitude is independent of the energy and thus
${\cal A}^{(0)}(2w)=\lambda_{K\pi\pi}$. The four terms in braces
in eq.~(\ref{eq:csimple}) are $T_1$ -- $T_4$ respectively. For
illustration we now consider the two cases $0 < t_1 = t_2\equiv
t$~\cite{LL} and $0 < t_2 \ll t_1$~\cite{dlinetal,cairns,MT}. In
the first case, $T_{3}=0$ and we have a factor \beq \frac{1}{L^3}
\sum_{\vec{k}} \frac{1}{(2w)^2} \l \{
\frac{e^{2(E-w)t}}{2(E+w)}+\frac{e^{2(E-w)t}-1}{2(E-w)}+\frac{1}{2(E+w)}
\r \}\, .\label{eq:LL} \eeq In the second case, since $t_1 \gg
t_2$ we can neglect $T_4$ since it is exponentially suppressed and
the corresponding factor is \beq \frac{1}{L^3} \sum_{\vec{k}}
\frac{1}{(2w)^2} \l \{
\frac{e^{2(E-w)t_2}}{2w}+\frac{e^{2(E-w)t_2}-1}{2(E-w)}+\frac{1}{2(E+w)}
\r \} \, . \label{eq:dlin} \eeq Here we are considering finite
volume corrections to correlation functions. As anticipated above,
we will show that the finite volume corrections to the amplitudes,
extracted after dividing by the appropriate source factor, are
identical in the two cases, in spite of the differences of the
expressions in \eqs{eq:LL}{eq:dlin}.

We now consider the sum over the momentum $\vec k$ and
separate the contribution corresponding to $w=E$, with the two
pions in the loop on the mass-shell, from the remaining terms.
For the two cases we are studying, the contribution
with $w=E$  is given by
\begin{equation}
\frac{1}{L^3} \frac{\nu}{(2E)^2} \l \{\frac{1}{4E} + t+\frac{1}{4E}
\r \} \ \  (t_1=t_2=t)  \label{eq:t1eqt2}
\end{equation}
and
\begin{equation}
\frac{1}{L^3} \frac{\nu}{(2E)^2} \l \{\frac{1}{2E} + t_2+\frac{1}{4E}
\r \} \ \  (t_1 \gg t_2)\,,\label{eq:t1ggt2}
\end{equation}
where $\nu=\sum_{\vec{k}:w=E}$. The term linear in time in
\eqs{eq:t1eqt2}{eq:t1ggt2}, after resummation, corresponds to a
shift of the two-pion energy due to the $\lambda \Phi^4$
interactions: \beq W= E_{T} + \Delta W =2E+\frac{\lambda \nu}{4L^3
E^2} \, .\label{eshift} \eeq It is straightforward to check that
the energy shift in eq.\,(\ref{eshift}) corresponds to that
obtained from the $O(\lambda)$ term in the perturbative expansion
of the L\"uscher quantization condition~\cite{L1,L2} using the
one-loop expression for the $s$-wave phase-shift.

Of the remaining terms in each of \eqs{eq:t1eqt2}{eq:t1ggt2}, the
first one comes from final-state pion rescattering and will be
cancelled by dividing by the matrix element of the source (this
will be demonstrated explicitly below). These final-state
interactions are different in the two cases $t_1 =  t_2$ and $t_2
\ll t_1$. The third term in each of \eqs{eq:t1eqt2}{eq:t1ggt2} is
part of the (finite volume) one-loop correction to the ``weak''
matrix element and, as expected, is the same in both cases, $t_1 =
t_2$ and $t_2 \ll t_1$.

It is convenient to group the terms with $w\neq E$ as follows:
\bea
\frac{1}{L^3} \sum_{\vec{k}: w \neq E} \frac{1}{(2w)^2} \l \{
 \frac{E}{(E^2-w^2)} e^{2(E-w)t} +\frac{w}{w^2-E^2} \r \} & &
(t_1=t_2=t) \label{eformTM1} \\
\frac{1}{L^3} \sum_{\vec{k}: w \neq E} \frac{1}{(2w)^2} \l \{
 \frac{E}{2w(E-w)} e^{2(E-w)t_2} +\frac{w}{w^2-E^2} \r \} & & (t_1
\gg t_2) \, .\label{eformTM2} \eea In the limit $t_2 \rightarrow
\infty$,  the terms proportional to $e^{2(E-w)t_2}$ with $w>E$ are
exponentially suppressed and only a finite number of contributions
with $w<E$ survive. As pointed out by Maiani and Testa, these
terms dominate the correlation function over the contribution one
is trying to isolate, i.e. the one with $w=E$.
We denote these terms as {\it Maiani-Testa terms}. They can be
eliminated by using the behaviour of the correlator as a function
of $t_2$ (at large $t_2$) and separating the different
exponentials, corresponding to the different
energies~\cite{LL,dlinetal,ciuchinietal}. We assume that this can
be done in the numerical simulations and do not consider these
contributions anymore. Up to exponentially small terms, the final
term in \eqs{eformTM1}{eformTM2} can be written as~\cite{L2}
\begin{equation}
\frac{1}{L^3} \sum_{\vec{k}:w \neq E } \frac{1}{(2w)^2}
\frac{w}{w^2-E^2} =  f(E)+\frac{z_n}{4(2\pi)^2EL}+\frac{\nu}{8E^3
L^3}  \, , \label{ewow} \eeq where $\vec n$ is defined by
$\vec{q}_1 = \frac{2\pi}{L}\vec{n}$,
\begin{equation}
z_n \equiv \sum_{\scriptsize{\begin{array}{c} |\vec{\imath}\,|
\neq|\vec n|\\ \vec{\imath}\in Z^3\end{array}}} \frac{1}
{(\vec{\imath}^{\,\,2}-\vec{n}^{\,2})} \,,
\hspace{0.2in}\textrm{and}\hspace{0.2in} f(E)={\cal
P}\int_{-\infty}^{\infty}\frac{d^{\,3}k}{(2\pi)^3}\,\frac{1}{4w}\,
\frac{1}{w^2-E^2}\,.\label{eq:fdef}
\end{equation}

It is instructive to rewrite $f(E)$ in a form which makes clear
the connection to the one-loop contribution to the physical
$K\to\pi\pi$ amplitude. We write $f(E)=$Re$\{-if_M(E)\}$ where
\begin{equation}
f_M(E)\equiv\int\,\frac{d^{\,4}k}{(2\pi)^4}\,\frac{1}{(k^2-m^2+i\varepsilon)
((q-k)^2-m^2+i\varepsilon)}\label{eq:fmdef}
\end{equation}
is the relevant integral in Minkowski space (hence the subscript
$M$). The relation $f(E)=$ Re$\{-if_M(E)\}$ can be readily
verified using contour integration in $k_0$-space. The real part
of the $K\to\pi\pi$ amplitude in Minkowski space is proportional
to $\textrm{Re}\{1+i\lambda\, f_M(E)\}=1-\lambda f(E)$, so that
one correctly obtains the real part of the Minkowski amplitude
from the Euclidean correlation function, in agreement with the
general arguments of ref.~\cite{MT}. (At this order in
perturbation theory the modulus of the amplitude is equal to its
real part.) We can also rewrite $f(E)$ as a one-loop four
dimensional integral in Euclidean space, $f(E)$=Re$\{f_E(iE)\}$,
where
\begin{equation}\label{eq:fedef}
\textrm{Re} f_E(iE) =
\textrm{Re}\left\{\int\,\frac{d^{\,4}k}{(2\pi)^4}\,\frac{1}{(k^2+m^2)
((q-k)^2+m^2)}\right\}_{q^2=-4E^2}\,.
\end{equation}
After the integration has been performed in Euclidean space, the
substitution $q^2=-4E^2$ is made.

From the above discussion we conclude that the correlation
function, at one-loop order in perturbation theory and keeping
only the terms with the required dependence on $t_{2}$, can be
written in the form (for the two cases being considered):
\bea
\ {\cal C}_{\qu -\qu}(t_K,t,t) &=& {\cal A}^{(0)}(W) \times {\cal
G}^{(0)}(t_K,t,t,W)   \times \label{eFVcor1} \\
&&\hspace{-1in}\l\{ 1- \lambda \l[ \Re
f_E(iW)+\frac{z_n}{4(2\pi)^2EL}+
\frac{\nu}{L^3} \frac{3}{16E^3} + \frac{\nu}{L^3} \frac{1}{16E^3}\r
]  \r\}
+ {\cal O}(\lambda^{2}) \quad\quad  (t_1=t_2=t) ,
\nn\eea
and
\bea {\cal C}_{\qu -\qu}(t_K,t_1,t_2) &=&
{\cal A}^{(0)}(W) \times {\cal G}^{(0)}(t_K,t_1,t_2,W) \times
\label{eFVcor2}\\
&&\hspace{-1in}\l\{ 1- \lambda \l[ \Re
f_E(iW)+\frac{z_n}{4(2\pi)^2EL}+
\frac{\nu}{L^3} \frac{3}{16E^3} + \frac{\nu}{L^3} \frac{1}{8E^3}\r
]  \r\}
+ {\cal O}(\lambda^{2}) \quad \quad  (t_1\gg t_2) \, .
\nn\eea
Although, in this case, the tree-level amplitude does not depend
on the energy, we  have nevertheless written it as ${\cal A}^{(0)}(W)$
in \eqs{eFVcor1} {eFVcor2}, in order to be able to generalise the
discussion to \ChPT in the following sections.

In order to determine the $K\to\pi\pi$ decay amplitudes we need to
divide the correlation functions by the appropriate factors
associated with the source (kaon) and the sink (two pions). For
single particle states this is straightforward, and there are no
finite volume power corrections,
\begin{equation}
\la 0\,|K_{\vec 0}(0) K_{\vec 0}^\dagger(t_K) |\,0\ra =
L^3\,Z_K^2\frac{e^{-m_K |t_K|}}{2m_K}\ ,
\label{eq:kk}\end{equation} where the subscript $\vec 0$ indicates
that the Fourier transform of the kaon field has been performed at
zero momentum.

The determination of the factors $\la 0|\pi\pi|n\ra$ associated
with the two-pion intermediate states $|n\ra$ is more subtle.
These matrix elements can be obtained by evaluating four-pion
correlation functions. However, one has to be careful to project
out the matrix element corresponding to the required state
$|n\ra$, which in the present case is a two-pion, cubically
invariant state which contains an s-wave component. Although one
can, in principle, try to use the behaviour of the correlation
function with $t_2$ to project out the state with the correct
energy, this may be difficult in practice if this state has an
energy close to one with no s-wave component. For this reason, it
is useful to project out the contribution of the required state
directly, by replacing exponential factors of the form $\exp(i\vec
k\cdot\vec x)$ in Fourier transforms by their s-wave projections,
$\sin(kr)/kr$ ($k=|\vec k|$ and $r=|\vec x|$). For our purposes it
will be sufficient instead, to average the correlation function
${\cal C}_{\vec{q}_1 -\vec{q}_1}$ in \eqs{eFVcor1}{eFVcor2} over
all $\vec q_1$'s which can be transformed into each other by
elements of the cubic group (in the formulae below we actually
choose to average over the $\nu$ momenta with the same modulus
$|\vec{q}_1|$). Such an average has a projection on the s-wave,
but no projection onto $l=1,2$ and 3 components of the states.
Since at one-loop order, both in the $\lambda\Phi^4$ theory and in
\ChPT, $l=4$ components (and higher partial waves) cannot be
generated, such an average over cubically equivalent states is
sufficient. We now carry out this procedure explicitly.

We start by evaluating the averaged correlation function (see
\eq{eq:first})
\begin{eqnarray}
\sum_{\vec{q}_1}C_{\vec{q}_1\,-\vec{q}_1}(t_K,t_1,t_2) & =
&\sum_{\vec{q}_1}\sum_{n,m}\ \la\,
0\,|\,\pi_{\vec{q}_1}(t_1)\,\pi_{-\vec{q}_1}(t_2)\,|\,n\,\ra
\,\la\, n\,|\,{\cal O}_W(0)\,|\,m\ra\,\la
m\,|\,K^\dagger(t_K)\,|\,0\,\ra\nn\\
&&\hspace{-1.5in}=\frac{Z_K}{2m_K}e^{-m_K|t_K|}e^{-Wt_2}
\sum_{\vec{q}_1} \la\,
0\,|\,\pi_{\vec{q}_1}(t_1-t_2)\,\pi_{-\vec{q}_1}(0)\,|\,W\,\ra
\,\la\, W\,|\,{\cal O}_W(0)\,|\,K\ra\, + \cdots\ ,
\label{eq:cav}\end{eqnarray} where $|W\ra$ represents the
cubically invariant two-pion state with energy $W$ which contains
an s-wave component. As explained above, the sum over $\vec{q}_1$,
runs over all the three-momenta with the same modulus,
$|\vec{q}_1|$. The ellipses in \eq{eq:cav} represent terms with a
different behaviour in $t_2$, and as always, we assume that we can
use the time dependence to isolate the first term on the left hand
side.

In order to obtain the $K\to\pi\pi$ matrix element from the
correlation function in \eq{eq:cav}, we need to divide by the
matrix element
\begin{equation}
\sum_{\vec{q}_1}
\la\,
0\,|\,\pi_{\vec{q}_1}(t_1-t_2)\,\pi_{-\vec{q}_1}(0)\,|\,W\,\ra\ .
\label{eq:sink}\end{equation}
We can determine this matrix element from the four-pion correlation
function:
\begin{eqnarray}
\sum_{\vec{q}_1}\sum_{\vec{p}}\
\la\,0\,|\,\pi_{\vec{q}_1}(t_1)\,\pi_{-\vec{q}_1}(t_2)
\,\pi^\dagger_{-\vec{p}\,}(-t_2)\,\pi^\dagger_{\vec{p}\,}(-t_1)\,|\,0\,\ra\nn\\
&&\hspace{-4in}=e^{-2Wt_2}\sum_{\vec{q}_1}\sum_{\vec{p}}\
\la\,0\,|\,\pi_{\vec{q}_1}(t_1-t_2)\,\pi_{-\vec{q}_1}(0)\,|\,W\,\ra
\,\la\,W\,|\pi^\dagger_{-\vec{p}}(0)\,\pi^\dagger_{\vec{p}}(-(t_1-t_2))\,|\,0\ra\,+\cdots\
, \label{eq:fourpion}\end{eqnarray} where the sum over $\vec p$ is
over the same range as the one over $\vec q_1$. Evaluating the
four-pion correlation function in perturbation theory we find:
\bea \sum_{ \vec{p},\vec{q_1}} \la\,
0\,|\pi_\qu(t_1)\pi_{-\qu}(t_2)\pi_{-\vec{p}\,}^{\dag}(-t_2)\pi_{\vec{p}\,}^{\dag}
(-t_1)|0\ra&=&\l \{ \barr{c c} \nu\frac{e^{-2
Wt_2}}{(2E)^2}L^6(1-\frac{\lambda \nu}{8E^3L^3}) & (t_1=t_2) \\
\hfill\textrm{and}&\\ \nu\frac{e^{-2
Wt_2-2E(t_1-t_2)}}{(2E)^2}L^6(1-\frac{\lambda \nu}{4E^3L^3}) &
(t_1 \gg t_2)\earr \r. \label{eq:4pioncf}\eea Combining the
results in \eq{eq:4pioncf} with those for the $K\to\pi\pi$
correlation functions in \eqs{eFVcor1}{eFVcor2} we find that the
one-loop contribution to the amplitude in both cases is
\begin{equation}
\frac{\la\, W\,|\,{\cal O}_W(0)\,|\,K\ra}{\sqrt{\nu}}=
{\cal A}^{(0)}(W)\l\{ 1- \lambda \l[ \Re
f_E(iW)+\frac{z_n}{4(2\pi)^2EL}+ \frac{\nu}{L^3}
\frac{3}{16E^3}\r]  \r\}\ . \label{eq:ampfv}\end{equation} %
The finite volume corrections in \eq{eq:ampfv} are precisely those
given by the Lellouch-L\"uscher factor (see
subsec.~\ref{subsubsec:ratios}). We stress that in the
perturbative example studied in this section we use the
conventional covariant normalization of states. The remaining
diagrams in \fig{kppgraph} also contribute to the infinite volume
result, but they do not have corrections which decrease as inverse
powers of the volume.

The above argument is a simple example of the discussion presented
in ref.~\cite{cairns}, in which it was shown that from the
correlation functions \eq{eq:first}, after dividing by the matrix
elements corresponding to the source and the sink, one obtains the
modulus of the amplitude, up to finite-volume corrections given by
the Lellouch-L\"uscher factor. The case $t_2\ll t_1$ was analysed
in ref.~\cite{dlinetal} where it was shown that the correlation
function \eq{eq:first} gives the real part of the decay amplitude
up to power corrections in the volume which are not those
described by the Lellouch-L\"uscher factor. This is before
dividing by the two-pion matrix element \eq{eq:sink}. Division by
this matrix element removes a factor of $\cos(\delta)$ (where
$\delta$ is the s-wave phase-shift), turning the real part of the
amplitude into its modulus, and modifies the finite-volume
corrections into the Lellouch-L\"uscher factor~\cite{cairns}. The
above exercise is an explicit one-loop demonstration of this
effect, although at this order of perturbation theory the real
part and modulus of the amplitude are equivalent
($\cos(\delta)=1$).

The remaining diagrams in fig.~\ref{kppgraph} do not have
singularities in the range of integration and hence only have
exponential corrections in the volume (i.e. terms which decrease
like, for example, $\exp(-mL)$). To illustrate this feature
consider the complete contribution from $T_1$ in \eq{eintts},
which is proportional to
\begin{equation} \frac{1}{L^3} \sum_{\vec{q}} \frac{1}{(2w)^2}
\frac{1}{2(w+E)} = \int \frac{d^3q}{(2\pi)^3} \frac{1}{(2w)^2}
\frac{1}{2(w+E)}+{\cal O}(e^{-m L})\, . \eeq

\subsubsection{Evaluation of the Decay Amplitude}
\label{subsubsec:ratios}

The one-loop toy-model study in \sec{skpplphi4} is an explicit
illustration of the general procedure for evaluating
infinite-volume amplitudes from correlation functions computed in
a finite volume. Before proceeding to develop the discussion of
\sec{skpplphi4} in the context of chiral perturbation theory, we
briefly summarise this general procedure for the determination of
the decay amplitudes. For illustration we consider the case where
the kaon is at rest and the total momentum of the two-pions is
also zero (see however sec.~\ref{subsubsec:moving}).
The procedure is as follows:
\begin{enumerate}
\item We evaluate the correlation function ${\cal C}_{\qu
-\qu}(t_K,t_1,t_2)$ (defined in \eq{eq:first}) averaged over all
$\qu$ with the same modulus. We envisage evaluating the matrix
element with the two-pion energy equal to $W=2E_{q_1}$ +
finite-volume corrections. Let ${\cal C}_{\qu
-\qu,W}(t_K,t_1,t_2)$ be the component of ${\cal C}_{\qu
-\qu}(t_K,t_1,t_2)$ whose behaviour with $t_2$ is given by
$\exp(-Wt_2)$.
\item We evaluate the four-pion correlation function (see
\eq{eq:fourpion})
\begin{equation}
{\cal C}^{4\pi}_{\qu \qd}(t_1,t_2)=
\la\,0\,|\,\pi_{\qu}(t_1)\,\pi_{-\qu}(t_2)
\,\pi^\dagger_{-\qd}(-t_2)\,\pi^\dagger_{\qd}(-t_1)\,|\,0\,\ra\, ,
\label{eq:fourpiongen}\end{equation} where again we average over
all $\qu$ and $\qd$ with modulus equal to $|\qu|$. As above, let
${\cal C}^{4\pi}_{\qu \qd,W}(t_1,t_2)$ be the component of ${\cal
C}^{4\pi}_{\qu \qd}(t_1,t_2)$ whose behaviour with $t_2$ is given
by $\exp(-2Wt_2)$.
\item Finally we calculate the kaon propagator,
${\cal C}^K_{\vec 0}(t_K)$
\begin{equation}
{\cal C}^K_{\vec 0}(t_K)\equiv\la 0|K_{\vec
0}(-t_K)K^\dagger_{\vec 0}(t_K)\,|0\ra
\end{equation}
for sufficiently large $|t_K|$ so that the correlation function is
dominated by the kaon-state.
\item The finite-volume matrix elements are given by:
\begin{equation}
\l|\,_V\la \pi\pi|\op_W(0)|K\ra_V\,\r|=\frac{{\cal C}_{\qu
-\qu,W}(t_K,t_1,t_2)} {\rule{0pt}{14pt}\sqrt{{\cal C}^{4\pi}_{\qu
\qd,W}(t_1,t_2)\,{\cal C}^K_{\vec 0}(t_K)}}\,,
\label{eq:ratios}\end{equation} where the subscripts $V$ on the
matrix element denotes that it is the finite-volume matrix element
(with the finite-volume normalization). Note that the $K\to\pi\pi$
matrix elements discussed in sec.~\ref{skpplphi4} in general and
eq.~(\ref{eq:ampfv}) in particular, which were presented using the
covariant normalization, should be divided by $\sqrt{8m_KE^2V^3}$ to
translate them into the finite-volume normalization in
eq.~(\ref{eq:ratios}).
\item \label{item:ll} To obtain the infinite-volume matrix elements
$\la\pi\pi|\op_W(0)|K\ra$ (with the conventional covariant
normalization) from the finite-volume ones, we multiply the latter
by the LL factor~\cite{LL,dlinetal}~\footnote{In
$K^+\to\pi^+\pi^0$ decays the final state particles are
distinguishable  and so the factor in \eq{eq:llfactor} differs by
a factor of 2 compared with the corresponding one presented for
indistinguishable final state particles~\cite{LL,dlinetal}.}
\begin{equation}
|\la\pi\pi|\op_W(0)|K\ra|^2= 4\pi
V^2\{q\phi^\prime(q)+k\delta^\prime(k)\}\l(\frac{m_K}{k}\r)^3\,
|\,_V\la \pi\pi|\op_W(0)|K\ra_V\,|^2\,,
\label{eq:llfactor}\end{equation} where $k=\sqrt{W^2/4-m_\pi^2}$,
$q=kL/(2\pi)$, $\delta(k)$ is the s-wave (and in our case $I=2$)
$\pi$-$\pi$ phase-shift and the geometrical function $\phi(q)$ is
given by
\begin{equation}
\tan(\phi(q))=-\frac{2\pi^2q}{\sum_{\,\vec{\imath}\in
Z^3}1/(\vec{\imath}^{\,2}-q^2)}\,. \label{eq:phidef}\end{equation}
\end{enumerate}

This procedure holds for the matrix elements discussed in this
paper in full QCD and also in the quenched approximation (this is
not the case in the quenched approximation for $\Delta I=1/2$
transitions~\cite{dt12quenched}). The example in \sec{skpplphi4}
is a model one-loop demonstration of the validity of the
procedure.

At present the LL factor has not been generalized to the situation
in which the two-pion final state has a non-zero total
three-momentum, and so we are unable to carry out step
\ref{item:ll}. We now discuss this point in a little more detail.

\subsubsection{Finite-Volume Corrections in a Moving Frame}
\label{subsubsec:moving}

In order to implement our strategy, which requires the computation
of matrix elements at unphysical kinematics in general and SPQR
kinematics in particular, simulations have to be performed with
the two pions having non-zero total three-momentum. Thus, if the
calculations are to be precise up to exponential corrections in
the volume, we need the generalization of the L\"uscher
quantization condition and the LL factor away from the
centre-of-mass frame. The quantization condition in a moving frame
has been presented in ref.~\cite{RuGo}, whereas the generalization
of the LL-factor is being studied and we will report the results
in a future publication. We anticipate that the generalization of
the LL-factor will be obtained from the quantization condition in
an analogous way to the derivation in the centre-of-mass frame
given in ref.~\cite{dlinetal}.

At this point we are able to check whether the energy shift
obtained in perturbation theory following the procedure of
sec.~\ref{skpplphi4} generalized to a moving frame, agrees with
the expansion of the quantization condition presented in
ref.\,\cite{RuGo}. This is indeed the case. Repeating steps
(\ref{eq:csimple}) -- (\ref{eshift}) for two particles with
energies $E_1$ and $E_2$ we find that the energy shift is equal to
$\lambda\nu^d/(4E_1E_2L^3)$, where $\nu^d$ is the number of terms
in the momentum sum with energy equal to $E_1+E_2$. This agrees
with the result obtained by expanding the quantization condition
in ref.~\cite{RuGo} and using the one-loop expression for the
phase-shift~\footnote{We thank Kari Rummukainen for helping us to
clarify this point.}.

We postpone the general discussion of finite-volume corrections at
non-zero total momentum until the corresponding LL-factors have
been derived and now proceed to study one-loop chiral perturbation
theory.

\subsection{$K\to\pi\pi$ Decays in \ChPT on a Finite Volume}
\label{subsec:kppchptfv}

In this section we generalise the previous discussion to \ChPT. In
this case the weak amplitude in general depends on the energy, and
hence other finite-volume correction terms appear in the
correlation function. To illustrate this, consider the term in
\eq{eintts} which contains $T_2$, but now generalized to allow for
the energy dependence in the lowest order weak and strong
amplitudes (the power corrections arise from singular terms as
explained in the previous section):
\begin{equation}
{\cal C}={\cal G}^{(0)}(t_K,t_1,t_2,2E) \l\{{\cal
A}(2E)-\frac{1}{L^3} \sum_{\vec{k}}
\frac{e^{2(E-w)t_2}-1}{2(E-w)}{\cal A}(2w) {\cal N}(w)\r\}
\end{equation}
where ${\cal A}(2E)$ is the tree level amplitude and ${\cal N}(w)$
is the strong interaction amplitude (in the model of
\sec{skpplphi4} ${\cal N}(w)=\lambda/4w^2$). Separating the
contribution with $w=E$ from the rest, we find
\begin{eqnarray}
{\cal C}&=&{\cal G}^{(0)}(t_K,t_1,t_2,2E) \l\{ {\cal A}(2E)\l[1-
\frac{\nu {\cal N}(E) }{L^{3}} t_{2}\r] -\frac{1}{L^3}
\sum_{\vec{k}:w\neq E} \frac{e^{2(E-w)t_2}-1}{2(E-w)}{\cal
A}(2w){\cal N}(w)\r\}\nn\\
&=& {\cal G}^{(0)}(t_K,t_1,t_{2},W) \l\{ {\cal A}(2E)
-\frac{1}{L^3} \sum_{\vec{k}:w\neq E}
\frac{g(w)}{2(w^{2}-E^{2})}+... \r\} \label{pretriu}
\end{eqnarray}
where
\begin{equation}
g(w)=(w+E){\cal A}(2w){\cal N}(w)\,,
\end{equation}
and the ellipses represent terms which have a different $t_2$
behaviour (as above, we assume that we can isolate the terms with
the required behaviour in $t_2$).

We now use the summation formula (for smooth functions $F$),
\begin{equation}
\frac{1}{L^3}{\sum_{\vec
k}}^{\,\prime}\frac{F(k^2)}{K^2-k^2}={\cal P}\int\frac{d^3
k}{(2\pi)^3}\,\frac{F(k^2)}{K^2-k^2}-\frac{z_K}{4\pi^2L}F(K^2)
+\frac{\nu}{L^3}F^\prime(K^2)\,,
\label{eq:summation}\end{equation} where the prime on the
summation indicates the omission of the $\nu$ terms with $k^2=K^2$
and $F^\prime(K^2)$ denotes the derivative of $F$ with respect to
$k^2$, evaluated at $k^2=K^2$. \Eq{ewow} is an example of this
summation formula with $F(k^2)=-1/(4w)$ and $w^2=k^2+m^2$.
Applying \eq{eq:summation} to \eq{pretriu} we find
\begin{eqnarray}
{\cal C}&=&{\cal G}^{(0)}(t_K,t_1,t_{2},W) \l\{ {\cal A}(2E)
\l(1-h(E)\r)+\frac{\nu{\cal N}(E)}{2L^3}\left.\frac{\partial{\cal
A}}{\partial w}\right|_{w=E}\ \r\}\nn\\
&&\hspace{-.3in}-{\cal G}^{(0)}(t_K,t_1,t_{2},W) {\cal A}(2E)\l\{
E{\cal N}(E)\frac{z_n}{4\pi^2L}-\frac{\nu}{4EL^3}
\left.\frac{\partial\{(w+E){\cal N}(w)\}}{\partial
w}\right|_{w=E}\ \r\}\,, \label{eq:triu}\end{eqnarray} where
$-h(E){\cal A}(2E)$ is the one-loop correction to the
infinite-volume amplitude with
\begin{equation}
h(E)=\frac{1}{2} {\cal P}
\int\frac{d^3k}{(2\pi)^3}\frac{g(w)}{w^2-E^2}\,.
\end{equation}
The one-loop results for the $\Delta I =3/2$ matrix elements of
$\op_{4,7,8}$ are presented in detail in Appendices~\ref{sec:full}
and \ref{sec:quenched} and on the web-site~\cite{quenchedfull}. We
now look at the finite-volume corrections on the right-hand side
of \eq{eq:triu}. The term proportional to $\partial{\cal A}/
\partial w$ has the r\^ole of shifting the argument of the
amplitude from $2E$ to $W=2E+\nu{\cal N}(E)/L^3$. In an earlier
study this term had been misinterpreted as a genuine finite-volume
correction, whereas it is present to shift the argument of ${\cal
A}$. For this reason the authors of ref.~\cite{GL2} concluded that
there are different finite-volume corrections for the two cases
when $m_K=m_\pi$ and $m_K=2m_\pi$, with the two pions at rest.
These corrections must instead be the same since they only depend
on the final state. Our explicit calculations, presented in
\sec{sec:kppFV} below, demonstrate that this is indeed the case.

The second line on the right-hand side of \eq{eq:triu} is a
universal correction which depends only on strong interactions and
gives the Lellouch-L\"uscher factor. This concludes our
demonstration of finite-volume effects in one-loop \ChPT.

\subsection{$K^{+} \rightarrow \pi^{+} \pi^{0}$ Decays
in \ChPT at Finite Volume}
\label{sec:kppFV}

We now briefly summarize the results obtained in full QCD in
finite volume for the physical amplitudes, i.e. for a kaon at rest
decaying into two pions with the insertion of a weak operator at
zero four-momentum transfer. Of course, our approach requires us
to perform lattice simulations (quenched and unquenched), with the
insertion of non-zero momentum at the weak operator and the
corresponding results will be presented below in \sec{sec:SPQR}.

For the calculation of the correlation functions needed to
determine the $K^{+} \rightarrow \pi^{+} \pi^{0}$ decay amplitude
at one-loop in \ChPT we need the following propagators:
\begin{eqnarray}
\int d^3x\, e^{i \vec{q}\cdot(\vec{x}-\vec{y})} \la\,
\phi(\vec{x},t_x)\, \phi^\dagger(\vec{y},t_y)\,
\ra&=&\frac{e^{-w_{\vec{q}}|t_x-t_y|}}{2w_{\vec{q}}}  \\ \int
d^3x\,e^{i \vec{q}\cdot(\vec{x}-\vec{y})} \,\partial_{\mu}^{y}
\la\, \phi(\vec{x},t_x)
\phi^\dagger(\vec{y},t_y)\,\ra&=&\frac{e^{-w_{\vec{q}}|t_x-t_y|}}{2w_{\vec{q}}}\l(i
q_{j}\delta_{j \, \mu}+\delta_{\mu \, 4}w_{\vec{q}}\,
\varepsilon(t_x-t_y) \r) \nn \eea where
$w^{2}_{\vec{q}}=m^{2}_{\pi}+|\vec{q}|^{2}$, $\varepsilon$
represents the sign function, and the index $j$ runs over the
spatial components $1,2$ and $3$.

For the evaluation of the tadpole diagrams we need the relations:
\begin{eqnarray}
\int \frac{d^4 k}{(2\pi)^4}
\frac{1}{k^2+m^2}&=&\frac{1}{L^3}\sum_{\vec{q}}\frac{1}{2w_{\vec{q}}}\
\ \textrm{and}
\nn \\ \int \frac{d^4 k}{(2\pi)^4}
\frac{k^2}{k^2+m^2}&=&\frac{1}{L^3}\sum_{\vec{q}}\l(
\int\frac{dk_4}{2\pi}-\frac{m^2}{2w_{\vec{q}}}\r),
\end{eqnarray}
which are valid up to terms which are exponentially small in the
volume~\footnote{The divergent terms proportional to $\int
dk_4/(2\pi)$ cancel when all contributions to the amplitude are
summed.}.

When evaluating the diagrams  $(b)$, $(c)$ and $(d)$  we follow
similar steps to those in  \eq{eintts}. In \ChPT a number of
integrals appear with different numerators. The complete set of
these integrals and the corresponding results, obtained upon
integration over $k_4$ and $l_4$, are given in
Appendix~\ref{saFV}.

The result for the matrix element of $\op_4$, extracted from the
correlation function with the kaon at rest and with $W=m_K$, where
$W$ is the finite-volume two-pion energy, is given by the
expression:
\begin{eqnarray}
\la\pi^+\pi^0|\op_4|K^+\ra & = &-
\frac{6\sqrt{2}\alpha^{(27,1)}}{f_K\,f_\pi^2}\l[(m_K^2-m_\pi^2)\Bigl[1+
\frac{m_K^2}{(4\pi f)^2}\Bigl(I_{zf}+F(\sqrt{s}
L)\Bigr)\Bigr]\r.\nn\\
&&\hspace{0in}\l.+\frac{m_K^4}{(4\pi)^2f^2}\l(I_{a}+I^{E}_{b}
+I_{c+d}\r)\r]\,+\,O(m_{K,\pi}^4)\textrm{ counterterms} \, ,
\end{eqnarray} where $\sqrt{s}$ is the invariant mass of the two-pion
final state. In this case $\sqrt{s}=m_K$, however since the
finite-volume corrections depend only on the energy of the
two-pion state, we write the answer in terms of $\sqrt{s}$. The
functions $I_{zf,a,b,c+d}$ are defined in \eqs{eq:izdef}{eq:isdef}
below and $I_b^E=$\,Re\,$I_b$. $\alpha^{(27,1)}$ is the leading
order low-energy constant, and new constants appear at order
$m_{\pi,K}^4$. As explained in the introduction, in our approach
it is envisaged that all the low energy constants up this order
will be determined by studying the behaviour of the matrix element
as a function of masses and momenta. $F(\sqrt{s} L)$ is the
centre-of-mass finite volume correction
\begin{equation}
F(\sqrt{s}L)=-\frac{2z_n}{\sqrt{s}L}\l(1-2\frac{m_\pi^2}{s}\r)-\frac{8\pi^2
\nu}{s^{\,3/2} L^3}\l(1-6\frac{m_\pi^2}{s}\r) \,. \label{ecorrkpp}
\end{equation} The shift in the energy of
the two-pion state due to the finite volume is given by
\begin{equation}
\Delta W=W-2E=\frac{\nu \l(1-2\frac{m_\pi^2}{s}\r)}{f^2 L^3}\, .
\label{eq:wexpr}\end{equation} From \eq{eq:wexpr} with the pions
at rest, using $\Delta W=-4\pi a_0^{I=2}/(m_\pi L^3)$~\cite{L2},
we can obtain the infinite-volume $I=2$, $\pi-\pi$ scattering
length
\begin{equation}
a_0^{I=2}=-\frac{m_\pi}{8\pi f^2}\ .
\end{equation}

We remain now with the limit $m_{K} \rightarrow 2m_{\pi}$,
corresponding to the two pions at rest, and compare our results
with those of ref.~\cite{GL2}. In this case $2 z_n\rightarrow 2
z_0 = -17.82726584$ and $\nu=1$ so that the finite volume
correction becomes
\begin{equation}
F(m_K L)=\frac{17.827}{4 m_\pi L}+\frac{\pi^2 }{2 m_\pi^3 L^3} \,.
\label{eLLfact} \eeq This result is in agreement with the general
LL formula, whereas it disagrees with the result in
ref.~\cite{GL2} (see eq.(3.3) in \cite{GL2}), where the
corresponding correction was found to be:
\begin{equation}
\frac{17.827}{4 m_\pi L}+\frac{5\pi^2 }{2 m_\pi^3 L^3} \,.
\end{equation}
We have explicitly verified that the discrepancy is due to the
fact that in ref.~\cite{GL2} it was not recognized that part of
the $1/L^{3}$ finite-volume correction is proportional to
${\partial {\cal A}}/{\partial w}$ and is absorbed by the shift of
the argument of the amplitude, ${\cal A}(2E) \to {\cal A}(W)$ (see
the discussion in section~\ref{subsec:kppchptfv} above). This is
also the origin for the erroneous conclusion of these authors that
the finite-volume effects are different for the two-pions at rest
with $m_K=2m_\pi$ and $m_K=m_\pi$~\cite{GL,GL2}.

For the electroweak penguin operators $\op_7$ and $\op_8$ we
obtain
\begin{eqnarray}
\la\pi^+\pi^0|\op_{7,8}|K^+\ra & =&
\frac{2\sqrt{2\,}\gamma^{(8,8)}}{f_K\,f_\pi^2}\l[1+\frac{m_K^2}{(4\pi
f)^2}(I_{zf}+F(\sqrt{s}
L))+\frac{m_K^2}{(4\pi)^2f^2}\l(J^{E}_a+J^{E}_b+J^{E}_{c+d}\r)\r]
\nn\\ && \hspace{1.5in}+\,O(m_{K,\pi}^2)\textrm{ counterterms}\, ,
\end{eqnarray} where in this case $\sqrt{s}=m_K$, $\gamma^{(8,8)}$
are the leading-order low-energy constants (for $\op_7$ and
$\op_8$) and the functions $J^{E}_i$ are the real part of the
functions $J_i$ defined  in \eq{eAmPhys78}. As expected, the
finite volume corrections encoded in $F(\sqrt{s} L)$ are the same
as for $\op_4$.

For $K\to\pi\pi$ matrix elements for which the two pions have zero
total momentum, the finite-volume corrections take the universal
form
\begin{equation}
\la\pi^+\pi^0|\op_{W}|K^+\ra= \textrm{LO}\,\l(1+\frac{s}{(4\pi
f)^2} F(\sqrt{s}L)+\cdots\r)\,,
\end{equation}
where LO is the lowest order contribution and the ellipses
represent NLO infinite-volume terms. For $\Delta I=3/2$
transitions this is also true for the quenched approximation,
since the relevant one-loop contributions are as above. For
$\Delta I=1/2$ transitions this is not the
case~\cite{GP,dt12quenched}. In the following section and in
appendices~\ref{sec:full} and \ref{sec:quenched} we present the
results also for unphysical kinematics at one-loop order in \ChPT
without explicitly exhibiting the finite-volume effects. As
discussed in sec.\,\ref{subsubsec:moving} the general theory of
finite-volume corrections to matrix elements with the two-pions at
non-zero total momentum has not yet been developed. We are
currently investigating this question, and will present the
perturbative finite-volume corrections within the general
framework.

\section{One-loop corrections in \ChPT}\label{sec:logs}

In the preceding section, we have demonstrated that, at one-loop
order in chiral perturbation theory, the finite-volume corrections
to the amplitudes extracted from correlators in which the pions
are annihilated at the same time or at different times are exactly
the same and correspond to the general LL formula. As explicit
examples, we have given the one-loop amplitudes for the $\Delta
I=3/2$ operators, $\op_{4}$, $\op_{7}$ and $\op_{8}$, in the
physical case (i.e. with no momentum insertion at the weak
operator), for a kaon at rest decaying into two pions in full QCD.
In order to implement the strategy described in \sec{sec:strategy}
however, we also need the amplitudes in other kinematical
situations for which there is an insertion of momentum at the weak
operator. In this section we discuss the evaluation of the
one-loop corrections in the chiral expansion, both in full QCD and
in the quenched approximation for arbitrary meson masses and
momenta. The general results are very long and complicated: for
full QCD they are presented in appendix~\ref{subsec:generalfull}
(and also on the web site~\cite{quenchedfull}) and for the
quenched approximation on the web site~\cite{quenchedfull}. In
presenting our results we do not exhibit explicitly all the NLO
counterterms, i.e. the terms proportional to the $\beta_i$'s in
sec.~\ref{subsec:chiralo4gen} and the $\delta_i$'s in
sec.~\ref{subsec:chiralo78gen}. Of course these terms must be
included in any comparison of lattice data with \ChPT.

We now consider separately the following special cases, which are
useful for current numerical simulations:
\begin{enumerate}
\item  \textit{physical kinematics} with the kaon at rest and with the
mesons having arbitrary masses $m_{K}$ and $m_{\pi}$. By
``physical" here we mean that there is no insertion of momentum at
the weak vertex, so that the two pions each move with momentum
$k=\sqrt{{m_K^2}/{4}-m_\pi^2}$. The results for this case are
presented fully in \sec{subsec:physical} below;
\item \textit{SPQR kinematics} with the kaon and one of the final-state
pions at rest and the other one moving with arbitrary energy
$E_\pi$. The final state is symmetrized to ensure that it
 has $I=2$. The results are presented in
appendix~\ref{subsec:spqrfull} for full QCD and
appendix~\ref{subsec:spqrquenched} for the quenched theory;
\item choices of the quark masses such that $m_K=m_\pi$ and $m_K=2m_\pi$,
in both cases with all the particles at rest. These are presented
in sections~\ref{subsubsec:mkeq2mpifull} --
\ref{subsubsec:mkeqmpiquenched}.
\end{enumerate}

Below we follow the notation introduced in the preceding sections
and assume $SU(2)$ isospin symmetry ($m_s \neq m_u=m_d$), together
with the following relation for the $\eta$ mass:
\begin{equation} m_\eta^2=\frac{4}{3}m_K^2-\frac{1}{3}m_\pi^2
\, . \label{eetaKpi} \eeq Ultra-violet divergences are regulated
using dimensional regularization and, following the conventions of
ref.~\cite{ga-le}, renormalized by performing the subtraction of
the term
\begin{equation}
\frac{1}{\ep}-\gamma_E+\ln{4\pi} + 1\,. \end{equation}

\subsection{Matrix Elements for the Decay $K^{+} \rightarrow
\pi^{+}\pi^{0}$ with Physical Kinematics} \label{subsec:physical}

In this section we  describe the calculation of the matrix
elements for the decay $K^{+} \rightarrow \pi^{+}\pi^{0}$, in both
full QCD (\sec{subsubsec:physicalfull}) and the quenched theory
(\sec{subsubsec:physicalquenched}). We underline again that by
physical we mean that there is no insertion of momentum at the
weak operator. The masses of the mesons are arbitrary.

\subsubsection{Calculations in Full QCD}
\label{subsubsec:physicalfull}

We start with the discussion in Minkowski space. At lowest order
the matrix element of $\op_4$, obtained using the chiral operator
defined in \eq{o271l}, is given by
\begin{equation}
\la\pi^+\pi^0|\op_4|K^+\ra=
-\frac{6\sqrt{2\,}\alpha^{(27,1)}}{f_K\,f_\pi^2}(m_K^2-m_\pi^2)\,.
\end{equation}
At one-loop order we have to compute the diagrams of
\fig{kppgraph}. In the tadpole diagram $(a)$, all pseudoscalar
mesons propagate in the loop.  In the remaining diagrams, because
of the symmetries of the Lagrangian, it is the combinations
indicated in \tab{tpartperm} which contribute.
\begin{table}[t]
\begin{center}
\begin{tabular}{c c}
\hline\hline Diagram & Particles \\ \hline\hline $(b)$
&\rule{0pt}{12pt} $\pi^+\pi^0$\\ $(c)$ & $K^+\pi^0$,\ $K^0\pi^+$,\
$K^+\eta$ \\ $(d)$ & $K^0\pi^0$,\ $K^+\pi^-$,\ $K^0\eta$ \\ \hline
\end{tabular}
\caption[]{\it {Combinations of mesons looping in the different
diagrams.}} \label{tpartperm}
\end{center}
\end{table}

Our results for this case agree with those of ref.~\cite{BPP} and
with those of Golterman and
Leung~\cite{GL}, except for the finite terms which depend on the
regularization/renormalization (and the sign of the imaginary part
of diagram (b) in fig.~\ref{kppgraph}). The matrix element is
given by
\begin{equation}
\la\pi^+\pi^0|\op_4|K^+\ra=-\frac{6\sqrt{2\,}\alpha^{(27,1)}}{f_K\,f_\pi^2}
\l[(m_K^2-m_\pi^2)\Bigl[1+\frac{m_K^2}{(4\pi
f)^2}I_{zf}\Bigr]+\frac{m_K^4}{(4\pi)^2f^2}\l(I_a+I_b+I_{c+d}\r)\r]
\, , \label{eAmPhys} \end{equation} plus the $O(p^4)$ counterterms
(i.e. the terms proportional to the $\beta_i$'s in
\eq{eq:chiralo4phys}). The integrals in \eq{eAmPhys} are given by
\begin{eqnarray}
I_{zf}&=&\frac{7}{3}\ln{y}-\l[\frac{2}{3}-\frac{1}{6}y\r]\ln{\frac{m_{\eta}^2}
{m_{\pi}^2}}-3[1+y]\ln{\frac{m_{\pi}^2}{\mu^2}}\,
,\label{eq:izdef}\\
I_a&=&\l(\frac{20}{3}y^2-y-\frac{17}{3}\r)\ln{\frac{m_\pi^2}{\mu^2}}-\l(\frac{1}{6}y^2
-y+\frac{4}{3}\r)\ln{\frac{m_\eta^2}{m_\pi^2}}+\l(\frac{13}{3}-4y\r)\ln{y}\,,\nn\\
I_b&=&\l(1-\frac{13}{3}y+\frac{14}{3}y^2\r)\ln{\frac{m_\pi^2}{\mu^2}}+(1-3y+2y^2)
\l[A(y)-i\pi\sqrt{1-4y}-1\r] \,, \nn \\
I_{c+d}&=&\frac{1}{3}\l(5-8y-y^2\r)\ln{\frac{m_\pi^2}{\mu^2}}+\l(\frac{10}{3}
-\frac{4}{3y}-2y\r)-A(y)\l(1+\frac{5}{8y^2}-\frac{13}{8y}\r)\nn \\
&-&\frac{1}{72}\l(\frac{1}{y^2}-\frac{1}{y}\r)B(y)
+\l(\frac{1}{3}y^2-\frac{19}{12}y+\frac{19}{18}-\frac{23}{72y}+\frac{1}{72y^2}\r)\ln{\frac{m_\eta^2}{m_\pi^2}}
\nn \\
&-&\l(\frac{11}{18y^2}-\frac{23}{9y}+\frac{65}{18}-\frac{7}{3}y\r)\ln{y}
\, , \label{eq:isdef} \end{eqnarray} where $y=m_\pi^2/m_K^2$,
$\mu$ is the renormalization scale,
\begin{eqnarray}
A(y)&=&\sqrt{1-4y}\ \log\l(
\frac{1+\sqrt{1-4y}}{1-\sqrt{1-4y}}\r)\ \ \textrm{\ \ and},\nn\\
B(y)&=&\sqrt{1-44y+16y^2}\
\log\l(\frac{7-4y+\sqrt{1-44y+16y^2}}{7-4y- \sqrt{1-44y+16y^2}}\r)
\nn \\ &=&-2\sqrt{44y-1-16y^2}\arctan
\l(\frac{\sqrt{44y-1-16y^2}}{7-4y}\r). \label{eAmPhyspiu}
\end{eqnarray} In deriving the above expressions we have used
\eq{eetaKpi} and the fact that, with physical meson masses, the
argument of the square root in  \eq{eAmPhyspiu}, $1-44y+16y^2$, is
negative.

From \eq{eAmPhys}, we can extract the phase of the amplitude,
$\delta$, defined by
\begin{equation}
{\cal A}=|{\cal A}|e^{i\delta} \, . \label{eq:deltadef}\eeq We
find
\begin{equation}
\delta(k=\sqrt{{m_K^2}/4-m_\pi^2})=-\frac{m_K^2}{16\pi
f^2}(1-2y)\sqrt{1-4y}\, .\label{eq:delta1}\end{equation} Rewriting
the right-hand side of \eq{eq:delta1} in terms of $k$ we recover
the expression for the $I=2$ strong interaction  $\pi$-$\pi$
scattering phase,
\begin{equation}
\delta(k)=-\frac{(2k^2+m_\pi^2)}{8\pi f^2}\frac{k}{E} \, ,
\label{eq:delta2} \end{equation} as required by Watson's theorem.
The $I=2$ scattering length is given by\beq
a_0^{I=2}=\lim_{k\rightarrow
0}\frac{\delta(k)}{k}=-\frac{m_\pi}{8\pi f^2} \, . \label{ea0}
\eeq

For the electroweak penguin operators $\op_7$ and $\op_8$ the
corresponding formulae are:
\begin{equation}
\la\pi^+\pi^0|\op_{7,8}|K^+\ra=\frac{2\sqrt{2\,}\gamma^{(8,8)}}{f_K\,f_\pi^2}
\l[1+\frac{m_K^2}{(4\pi
f)^2}I_{zf}+\frac{m_K^2}{(4\pi)^2f^2}\l(J_a+J_b+J_{c+d}\r)\r] \, ,
\label{eAmPhys78} \eeq where $I_{zf}$ is given in \eq{eq:izdef}
plus the $O(p^2)$ counterterms (i.e. the terms proportional to the
$\delta_i's$ in \eq{eq:chiralo78phys}). The integrals appearing on
the right-hand side of \eq{eAmPhys78} are
\begin{eqnarray}
J_a&=&-\l(\frac{10}{3}y+\frac{11}{3}\r)\ln{\frac{m_\pi^2}{\mu^2}}-
\l(\frac{2}{3}-\frac{1}{6}y\r)\ln{\frac{m_\eta^2}{m_\pi^2}}+3\,
\ln{y}\,,\nn  \\
J_b&=&\l(1-\frac{8}{3}y\r)\ln{\frac{m_\pi^2}{\mu^2}}+(1-2y)\l[A(y)-i\pi\sqrt{1-4y}-1\r]
\,, \nn \\
J_{c+d}&=&\l(\frac{13}{6}-y\r)\ln{\frac{m_\pi^2}{\mu^2}}+\l(2-\frac{3}{2y}\r)
-A(y)\l(\frac{5}{8y^2}-\frac{1}{y}\r)\nn \\ &-&\frac{1}{24y^2}
B(y) +\l(-\frac{1}{3}y+\frac{19}{12}-\frac{11}{12y}+
\frac{1}{24y^2}\r)\ln{\frac{m_\eta^2}{m_\pi^2}}+ \nn \\
&-&\l(\frac{7}{12y^2}-\frac{4}{3y}+\frac{4}{3}\r)\ln{y} \, .
\label{eq:jsdef} \eea $A(y)$ and $B(y)$ have been defined in
\eq{eAmPhyspiu}. The strong interaction phase is of course, the
same as for $\op_4$, since it only depends on the isospin of the
two-pion state. For this case our results agree with those of
ref.~\cite{PPS}. In ref.~\cite{CG} the one-loop contributions to
the matrix elements are presented numerically and we agree with
these results.

\paragraph{Calculations in Euclidean Space}
The calculation of the Euclidean matrix elements in terms of the
low-energy constants at NLO in \ChPT proceeds along similar lines.
In the Euclidean theory the imaginary part is zero and the
arguments of the logarithms correspond to the absolute values of
the arguments in the Minkowski case. For a generic (Euclidean)
kaon momentum $p_K$ (which is also therefore the momentum of the
two-pion state) the logarithms take the form
\begin{equation}
\log\l(\frac{\ \l|(1+\sqrt{1+4\frac{m_\pi^2}{p_K^2}})\right|\ } {\
\left|(1-\sqrt{1+4\frac{m_\pi^2}{p_K^2}})\r|\ }\r) \, . \eeq
Setting $p_K^2\rightarrow-m_K^2$ we recover the Minkowski
amplitude without the phase
\begin{equation}
\la\pi^+\pi^0|\op^E_4|K^+\ra_E=-\frac{6\sqrt{2\,}\alpha^{(27,1)}}{f_K\,f_\pi^2}
\l[(m_K^2-m_\pi^2)\Bigl[1+\frac{m_K^2}{(4\pi
f)^2}I_{zf}\Bigr]+\frac{m_K^4}{(4\pi)^2f^2}\l(I_a+I^E_b+I_{c+d}\r)\r]
\, . \label{eAmPhysE} \eeq The functions  $I_x$ are those defined
in \eqs{eq:izdef}{eq:isdef} and $I^E_b=\mbox{Re} \, I_b$.

\subsubsection{Calculations in Quenched QCD}
\label{subsubsec:physicalquenched}

In this section we present the results for the matrix elements in
quenched QCD. As above we do not exhibit the NLO counterterms
given in \eqs{eq:chiralo4phys}{eq:chiralo78phys} explicitly, but
of course they must be included in any comparison with lattice
data.

The matrix element of the operator $\op_4$ takes the form:
\begin{equation}
\la\pi^+\pi^0|\op_4|K^+\ra=-\frac{6\sqrt{2\,}\alpha^{(27,1)}}{f_\pi^2f_K}\l[(m_K^2-m_\pi^2)
\Bigl[1+\frac{m_K^2}{(4\pi f)^2}I_{zf}^q
\Bigr]+\frac{m_K^4}{(4\pi)^2f^2_q}\l(I_a^q+I_b^q+I_{c+d}^q\r)\r]\,,
\label{eqn:AmpQ} \eeq where the integrals are given by
\begin{eqnarray}
I_{zf}^q&=&-\frac{2m_0^2}{9m_K^2}\l[1+\frac{1}{2(1-y)}\ln{\frac{y}{2-y}}\r]+
\frac{2\alpha}{9}\l[1+\frac{y(2-y)}{2(1-y)}\ln{\frac{y}{2-y}} \r]
\, \label{eq:izfqdef} \\ I_a^q&=& -\l(\frac{2}{3}+y-\frac{5}{3}y^2
\r)\ln{\frac{m_\pi^2}{\mu^2}}+\frac{2}{3}\l(1-y\r)\ln{y}+\nn \\ &
&+
\frac{m_0^2}{9m_K^2}\l[-2\l(2-y\r)+\frac{2-y}{1-y}\ln{\frac{2-y}{y}}
\r]+ \nn \\ & &+ \frac{\alpha}{9}\l[
2\l(2-y\r)-\frac{y(2-y)^2}{(1-y)}\ln{\frac{2-y}{y}}\r] \, , \nn \\
I_b^q&=&\l(1-\frac{13}{3}y+\frac{14}{3}y^2\r)\log\l(\frac{m_\pi^2}{\mu^2}\r)+(1-3y+2y^2)\l[A(y)-i\pi\sqrt{1-4y}-1\r]
\,, \nn \\
I_{c+d}^q&=&-\frac{5+y+2y^2}{6}\ln{\frac{m_\pi^2}{\mu^2}}+\l(\frac{3}{2}-\frac{1}{2y}-y\r)
\nn \\ & & -\l(\frac{1}{2}+\frac{1}{4y^2}-\frac{3}{4y}\r)A(y)
+\l(-\frac{2}{3}-\frac{1}{4y^2}+\frac{5}{4y}+\frac{2}{3}y\r)\ln{y}
\nn \\ &+& \frac{m_0^2}{m_K^2}\l[
\l(\frac{1}{6y^2}-\frac{1}{3y}\r)A(y)-\frac{1-8y+18y^2-8y^3}{6y^2(1-8y+4y^2)}C(y)-\frac{4}{9}y
\r. \nn \\ & &
+\l.\frac{3-15y+24y^2-8y^3}{18y^2(1-y)}\ln{\frac{2-y}{y}}+\l(1+\frac{1}{3y^2}-\frac{4}{3y}\r)\ln{y}
\r]+ \nn \\ &+& \alpha \l[
\l(\frac{2}{3}+\frac{1}{12y^2}-\frac{7}{12y} \r)A(y)+
\l(\frac{5}{12y^2}-\frac{15}{4y}+\frac{59}{6}-8y+2y^2
\r)\frac{C(y)}{1-8y+4y^2} \r. + \nn \\ & &
+\frac{1}{1-y}\l(-\frac{65}{12}-\frac{5}{12y^2}+\frac{5}{2y}+\frac{9}{2}y-\frac{29}{18}y^2+\frac{2}{9}y^3
\r)\ln{\frac{2-y}{y}} \nn \\ & & \l.
-\frac{(1-y)^2(1-2y)}{3y^2}\ln{y}+\l(\frac{1}{y}-4+\frac{31}{9}y
\r) \r] \, . \eea with
\begin{equation}
C(y)=\sqrt{1-8y+4y^2}\ln{\frac{3-2y+\sqrt{1-8y+4y^2}}{3-2y-\sqrt{1-8y+4y^2}}}
\, . \label{eq:Cdef}\end{equation}

For the EWP operators the corresponding results are
\begin{equation}
\la\pi^+\pi^0|\op_{7,8}|K^+\ra=\frac{2\sqrt{2\,}\gamma^{(8,8)}}{f_\pi^2f_K}\l[1+\frac{m_K^2}{(4\pi
f)^2}I_{zf}^q
+\frac{m_K^2}{(4\pi)^2f^2}\l(J_a^q+J_b^q+J_{c+d}^q\r)\r]
\label{eAmpQ78} \eeq where $I_{zf}^q$ is defined in
\eq{eq:izfqdef} and the remaining integrals are given by
\begin{eqnarray}
J_a^q&=&-\frac{1-y}{3}\ln{\frac{m_\pi^2}{\mu^2}}+\frac{1}{3}\ln{y}+
\frac{m_0^2}{9m_K^2}\l[-2+\frac{1}{1-y}\ln{\frac{2-y}{y}} \r]+ \nn
\\ & &+ \frac{\alpha}{9}\l[ 2
-\frac{y(2-y)}{1-y}\ln{\frac{2-y}{y}}\r] \, , \nn \\
J_b^q&=&\l(1-\frac{8}{3}y\r)\ln{\frac{m_\pi^2}{\mu^2}}+(1-2y)\l[A(y)-i\pi\sqrt{1-4y}-1\r]
\,, \nn \\ J_{c+d}^q&=&
\l(\frac{4}{3}+\frac{y}{3}\r)\ln{\frac{m_\pi^2}{\mu^2}}-1-\frac{1}{2y}A(y)-\l(\frac{1}{3}+\frac{1}{2y}
\r)\ln{y} \nn \\ &+& \frac{m_0^2}{m_K^2}\l[
-\frac{A(y)}{12y^2(1-y)}+\frac{-1+6y-16y^2+8y^3}{12y^2(1-y)(1-8y+4y^2)}C(y)
\r. \nn \\ & &\l.+\frac{4}{9}-\frac{1}{3y} +
\frac{3-6y-2y^2}{36y^2(1-y)}\ln{\frac{2-y}{y}} \r]+ \nn \\ &+&
\alpha \l[
-\l(\frac{1-6y+4y^2}{12y^2(1-y)}\r)A(y)+\l(\frac{1}{3y}-\frac{4}{9}\r)
\r. \nn \\ & &
+\l(\frac{17}{3}+\frac{1}{4y^2}-\frac{2}{y}-5y+\frac{4}{3}y^2\r)\frac{C(y)}{(1-y)(1-8y+4y^2)}+
\nn \\ & &
-\frac{9-36y+42y^2-28y^3+8y^4}{36y^2(1-y)}\ln{\frac{2-y}{y}} \nn
\\ & & \l. -\frac{(1-y)(1-3y)}{3y^2}\ln{y} \r] \, .\label{eq:jqsdef}\eea

\subsection{$K^{+} \rightarrow \pi^{+}\pi^{0}$ Matrix Elements
with the SPQR Kinematics} \label{sec:SPQR}

As explained in the Introduction and section~\ref{sec:strategy},
our aim is to extract the physical $K\to\pi\pi$ matrix elements
from simulations at unphysical kinematics using NLO \ChPT. This
requires the calculation of the one-loop chiral corrections to the
$K^{+} \rightarrow \pi^{+}\pi^{0}$ amplitudes with the insertion
of weak operators carrying four-momentum. In this section we
consider the matrix elements for the SPQR kinematics, in which the
kaon and one of the final-state pions are at rest and the second
pion has an energy $E_\pi$. We have seen in
section~\ref{sec:strategy} that the computation of these matrix
elements are sufficient to determine all the required low-energy
constants at NLO in the chiral expansion.

In order to ensure that the final state has $I=2$, it is necessary
to symmetrise over the momenta of the two pions. Note also that
the only diagram which contributes to the rescattering phase is
diagram $(b)$.

The results for full QCD are presented in
appendix~\ref{subsec:spqrfull} and for the quenched theory in
appendix~\ref{subsec:spqrquenched}. Here we present the results
for the two choices of quark masses corresponding to $m_K=2 m_\pi$
or $m_K=m_\pi$ in each case with both pions at rest.

\subsubsection{Matrix Elements with \boldmath{$m_K=2m_\pi$} and both Pions at
Rest in Full QCD}\label{subsubsec:mkeq2mpifull}

Consider first the operator $\op_4$. In the limit where both pions
are at rest ($\omega=1$) and $m_K=2m_\pi$ ($z=1/2$) from
eqs.(\ref{eq:spqrfull271}) -- (\ref{eq:spqrfullicd}) we can
readily recover the corresponding result ($y=1/4$) in
eqs.~(\ref{eAmPhys}) -- (\ref{eq:isdef}):
\begin{eqnarray}
\la\pi^+\pi^0|\op_4|K^+\ra &=&
-\frac{18\sqrt{2\,}\alpha^{(27,1)}m_\pi^2}{f_K\,f_\pi^2}\bigg[ 1+
\frac{m_\pi^2}{4\pi^2f^2}\bigg(-\frac{19}{2}\log\l(\frac{m_\pi^2}{\mu^2}\r)-\frac{23}{6}\nn\\
&&\hspace{0.5in}+\frac43\arctan\frac12-\frac{10}{3}\log(4)-\frac{31}{12}\log(5)\bigg)\bigg]\
. \label{eAmPhyslim} \end{eqnarray}

The corresponding result for the EWP operators is:
\begin{equation}
\la\pi^{+}\pi^{0}|\op_{7,8}|K^{+}\ra =
 \frac{2\sqrt{2}}{f_{K}f^{2}_{\pi}}
 \gamma^{(8,8)}\,\Bigg[1 +
\frac{m_\pi^2}{4\pi^2 f^2}
\l(-6\ln{\frac{m_\pi^2}{\mu^2}}-\frac{9}{2}+4\arctan{\frac{1}{2}}-\frac{11}{4}\ln{5}
\r)\Bigg] \, . \end{equation}

\subsubsection{Matrix Elements in the \boldmath{$SU(3)$}
limit, \boldmath{$m_K=m_\pi$}, and with both Pions at
Rest in Full QCD}\label{subsubsec:mkeqmpifull}

In the SU(3) limit $m_s=m_d=m_u$, for which $m_K=m_\pi$, with the
two pions both at rest eqs.(\ref{eq:spqrfull271}) --
(\ref{eq:spqrfullicd}) reduce to
\begin{equation}
\la\pi^+\pi^0|\op_4|K^+\ra =
-\frac{12\sqrt{2}\alpha^{(27,1)}m_\pi^2}{f_K\,f_\pi^2}
\l[1+\frac{m_\pi^2}{(4\pi)^2f^2}\l(-15\,
\ln{\frac{m_\pi^2}{\mu^2}}-3\r)\r] \, , \eeq in agreement with the
result given in eq.~(82) of ref.~\cite{GL}, up to scheme-dependent
finite terms.

For $\op_{7,8}$ the corresponding result is:
\begin{equation}
\la\pi^{+}\pi^{0}|\op_{7,8}|K^{+}\ra =
 \frac{2\sqrt{2}}{f_{K}f^{2}_{\pi}}
 \gamma^{(8,8)}\,\Bigg[1 +\frac{m_\pi^2}{(4\pi
f)^2}\l(-8\ln{\frac{m_\pi^2}{\mu^2}}\r)\Bigg] \, . \end{equation}

\subsubsection{Matrix Elements with
\boldmath{$m_K=2m_\pi$} and both Pions at Rest in Quenched
QCD}\label{subsubsec:mkeq2mpiquenched}

The results for the matrix elements with the SPQR kinematics in
the quenched theory are presented in detail in
appendix~\ref{subsec:spqrquenched}. In this and the following
subsection we present the results for the special cases
$m_K=2m_\pi$ and $m_K=m_\pi$ respectively, in each case with both
the pions at rest.

For the matrix element of $\op_4$ in the quenched theory with
$m_K=2m_\pi$ We find
\begin{eqnarray}
\la\pi^{+}\pi^{0}|\op_4|K^{+}\ra &=&
-\frac{18\sqrt{2\,}\alpha^{(27,1)}m_\pi^2}{f_\pi^2f_K}\l\{1+\frac{m_\pi^2}{4\pi^2
f^2}\l[-2\ln{\frac{m_\pi^2}{\mu^2}}-\frac{3}{2}-\frac{4}{3}\ln{4}+
\r.\r.\nn \\ & & \l.\l.
+\frac{m_0^2}{4m_\pi^2}\l(-\frac{8}{9}+\frac{40}{27}\ln{7}-\frac{4}{3}\ln{4}
\r)+\r.\r. \nn \\ & & \l.\l.
+\alpha\l(\frac{17}{9}+2\,\ln{4}-\frac{113}{54}\ln{7}-\frac{2}{\sqrt{3}}\arctan{\frac{\sqrt{3}}{5}}
\r)\r]\r\} \, .\label{eq:mkeq2mpires}\eea We are grateful to
M.~Golterman for revisiting this calculation and confirming his
agreement with the expression in
eq.~(\ref{eq:mkeq2mpires})~\cite{Mgprivate} (the result in
eq.~(\ref{eq:mkeq2mpires}) is in disagreement with that presented
in eqs.~(3.1) and (3.2) of ref.~\cite{GL2}). The corresponding
result for the EWP operators is:
\begin{eqnarray}
\la\pi^{+}\pi^{0}|\op_{7,8}|K^{+}\ra &=&
\frac{2\sqrt{2\,}\gamma^{(8,8)}}{f_\pi^2f_K}\l\{1+\frac{m_\pi^2}{4\pi^2
f^2}\l[\frac{3}{2}\ln{\frac{m_\pi^2}{\mu^2}}-\frac{3}{2}+2\ln{4}+
\r.\r.\nn \\ & & \l.\l.
+\frac{m_0^2}{4m_\pi^2}\l(-\frac{4}{3}+\frac{10}{9}\ln{7}-\frac{8\sqrt{3}}{9}\arctan{\frac{\sqrt{3}}{5}}
\r)+\r.\r. \nn \\ & & \l.\l.
+\alpha\l(\frac{4}{3}+\ln{4}-\frac{13}{9}\ln{7}+\frac{8\sqrt{3}}{9}\arctan{\frac{\sqrt{3}}{5}}
\r)\r]\r\} \, . \eea

\subsubsection{Matrix Elements in the \boldmath{$SU(3)$}
 limit, \boldmath{$m_K=m_\pi$}, and with both Pions at
Rest in Quenched QCD}\label{subsubsec:mkeqmpiquenched}

Finally we present the results with $m_K=m_\pi$ in the quenched
theory with the pions at rest. In this case for the matrix element
of $\op_4$ we do agree with the result in ref.~\cite{GL} finding
\begin{equation}
\la\pi^+\pi^0|\op_4|K^+\ra =
-\frac{12\sqrt{2\,}\alpha^{(27,1)}m_\pi^2}{f_K\,f_\pi^2}\l[1+\frac{m_\pi^2}{(4\pi)^2f^2}\l(-3
\log\l(\frac{m_\pi^2}{\mu^2}\r)-3\r)\r]\,.
\end{equation}
For the matrix elements of $\op_{7,8}$ the corresponding result
is:
\begin{equation}
\la\pi^{+}\pi^{0}|\op_{7,8}|K^{+}\ra =
\frac{2\sqrt{2\,}\gamma^{(8,8)}}{f_\pi^2f_K}\l\{1+\frac{m_\pi^2}{(4\pi
f)^2}\ 4\log\l(\frac{m_\pi^2}{\mu^2}\r)\r\}\ .
\end{equation}

\section{Conclusions}\label{sec:concs}

The precise evaluation of $K\to\pi\pi$ matrix elements in lattice
simulations is a major long-term challenge. In this paper we have
explained our strategy for the reduction of an important source of
systematic error in current studies, viz. the use of \ChPT at
leading order in which final state interactions are neglected. We
propose to use lattice computations of $K\to\pi\pi$ matrix
elements with meson masses and momenta which are unphysical (but
accessible to lattice simulations) to determine all the low energy
constants necessary to obtain the physical amplitudes at NLO in
\ChPT. We have presented all the necessary ingredients to
implement this strategy for $\Delta I=3/2$ transitions. In
particular we have evaluated all the chiral logarithms at one-loop
order for arbitrary masses and momenta.

We have shown in \sec{sec:strategy} that it is possible to
determine all the necessary low-energy constants from matrix
elements with the SPQR kinematics, i.e. with the kaon and one of
the pions at rest and with the second pion having an arbitrary
momentum. We present the results for this choice of kinematics
separately, but stress that of course this is not the only choice
for the determination of the low-energy constants (for this reason
we present the very lengthy results for general kinematics).

We incorporate into our work the recent progress in understanding
finite-volume effects in $K\to\pi\pi$ decays. For $\Delta I=3/2$
transitions we find that these effects are given by the
Lellouch-L\"uscher factor (for zero total three-momentum), even in
the quenched approximation. This is not the case for $\Delta
I=1/2$ transitions, where the absence of unitarity in the quenched
theory leads to major subtleties and difficulties, such as the
dependence of the final state interactions on the choice of
$\Delta I=1/2$ weak operator and non-standard behaviour of the
correlation functions with time and volume. This is the subject of
a paper under preparation. We are also investigating the
generalization of the LL-factor to the case in which the total
three-momentum of the two-pions is not equal to zero.

In ref.~\cite{laiho} a particular implementation of our proposed
strategy is discussed. The conclusion of ref.~\cite{laiho} is that
it is possible to determine all the low-energy constants necessary
to obtain the physical matrix elements for (27,1) and (8,1)
operators from the knowledge of:
\begin{enumerate}
\item[i)] $K\to\pi\pi$ matrix elements with both the pions at rest
for $m_K=2m_\pi$ and $m_K=m_\pi$;
\item[ii)] $K\to\pi$ matrix elements at non-zero momentum;
\item[iii)] the matrix elements for $K^0$--$\bar K^0$ mixing.
\end{enumerate}
We agree with this conclusion for the $\Delta I=3/2$, (27,1)
operators studied above. Notice however that our complete basis
for the $\chi$PT weak operators at order $p^4$ differs in the
replacement of the operator $O_{20}$ with our operator
$O_{22}$\footnote{The numbering of operators follow ref.~\cite{KMW}
in both cases.}. The two choices turn out to be
equivalent also in the most general case with unphysical
kinematics (i.e. with an injection of momentum by the weak
operator).

For $\Delta I=3/2$ decays our strategy can also be implemented for
partially quenched lattice QCD, and we will present the
corresponding one-loop results in a future paper. (For the case
$m_K=m_\pi$, with all the mesons at rest and with degenerate
sea-quarks, the matrix element of ${\cal O}_4$ has been evaluated
in partially quenched QCD in ref.~\cite{Glpart}.)

\subsection*{Acknowledgements}
We thank Sebastien Descotes-Genon, Jonathan Flynn, Maarten
Golterman, Naruhito Ishizuka, Mauro Papinutto, Douglas Ross,
Martin Savage and Massimo Testa for many helpful discussions. CJDL
and CTS thank the organisers and participants of the Programme
\textit{Lattice QCD and Hadron Phenomenology} at the Institute for
Nuclear Theory, Univ. of Washington, Seattle, for their
hospitality during the early stages of this project.

This work was supported by European Union grant
HTRN-CT-2000-00145. CJDL and CTS acknowledge support from PPARC
through grants PPA/G/S/1998/00529 and \\ PPA/G/O/2000/00464. EP
was supported in part by the Italian MURST under the program
\textit{Fenomenologia delle Interazioni Fondamentali}.

\appendix
\section{Integrals which appear in the evaluation of the correlation functions}
\label{saFV}

In this appendix we list the energy integrals which appear in the
evaluation of the contributions of diagrams $(b)$, $(c)$ and $(d)$
of Fig.\ref{kppgraph} and present the corresponding results. We
write each integral in the generic form:
\begin{equation}
I({\cal N})=\int \frac{dk_4 dl_4}{(2\pi)^2}\frac{{\cal N}\, e^{- i
(k_4+l_4)t}}{(k_4^2+w_1^2)(l_4^2+w_2^2)}
\label{eq:indef}\end{equation} where ${\cal N}$ is a polynomial in
$k_4$ and $l_4$, $w_1^2=m_1^2+\vec{k}^2$ and
$w_2^2=m_2^2+\vec{l}^2$.

Using the identity
\begin{equation}
\int \frac{dx}{2\pi} \frac{x^{n}
e^{-ixt}}{x^2+w^2}=\l(i\frac{\partial}{\partial t}\r)^{n} \int
\frac{dx}{2\pi} \frac{e^{-ixt}}{x^2+w^2} \end{equation} we obtain
\begin{eqnarray}
I(1)&=&\frac{1}{(2w_1)(2w_2)}e^{-(w_1+w_2)|t|} \nn\\
I(k_4)&=&-i\, \varepsilon(t)\,\frac{1}{4w_2}e^{-(w_1+w_2)|t|} \nn
\\
I(l_4)&=&-i\, \varepsilon(t)\,\frac{1}{4w_1}e^{-(w_1+w_2)|t|} \nn
\\
I(k_4 l_4)&=&-\frac{1}{4}e^{-(w_1+w_2)|t|} \nn \\
I(k_4^2)&=&\frac{\delta(t)}{2w_2}-\frac{w_1}{4w_2}e^{-(w_1+w_2)|t|}
\nn \\
I(l_4^2)&=&\frac{\delta(t)}{2w_1}-\frac{w_2}{4w_1}e^{-(w_1+w_2)|t|}
\nn \\
I(k_4^2 l_4)&=&i\,
\varepsilon(t)\frac{w_1}{4}e^{-(w_1+w_2)|t|}\nn\\
I(k_4 l_4^2)&=&i\, \varepsilon(t)\frac{w_2}{4}\,e^{-(w_1+w_2)|t|}
\nn\\
I(k_4^2 l_4^2)&=&\delta(t)\,\int \frac{dk_4}{2\pi}
-\frac{w_1+w_2}{2}\delta(t)+\frac{w_1 w_2}{4}e^{-(w_1+w_2)|t|} \nn
\eea where $\varepsilon$ represents the sign function.

\section{Results in Full QCD}\label{sec:full}

In this appendix we present the results for matrix elements in
full QCD. The results for general kinematics, i.e. for arbitrary
quark masses and momenta, with the two pions in an I=2 state, are
given in subsection~\ref{subsec:generalfull}. We start however,
with subsection~\ref{subsec:spqrfull} which contains the results
for SPQR kinematics.

The integral $I_{zf}$, which corresponds to the one-loop
contributions from the renormalization of the mesonic
wave-functions and from the replacement of the factor $1/f^3$
which appears at lowest-order in the chiral expansion, by
$1/(f_Kf_\pi^2)$ is given by ($y = m^{2}_{\pi}/m^{2}_{K}$):
\begin{equation}
 I_{zf} = -3 \left ( 1+ y\right )
  \log \left ( \frac{m^{2}_{\pi}}{\mu^{2}}\right ) + \frac{7}{3}
  \log(y) + \frac{1}{6} \left ( y- 4\right )
  \log \left ( \frac{m^{2}_{\eta}}{m^{2}_{\pi}}\right ) .
\label{eq:izffull}\eeq

The results are given in terms of the variables, $y =
m^{2}_{\pi}/m^{2}_{K}$, $z=m_\pi/m_K$ and $\omega=E_\pi/m_\pi$,
where $E_\pi$ is the energy of the pion whose momentum is (in
general) not equal to zero.

\subsection{$K^+\to\pi^+\pi^0$ Decays with SPQR Kinematics in Full QCD}
\label{subsec:spqrfull}

\subsubsection{Results for the matrix elements of $\op_4$ in full
QCD with SPQR kinematics}

The matrix elements of $\op_4$ with SPQR kinematics is given by:
\begin{eqnarray}
\la\pi^{+}\pi^{0}|\op_4|K^{+}\ra &=&
\frac{-6\sqrt{2}\,m_K^2}{f_{K}f^{2}_{\pi}}\alpha^{(27,1)}\,\Bigg\{
 \frac{z(1+\omega+2z)}{2\omega}
 \left( 1 + \frac{m^{2}_{K}}{16\pi^{2}f^{2}}
 \left ( I_{zf}\right )\right ) \nonumber\\
 &&\hspace{1.2in}+ \frac{m^{2}_{K}}{16\pi^{2}f^{2}}\, \left (
  I_{a} + I_{b} + I_{c+d}\right )\Bigg\}\,,\label{eq:spqrfull271}\eea
where $I_{zf}$ is given in \eq{eq:izffull} and $I_{a,b,c+d}$ are
as follows:
\begin{eqnarray}
%
% tadpole diagram
%
I_a &=& \left ( \frac{1}{12\omega}\right)\Bigg\{4\,\left( 12\,z^2
- \omega  + 8\,z\,\left( 1 + \omega \right) \right) \,
  \log (z^2)\nonumber\\
& & \hspace{-0.5in}-\left( 56\,z^2 - 4\,\omega  + 44\,z\,\left( 1
+ \omega \right) +
    26\,z^3\,\left( 1 + \omega  \right)  +
    4\,z^4\,\left( 21 + \omega  \right)  \right) \,
  \log \left(\frac{m^{2}_{\pi}}{\mu^{2}}\right)\nonumber\\
& &\mbox{ }\mbox{ }+ z\,\left( -4 + z^2 \right) \,\left( 3 + 2\,z
+ 3\,\omega  \right) \,
  \log \left(\frac{m^{2}_{\eta}}{m^{2}_{\pi}}\right)
\Bigg \} ,\label{eq:spqrfullia}
\end{eqnarray}
\begin{eqnarray}
%
% scattering diagram
%
I_b &=& \left (\frac{-z^3}
  {3\omega^{2}}\right )
\Bigg \{ 3(1+2z+\omega) + 3 \sqrt{\frac{1-\omega}{1+\omega}}
(1+2z+\omega)
 {\mathrm{log}}\left (
  \frac{-1+\sqrt{\frac{1-\omega}{1+\omega}}}
  {1+\sqrt{\frac{1-\omega}{1+\omega}}}\right )\nonumber\\
& &\mbox{ }\mbox{ }\mbox{ }+
 \left [ -3-2\omega+\omega^{2}-2z(3-3\omega+\omega^{2})\right ]
   {\mathrm{log}}\left (\frac{m^{2}_{\pi}}{\mu^{2}}\right ) \Bigg \} \mbox{
},\label{eq:spqrfullib}
\end{eqnarray}
and
\begin{equation}\label{eq:spqrfullicd}
I_{c+d}=I_{c+d}^1 +
I_{c+d}^2\log\left(\frac{m_\pi^2}{\mu^2}\right) +
I_{c+d}^3\,\log(z^2) +
I_{c+d}^4\,\log\l(\frac{m_\eta^2}{m_\pi^2}\r) + I_{c+d}^5.
\end{equation} The separate components in \eq{eq:spqrfullicd} are
as follows:
\begin{eqnarray}
I_{c+d}^1 &=& \left ( \frac{z^2}{9\,{\omega }^2\,{\left( -2\,z +
\omega + z^2\,\omega  \ \right) }^2} \right ) \bigg\{
2\,z^5\,{\omega }^2\,\left( -16 - 15\,\omega  + {\omega }^2
\right)  +
  3{\omega }^2\,\left( 3 - 26\,\omega  + 3\,{\omega }^2 \right)
\nonumber\\ &&+
  2\,z\,\omega \,\left( -18 + 140\,\omega  - 33\,{\omega }^2 +
     {\omega }^3 \right)  + z^4\,\omega \,
   \left( 43 + 129\,\omega  - {\omega }^2 + 9\,{\omega }^3 \right)
\\&&\hspace{-0.7in}+
  z^2\left( 36 - 269\,\omega  + 174\,{\omega }^2 - 79\,{\omega }^3 +
     18\,{\omega }^4 \right)
   -
  2z^3\left( 11 + 78\,\omega  - 107\,{\omega }^2 + 48\,{\omega }^3 +
     30\,{\omega }^4 \right)
\bigg \}\,,\nonumber\end{eqnarray}
%%%%%%%
\begin{eqnarray}
I_{c+d}^2&=&\left(\frac{1}{6\omega^{2}}\right)\bigg \{ -2\,{\omega
}^2 - 2\,z^4\,{\omega }^2 +
  4\,z\,\omega \,\left( 1 + \omega  \right)  +
  z^2\,\left( -6 + 28\,\omega  - 6\,{\omega }^2 \right)\nonumber\\
& &\hspace{1in} -
  z^3\,\left( -1 + {\omega }^2 \right)\bigg \},
\end{eqnarray}
%%%%%%%%%%%%%%%%%%
\begin{eqnarray}
I_{c+d}^3& = &\left(\frac{1}{54\,\left( -1 + z \right)\,{\omega
}^2\,
    {\left( -2\,z + \omega  + z^2\,\omega  \right)
    }^3}\right)\times\nonumber\\ %
&&\bigg\{ -18\,{\omega }^5 + 18\,z\,{\omega }^4\,\left( 7 +
2\,\omega  \right)  +
  4\,z^{10}\,{\omega }^3\,\left( -7 - 3\,\omega  + 4\,{\omega }^2 \right)
\nonumber\\ %
& &\hspace{-0.5in}-
  6\,z^2\,{\omega }^3\,\left( 63 - 11\,\omega  + 21\,{\omega }^2 \right)  -
  2\,z^9\,{\omega }^2\,\left( -135 - 50\,\omega  + 12\,{\omega }^2 +
     8\,{\omega }^3 \right)
\nonumber\\ & &\hspace{-0.7in} +
  5\,z^3\,{\omega }^2\,\left( 126 - 179\,\omega  + 129\,{\omega }^2 +
     8\,{\omega }^3 \right)  +
  3\,z^8\,\omega \,\left( -90 - 156\,\omega  - 265\,{\omega }^2 +
     71\,{\omega }^3 \right) \nonumber\\
& & + z^4\,\omega \,\left( -576 + 1926\,\omega  - 1373\,{\omega
}^2 + 75\,{\omega }^3 - 202\,{\omega }^4 \right) \\ %
& & + z^7\,\left( 36 + 582\,\omega  + 1116\,{\omega }^2 -
357\,{\omega }^3 + 1047\,{\omega }^4 - 36\,{\omega }^5 \right)
\nonumber\\ %
& &+ 6\,z^5\,\left( 36 - 179\,\omega  + 222\,{\omega }^2 -
210\,{\omega }^3 + 197\,{\omega }^4 + 56\,{\omega }^5 \right)
\nonumber\\ & & -  6\,z^6\,\left( 42 + 85\,\omega  - 195\,{\omega
}^2 + 309\,{\omega }^3 + 29\,{\omega }^4 + 77\,{\omega }^5 \right)
\bigg\}\,,\nonumber
\end{eqnarray}
\begin{eqnarray}
I_{c+d}^4&= &\left(\frac{z}{108\,\left( -1 + z \right)\,{\omega
}^2\,
    {\left( -2\,z + \omega  + z^2\,\omega  \right) }^3}\right)\nonumber\\
& &\times\bigg\{ -36\,z^{10}\,{\omega }^4 - 36\,{\omega
}^4\,\left( 1 + \omega  \right)  +
  z^9\,{\omega }^3\,\left( 187 + 3\,\omega  - 4\,{\omega }^2 \right)
\nonumber\\ & &
  +
  12\,z\,{\omega }^3\,\left( 18 + 29\,\omega  + 3\,{\omega }^2 \right)  +
  z^8\,{\omega }^2\,\left( -378 + 11\,\omega  + 141\,{\omega }^2 +
     4\,{\omega }^3 \right)\nonumber\\
& &  -
  z^2\,{\omega }^2\,\left( 432 + 1052\,\omega  + 303\,{\omega }^2 +
     163\,{\omega }^3 \right)\nonumber\\
& &  -
  3\,z^7\,\omega \,\left( -114 + 6\,\omega  + 188\,{\omega }^2 +
     29\,{\omega }^3 + 21\,{\omega }^4 \right)\\
& &  -
  3\,z^4\,\omega \,\left( 230 - 30\,\omega  + 363\,{\omega }^2 +
     151\,{\omega }^3 + 28\,{\omega }^4 \right)\nonumber\\
& &  +
  z^3\,\omega \,\left( 288 + 1314\,\omega  + 566\,{\omega }^2 +
     771\,{\omega }^3 + 163\,{\omega }^4 \right)\nonumber\\
& &  +
  3\,z^6\,\left( -36 - 26\,\omega  + 288\,{\omega }^2 + 14\,{\omega }^3 +
     89\,{\omega }^4 + 21\,{\omega }^5 \right)\nonumber\\
& &  +
  3\,z^5\,\left( 36 - 178\,\omega  + 192\,{\omega }^2 - 111\,{\omega }^3 +
     19\,{\omega }^4 + 28\,{\omega }^5 \right)\bigg\}\nonumber
\end{eqnarray}
and
\begin{eqnarray}
I_{c+d}^5&= &\left(\frac{z^3}{6\,{\omega }^2
 \,{\left( -2\,z + \omega  + z^2\,\omega  \
\right) }^3}\right) {\sqrt{1 - {\omega }^2}}\,\log \left(\frac{1 -
{\sqrt{1 - {\omega }^2}}}
    {1 + {\sqrt{1 - {\omega }^2}}}\right)\times\nonumber\\ &
&\hspace{-0.6in}\bigg\{ 18\,z^5\,{\omega }^2 + 6\,{\omega
}^2\,\left( -1 + 4\,\omega  \right)  -
  3\,z^4\,\omega \,\left( 5 + 2\,\omega  + 11\,{\omega }^2 \right)  -
  3\,z\,\omega \,\left( -8 + 21\,\omega  + 5\,{\omega }^3 \right)
\nonumber\\ & &
  +
  3\,z^2\,\left( -8 + 17\,\omega  - 4\,{\omega }^2 + 7\,{\omega }^3 \
\right)  + z^3\,\left( -2 + 24\,\omega  - 11\,{\omega }^2 +
     25\,{\omega }^4 \right)
\bigg\}\nonumber\\ & &+\left( \frac{2z^2\,\left( 1 + z \right)
}{9\,{\left( -1 + z \right) }^2
 \,\omega }\right)\,
{\sqrt{{\left( -1 + z \right) }^2\,\left( -2 - z + z^2 \right)
}}\times\label{eq:icd5def}
\\ & & \log \left(
 \frac{2 + 3\,z - 2\,z^2 -
     2\,{\sqrt{{\left( -1 + z \right) }^2\,\left( -2 - z + z^2 \right) }}}
{2 + 3\,z - 2\,z^2 + 2\,{\sqrt{{\left( -1 + z \right) }^2\,
          \left( -2 - z + z^2 \right) }}}\right)+\nonumber\\
& &\hspace{-0.5in}\left(\frac{z^2}{54\,{\omega }^2\,
    {\left( -2\,z + \omega  + z^2\,\omega  \right) }^3}\right)
{\sqrt{12\,z\,\omega  - 12\,z^3\,\omega  - 8\,{\omega }^2 +
     4\,z^4\,{\omega }^2 + z^2\,\left( 9 - 5\,{\omega }^2 \right) }}\times
    \nonumber\\
& &\hspace{-0.2in}\bigg\{ 6\,{\omega }^3 + 3\,z^4\,\omega \,\left(
-11 + 5\,{\omega }^2 \right)  +
  z\,{\omega }^2\,\left( -31 + 13\,{\omega }^2 \right)  +
  z^2\,\left( 33\,\omega  - 21\,{\omega }^3 \right)
\nonumber\\ & &
 +
  z^5\,\left( 14\,{\omega }^2 - 8\,{\omega }^4 \right)  +
  z^3\,\left( 18 - 19\,{\omega }^2 + 13\,{\omega }^4 \right)
\bigg\}\nonumber\\
&& \hspace{-0.5in}\log \left(\frac{3\,z +
2\,\omega  - 2\,z^2\,\omega
-
     {\sqrt{12\,z\,\omega  - 12\,z^3\,\omega  - 8\,{\omega }^2 +
         4\,z^4\,{\omega }^2 + z^2\,\left( 9 - 5\,{\omega }^2 \right) }}}
{3\,z + 2\,\omega  - 2\,z^2\,\omega  +
     {\sqrt{12\,z\,\omega  - 12\,z^3\,\omega  - 8\,{\omega }^2 +
         4\,z^4\,{\omega }^2 + z^2\,\left( 9 - 5\,{\omega }^2
\right) }}}\right) . \nonumber\eea

\subsubsection{Results for the matrix elements of $\op_{7,8}$ in
full QCD with SPQR kinematics}

The matrix elements of the EWP operators are given by:
\begin{equation} \la\pi^{+}\pi^{0}|\op_{7,8}|K^{+}\ra =
 \frac{2\sqrt{2}}{f_{K}f^{2}_{\pi}}
 \gamma^{(8,8)}\,\left[1+\frac{m^{2}_{K}}{16\pi^{2}f^{2}}
  \left ( I_{zf} +
  J_{a} + J_{b} +
  J_{c+d}\right)\right],
\label{eq:spqrfull88}\end{equation}
where $I_{zf}$, the one-loop contribution from the renormalization
of the mesons' wavefunctions and the replacement of $1/f^3$ by
$1/f_Kf_\pi^2$ is given in \eq{eq:izffull} and the $J_{a,b,c+d}$
are as follows:
\begin{equation}
J_a = \frac16\l\{18\,\log (z^2) - 2\,
       \left( 11 + 10\,z^2 \right) \,
     \log\l(\frac{{{m_{\pi }}}^2}{{\mu }^2}\r) +
    \left( -4 + z^2 \right) \,
   \log\l(\frac{{{m_{\eta }}}^2}{{{m_{\pi }}}^2}\r)\r\}\,
\end{equation}
\begin{equation}
%
% scattering diagram
%
J_b =
 \frac{-2z^2}{3\omega}\left\{3+3\sqrt{\frac{1-\omega}{1+\omega}}
   {\mathrm{log}}
  \left (\frac{-1+\sqrt{\frac{1-\omega}{1+\omega}}}
   {1+\sqrt{\frac{1-\omega}{1+\omega}}} \right )
   + (-3+\omega){\mathrm{log}}\l(\frac{m^{2}_{\pi}}{\mu^{2}}\r)
    \right\}
     \mbox{ },
\end{equation}
and
\begin{eqnarray}
%
% hook diagrams
%
J_{c+d}\hspace{-4.5pt} &=&\hspace{-5pt}
\frac{-(1+\omega)z\bigg(4\omega+4z^{2}\omega-z(5+3\omega)\bigg)}
{\omega(-2z+\omega+z^{2}\omega)} +\left (\frac{-2\,\omega
-
 6\,z^2\,\omega  + 15\,z\,\left( 1 + \omega  \right) }
  {6\,\omega }\right )\log\left(
\frac{m^{2}_{\pi}}{\mu^{2}}\right)\nonumber
\\ & &\hspace{-0.5in}+\left
(\frac{1}{6(-1+z)\omega (-2z+\omega+z^{2}\omega)^{2}}
  \right )
\,\Bigg \{ 2\,{\omega }^3 - 4\,z^6\,{\omega }^2\,\left( 1 + \omega
\right)  +
  3\,z\,{\omega }^2\,\left( 3 + 5\,\omega  \right)  +
\nonumber\\ &&\hspace{-0.4in}
  2\,z^5\,\omega \,\left( 15 + 10\,\omega  + 7\,{\omega }^2 \right)-
  z^2\,\omega \,\left( 34 + 77\,\omega  + 13\,{\omega }^2 \right)  -
  z^4\,\left( 30 + 46\,\omega  + 93\,{\omega }^2 + 5\,{\omega }^3 \right)
\nonumber\\ & &\mbox{ }\mbox{ }\mbox{ }\mbox{ }
 +
  z^3\,\left( 30 + 102\,\omega  + 41\,{\omega }^2 + 43\,{\omega }^3 \right)
\bigg \} \,\log(z^{2})+ \nonumber\\ %%
& &\hspace{-0.7in}\left ( \frac{1}{12\,\left( -1 + z
\right)\,\omega \,
    {\left( -2\,z + \omega  + z^2\,\omega  \right) }^2}
 \right )\Bigg \{ 8\,{\omega }^3 - 4\,z^7\,{\omega }^3 -
  4\,z\,{\omega }^2\,\left( 7 + \omega  \right)  +
  z^6\,{\omega }^2\,\left( 17 + 5\,\omega  \right)  +
\nonumber\\
  &&\hspace{-0.45in}
  z^5\,\omega \,\left( -28 - 21\,\omega  + {\omega }^2 \right)+
  2\,z^2\,\omega \,\left( 16 + 6\,\omega  + 5\,{\omega }^2 \right)  -
  z^3\,\left( 18 + 16\,\omega  + 25\,{\omega }^2 + {\omega }^3
\right)+\nonumber\\ & &\mbox{ }\mbox{ }\mbox{ }\mbox{ }
  z^4\,\left( 18 + 32\,\omega  + 5\,{\omega }^2 + 5\,{\omega }^3 \right)
\Bigg\}\,\log\left
(\frac{m^{2}_{\eta}}{m^{2}_{\pi}}\right)\label{eq:jcdfulldef}\\
%%%%%%%%%%%%%%
&&+z^2\left ( \frac{6\,\omega  + 6\,z^2\,\omega  -
   z\,\left( 7 + 5\,{\omega }^2 \right) }
  {2\,\omega \,{\left( -2\,z + \omega  + z^2\,\omega  \right) }^2}\right )
  \,{\sqrt{1 - {\omega }^2}}\,\log
   \left(\frac{1 - {\sqrt{1 - {\omega }^2}}}
    {1 + {\sqrt{1 - {\omega }^2}}}\right)+
\nonumber\\
%%%%%%%%
& &\hspace{-0.92in}\left ( \frac{2z}{3\,{\left( -1 + z \right)
}^2}\right )\, {\sqrt{{\left( -1 + z \right) }^2\,\left( -2 - z +
z^2 \right) }}\,\log \left(\frac{2 + 3\,z - 2\,z^2
-2\,{\sqrt{{\left( -1 + z \right) }^2\,\left( -2 - z + z^2 \right)
}}} {2 + 3\,z - 2\,z^2 + 2\,{\sqrt{{\left( -1 + z \right) }^2\,
           \left( -2 - z + z^2 \right) }}}\right)
\nonumber\\ & &\hspace{-0.5in}+z\left (\frac{2\,\omega  +
2\,z^2\,\omega  - z\, \left( 3 + {\omega }^2 \right) }
  {6\,\omega \,{\left( -2\,z + \omega  + z^2\,\omega
 \right) }^2} \right )\times
{\sqrt{12\,z\,\omega  - 12\,z^3\,\omega  - 8\,{\omega }^2 +
      4\,z^4\,{\omega }^2 + z^2\,\left( 9 - 5\,{\omega }^2 \right) }}\,
\nonumber\\ &&\hspace{-0.4in}
  \times\log \left(\frac{3\,z + 2\,\omega  - 2\,z^2\,\omega  -
      {\sqrt{12\,z\,\omega  - 12\,z^3\,\omega  - 8\,{\omega }^2 +
          4\,z^4\,{\omega }^2 + z^2\,\left( 9 - 5\,{\omega }^2 \right) }}}
{3\,z + 2\,\omega  - 2\,z^2\,\omega  +
      {\sqrt{12\,z\,\omega  - 12\,z^3\,\omega  - 8\,{\omega }^2 +
          4\,z^4\,{\omega }^2 + z^2\,\left( 9 - 5\,{\omega }^2 \right) }}}
\right) .\nonumber\eea

\subsection{Results for the Matrix Elements in Full QCD for General
Kinematics}\label{subsec:generalfull}

In this section we present the results for the matrix elements in
full QCD for general kinematics, i.e. with arbitrary masses and
momenta for the mesons. The results are rather long and
complicated so we also present them on the web
site~\cite{quenchedfull}. The results are presented in terms of
the following functions:
\begin{eqnarray} d(p)&=&\frac{1}{p^2} \,
, \nn
\\ \textrm{div}(p)&=&-1-\log\l(\frac{p^2}{\mu^2}\r) \, , \nn \\
B(p,m_1,m_2)&=&\sqrt{(1-m_1^2 d(p)+m_2^2 d(p))^2-4m_2^2d(p)} \, ,
\label{eq:def} \\
L1(p,m)&=&-\log\l(\frac{B(p,m,m)-1}{B(p,m,m)+1}\r) \, , \nn \\
LL(p,m_1,m_2)&=&\log\l(\frac{1-(m_1^2+m_2^2)d(p)+B(p,m_1,m_2)}{1-(m_1^2+m_2^2)d(p)-B(p,m_1,m_2)}\r)
\, . \nn \eea

\subsubsection{Result for the matrix elements of $\op_4$ in Full
QCD with General kinematics:}

We start with the matrix elements of $\op_4$ which we write in the
form
\begin{equation} \la \pi^+ \pi^0|\op_4|K^+\ra=
O_4^{\textrm{\scriptsize{tree}}}\,[1+\frac{m_K^2}{16 \pi^2
f^2}I_{zf}]+O_4^a+O_4^b+O_4^{c+d}+\textrm{counterterms}\, ,
\label{eq:o4fullgen}\eeq where the counterterms are given in
\eq{eq:chiralo4gen}. The two-pion state is symmetrised over the
two-momenta $p_1$ and $p_2$ to ensure that it has isospin 2.
$I_{zf}$ is given in \eq{eq:izffull} and we now present the
remaining ingredients of the right-hand side of \eq{eq:o4fullgen}.
\begin{equation}
%
% lowest order
%
O_{4}^{\mathrm{tree}} = -\frac{\alpha^{(27,1)}\times
   \left\{ 3\sqrt{2}\left(p_{K}\cdot p_{1}+p_{K}\cdot p_{2}
   +2 p_{1}\cdot p_{2}\right)\right\}}{f_{\pi}^{2}f_{K}}\, ,
\end{equation}
\begin{eqnarray}
%
% tadpole diagram
%
O_{4}^{a} &=& \mbox{ }\mbox{ } \frac{\alpha^{(27,1)}\log
\left(\frac{m_{K}^{2}}{{\mu }^2}\right)m_{K}^{2}
    \left( 12{p_1}\cdot {p_2} + 8{p_K}\cdot {p_1} +
      8{p_K}\cdot {p_2} - m_{K}^{2} \right)}{4
    {\sqrt{2}}f^{2}\,f_{\pi}^{2}f_{K}{\pi }^2}\nonumber\\
& &+ \frac{\alpha^{(27,1)}
    \log \left(\frac{m_{\eta}^2}{{\mu }^2}\right)
    \left( 4m_{K}^2 - m_{\pi}^{2} \right)
    \bigg( 2{p_1}\cdot {p_2} +
      3\left( {p_K}\cdot {p_1} + {p_K}\cdot {p_2} \right)  \bigg)
     }{16{\sqrt{2}}f^{2}\,f_{\pi}^{2}f_{K}{\pi }^2}\nonumber\\
& &+ \frac{\alpha^{(27,1)}
  \log \left(\frac{m_{\pi}^{2}}{\mu^2}\right)m_{\pi}^{2}
    \left( 86{p_1}\cdot {p_2} + 29{p_K}\cdot {p_1} +
      29{p_K}\cdot {p_2} + 4m_{\pi}^{2} \right) }{16{\sqrt{2}}
    f^{2}\,f_{\pi}^{2}f_{K}{\pi}^2}\,,
\end{eqnarray}
\begin{eqnarray}
%
% scattering diagram
%
O_{4}^{b} &=& \mbox{
}\frac{\alpha^{(27,1)}}{4\sqrt{2}f^{2}\,f_{\pi}^{2}f_{K}\pi^{2}}
 \times\bigg[
\left( {\textrm{div}}({p_1} + {p_2}) -
    \log (d({p_1} + {p_2})m_{\pi}^2) \right) \nonumber\\
& &\mbox{ }\mbox{ }\mbox{ }\mbox{ }\mbox{ }\mbox{ }\mbox{ }\mbox{
}
   \mbox{ }\mbox{ }\mbox{ }\mbox{ }\mbox{ }\mbox{ }\mbox{ }\mbox{ }
  \times\bigg( 6{\left( {p_1}\cdot {p_2} \right) }^2 +
    3({p_1}\cdot {p_2})\left( {p_K}\cdot {p_1} + {p_K}\cdot {p_2} -
       2m_{\pi}^2 \right)
\nonumber\\ & &\mbox{ }\mbox{ }\mbox{ }\mbox{ }\mbox{ }\mbox{
}\mbox{ }\mbox{ }
   \mbox{ }\mbox{ }\mbox{ }\mbox{ }\mbox{ }\mbox{ }\mbox{ }\mbox{ }
   \mbox{ }\mbox{ }\mbox{ }\mbox{ }\mbox{ }
+
    m_{\pi}^2\left( -{p_K}\cdot {p_1} - {p_K}\cdot {p_2} +
       2m_{\pi}^2 \right)  \bigg)
\mbox{ }\bigg]\nonumber\\ &
&+\frac{\alpha^{(27,1)}}{12\sqrt{2}f^{2}\,f_{\pi}^{2}f_{K}\pi^{2}}
 \times\bigg\{
36{\left( {p_1}\cdot {p_2} \right) }^2 +
  ({p_1}\cdot {p_2})\left( 18{p_K}\cdot {p_1} + 18{p_K}\cdot {p_2} -
     50m_{\pi}^2 \right)
\nonumber\\ & &\mbox{ }\mbox{ }\mbox{ }\mbox{ }\mbox{ }\mbox{
}\mbox{ }\mbox{ }
   \mbox{ }\mbox{ }\mbox{ }\mbox{ }\mbox{ }\mbox{ }\mbox{ }\mbox{ }
   \mbox{ }\mbox{ }\mbox{ }\mbox{ }\mbox{ }
+
  m_{\pi}^2\left( -9{p_K}\cdot {p_1} - 9{p_K}\cdot {p_2} +
     22m_{\pi}^2 \right)
\nonumber\\ & &\mbox{ }\mbox{ }\mbox{ }\mbox{ }\mbox{ }\mbox{
}\mbox{ }\mbox{ }
   \mbox{ }\mbox{ }\mbox{ }\mbox{ }\mbox{ }\mbox{ }\mbox{ }\mbox{ }
   \mbox{ }\mbox{ }\mbox{ }\mbox{ }\mbox{ }
+
  4d({p_1} + {p_2})m_{\pi}^2
\nonumber\\ & &\mbox{ }\mbox{ }\mbox{ }\mbox{ }\mbox{ }\mbox{
}\mbox{ }\mbox{ }
   \mbox{ }\mbox{ }\mbox{ }\mbox{ }\mbox{ }\mbox{ }\mbox{ }\mbox{ }
   \mbox{ }\mbox{ }\mbox{ }\mbox{ }\mbox{ }\mbox{ }\mbox{ }\mbox{ }
   \times\bigg[ 8{\left( {p_1}\cdot {p_2} \right) }^2
      \left( 2 + 3d({p_1} + {p_2})
         m_{\pi}^2 \right)
\nonumber\\ & &\mbox{ }\mbox{ }\mbox{ }\mbox{ }\mbox{ }\mbox{
}\mbox{ }\mbox{ }
   \mbox{ }\mbox{ }\mbox{ }\mbox{ }\mbox{ }\mbox{ }\mbox{ }\mbox{ }
   \mbox{ }\mbox{ }\mbox{ }\mbox{ }\mbox{ }\mbox{ }\mbox{ }\mbox{ }
   \mbox{ }\mbox{ }\mbox{ }
+
     m_{\pi}^2\bigg( 3{p_K}\cdot {p_1} + 3{p_K}\cdot {p_2} +
        4m_{\pi}^2\left( -5 +
           6d({p_1} + {p_2})m_{\pi}^2 \right)\bigg)
\nonumber\\ & &\mbox{ }\mbox{ }\mbox{ }\mbox{ }\mbox{ }\mbox{
}\mbox{ }\mbox{ }
   \mbox{ }\mbox{ }\mbox{ }\mbox{ }\mbox{ }\mbox{ }\mbox{ }\mbox{ }
   \mbox{ }\mbox{ }\mbox{ }\mbox{ }\mbox{ }\mbox{ }\mbox{ }\mbox{ }
   \mbox{ }\mbox{ }\mbox{ }
+ ({p_1}\cdot {p_2})
      \bigg( 3{p_K}\cdot {p_1} + 3{p_K}\cdot {p_2} +
        4m_{\pi}^2\left( -1 +
           12d({p_1} + {p_2})m_{\pi}^2 \right)  \bigg)  \bigg]
\mbox{ }\bigg\}\nonumber\\ &
&+\frac{\alpha^{(27,1)}}{8f^{2}\,f_{\pi}^{2}f_{K}\pi^{2}}
\times\bigg\{ {L1}({p_1} + {p_2},{m_{\pi }})
  {\sqrt{2 - 8d({p_1} + {p_2}){{m_{\pi }}}^2}}\nonumber\\
& &\mbox{ }\mbox{ }\mbox{ }\mbox{ }\mbox{ }\mbox{ }\mbox{ }\mbox{
}
   \mbox{ }\mbox{ }\mbox{ }\mbox{ }\mbox{ }\mbox{ }
\times
  \bigg[
m_{\pi}^2\bigg( -1 +
       2d({p_1} + {p_2})m_{\pi}^2 \bigg)
     \bigg( {p_K}\cdot {p_1} + {p_K}\cdot {p_2} - 2m_{\pi}^2 +
       8d({p_1} + {p_2})m_{\pi}^4 \bigg)
\nonumber\\ & &\mbox{ }\mbox{ }\mbox{ }\mbox{ }\mbox{ }\mbox{
}\mbox{ }\mbox{ }
   \mbox{ }\mbox{ }\mbox{ }\mbox{ }\mbox{ }\mbox{ }\mbox{ }\mbox{ }
   \mbox{ }
+
    2{\left( {p_1}\cdot {p_2} \right) }^2
     \bigg( 3 + 6d({p_1} + {p_2})m_{\pi}^2 +
       8{d({p_1} + {p_2})}^2m_{\pi}^4 \bigg)
\nonumber\\ & &\mbox{ }\mbox{ }\mbox{ }\mbox{ }\mbox{ }\mbox{
}\mbox{ }\mbox{ }
   \mbox{ }\mbox{ }\mbox{ }\mbox{ }\mbox{ }\mbox{ }\mbox{ }\mbox{ }
   \mbox{ }
    + ({p_1}\cdot {p_2})
   \bigg( 3{p_K}\cdot {p_2} -
       6m_{\pi}^2 + 2d({p_1} + {p_2})
        {p_K}\cdot {p_2}m_{\pi}^2 +
       32{d({p_1} + {p_2})}^2m_{\pi}^6 \nonumber\\
& &\mbox{ }\mbox{ }\mbox{ }\mbox{ }\mbox{ }\mbox{ }\mbox{ }\mbox{
}
   \mbox{ }\mbox{ }\mbox{ }\mbox{ }\mbox{ }\mbox{ }\mbox{ }\mbox{ }
   \mbox{ }\mbox{ }\mbox{ }\mbox{ }\mbox{ }\mbox{ }\mbox{ }\mbox{ }
   \mbox{ }\mbox{ }\mbox{ }\mbox{ }\mbox{ }\mbox{ }\mbox{ }
+
       ({p_K}\cdot {p_1})\left( 3 +
          2d({p_1} + {p_2})m_{\pi}^2 \right) \bigg)
\mbox{ }\bigg] \mbox{ }\bigg\} \,,
\end{eqnarray}
and finally the monster
\begin{eqnarray}
%
% hook diagram
%
O_{4}^{c+d} &=&
\frac{\alpha^{(27,1)}}{864\sqrt{2}f^{2}\,f_{\pi}^{2}f_{K}\pi^{2}}\nonumber\\
& &\times\bigg\{ 3B(-p_1 + p_K,{m_{\eta }},{m_K})\times LL(-p_1 +
p_K,{m_{\eta }},{m_K}) \nonumber\\ & &\mbox{ }\mbox{ }\times\bigg[
%%% SimpA
\left({p_1}\cdot {p_2}\right) \big[  -9\left( 3{p_K}\cdot {p_1} +
8m_{K}^2 -
        2m_{\pi}^2 \right)\nonumber\\
& &\mbox{ }\mbox{ }\mbox{ }\mbox{ }\mbox{ }\mbox{ }\mbox{ }\mbox{
}
  \mbox{ }\mbox{ }\mbox{ }\mbox{ }\mbox{ }\mbox{ }\mbox{ }\mbox{ }
  \mbox{ }\mbox{ }\mbox{ }
    +2{d(-{p_1} + {p_K})}^3
      {\left( m_{K}^2 - m_{\pi}^2 \right) }^3
      \left( -2{p_K}\cdot {p_1} + m_{K}^2 + m_{\pi}^2 \right)\nonumber\\
& &\mbox{ }\mbox{ }\mbox{ }\mbox{ }\mbox{ }\mbox{ }\mbox{ }\mbox{
}
  \mbox{ }\mbox{ }\mbox{ }\mbox{ }\mbox{ }\mbox{ }\mbox{ }\mbox{ }
  \mbox{ }\mbox{ }\mbox{ }  -
     {d(-{p_1} + {p_K})}^2
      \left( m_{K}^2 - m_{\pi}^2 \right)\nonumber\\
& &\mbox{ }\mbox{ }\mbox{ }\mbox{ }\mbox{ }\mbox{ }\mbox{ }\mbox{
}
  \mbox{ }\mbox{ }\mbox{ }\mbox{ }\mbox{ }\mbox{ }\mbox{ }\mbox{ }
  \mbox{ }\mbox{ }\mbox{ }\mbox{ }\mbox{ }\mbox{ }\mbox{ }\mbox{ }\mbox{ }
      \times\bigg( 41m_{K}^4 + 44m_{K}^2m_{\pi}^2 -
        13m_{\pi}^4 + 18({p_K}\cdot {p_1})
         \left( -5m_{K}^2 + m_{\pi}^2 \right)  \bigg)\nonumber\\
& &\mbox{ }\mbox{ }\mbox{ }\mbox{ }\mbox{ }\mbox{ }\mbox{ }\mbox{
}
  \mbox{ }\mbox{ }\mbox{ }\mbox{ }\mbox{ }\mbox{ }\mbox{ }\mbox{ }
  \mbox{ }\mbox{ }\mbox{ }  -
     3d(-{p_1} + {p_K})
      \bigg( -5m_{K}^4 - 14m_{K}^2m_{\pi}^2 +
        m_{\pi}^4 + 3({p_K}\cdot {p_1})
         \left( 5m_{K}^2 + m_{\pi}^2 \right)  \bigg)
\mbox{ }\big ]\nonumber\\ & &\mbox{ }\mbox{ }\mbox{ } +
\left({p_K}\cdot {p_2}\right) \big [ -9
      \left( 3{p_K}\cdot {p_1} + 10m_{K}^2 - 4m_{\pi}^2\right)\nonumber\\
& &\mbox{ }\mbox{ }\mbox{ }\mbox{ }\mbox{ }\mbox{ }\mbox{ }\mbox{
}
  \mbox{ }\mbox{ }\mbox{ }\mbox{ }\mbox{ }\mbox{ }\mbox{ }\mbox{ }
  \mbox{ }\mbox{ }\mbox{ }
- 2{d(-{p_1} + {p_K})}^3
      {\left( m_{K}^2 - m_{\pi}^2 \right) }^3
      \left( -2{p_K}\cdot {p_1} + m_{K}^2 + m_{\pi}^2 \right)\nonumber\\
& &\mbox{ }\mbox{ }\mbox{ }\mbox{ }\mbox{ }\mbox{ }\mbox{ }\mbox{
}
  \mbox{ }\mbox{ }\mbox{ }\mbox{ }\mbox{ }\mbox{ }\mbox{ }\mbox{ }
  \mbox{ }\mbox{ }\mbox{ }
 +3d(-{p_1} + {p_K})
      \bigg( 19m_{K}^4 - 2m_{K}^2m_{\pi}^2 +
        m_{\pi}^4 + 9({p_K}\cdot {p_1})
         \left( -3m_{K}^2 + m_{\pi}^2 \right)  \bigg)\nonumber\\
& &\mbox{ }\mbox{ }\mbox{ }\mbox{ }\mbox{ }\mbox{ }\mbox{ }\mbox{
}
  \mbox{ }\mbox{ }\mbox{ }\mbox{ }\mbox{ }\mbox{ }\mbox{ }\mbox{ }
  \mbox{ }\mbox{ }\mbox{ }
 + {d(-{p_1} + {p_K})}^2
      \left( m_{K}^2 - m_{\pi}^2 \right)\nonumber\\
& &\mbox{ }\mbox{ }\mbox{ }\mbox{ }\mbox{ }\mbox{ }\mbox{ }\mbox{
}
  \mbox{ }\mbox{ }\mbox{ }\mbox{ }\mbox{ }\mbox{ }\mbox{ }\mbox{ }
  \mbox{ }\mbox{ }\mbox{ }\mbox{ }\mbox{ }\mbox{ }\mbox{ }\mbox{ }\mbox{ }
      \times
      \bigg( 35m_{K}^4 + 44m_{K}^2m_{\pi}^2 -
        7m_{\pi}^4 + ({p_K}\cdot {p_1})
         \left( -78m_{K}^2 + 6m_{\pi}^2 \right)  \bigg)
\mbox{ }\big ]
%%% end of SimpA
\mbox{ }\bigg]\nonumber\\ & &+2\log \left(d(-p_1 +
p_K)m_{K}^2\right)\nonumber\\ & &\mbox{ }\mbox{ }\times\bigg[
%%% SimpC1
108{d(-{p_1} + {p_K})}^5
   {\left( m_{K}^2 - m_{\pi}^2 \right) }^5
   {\left( -2{p_K}\cdot {p_1} + m_{K}^2 + m_{\pi}^2 \right) }^2
\nonumber\\ & &\mbox{ }\mbox{ }\mbox{ }\mbox{ }\mbox{ }\mbox{
}\mbox{ }
 -
  2{d(-{p_1} + {p_K})}^4
   \left( 2{p_K}\cdot {p_1} - m_{K}^2 - m_{\pi}^2 \right)
   {\left( m_{K}^2 - m_{\pi}^2 \right) }^3\nonumber\\
& &\mbox{ }\mbox{ }\mbox{ }\mbox{ }\mbox{ }\mbox{ }\mbox{ }\mbox{
}\mbox{ }
   \mbox{ }\mbox{ }\mbox{ }\mbox{ }\mbox{ }\mbox{ }
   \times\bigg( 121({p_1}\cdot {p_2})m_{K}^2 - 216m_{K}^4 -
     121({p_1}\cdot {p_2})m_{\pi}^2 -
     108m_{K}^2m_{\pi}^2 - 216m_{\pi}^4\nonumber\\
& &\mbox{ }\mbox{ }\mbox{ }\mbox{ }\mbox{ }\mbox{ }\mbox{ }\mbox{
}\mbox{ }
   \mbox{ }\mbox{ }\mbox{ }\mbox{ }\mbox{ }\mbox{ }\mbox{ }\mbox{ }
 - 121{p_K}\cdot {p_2}\left( m_{K}^2 - m_{\pi}^2 \right)  +
     270{p_K}\cdot {p_1}\left( m_{K}^2 + m_{\pi}^2 \right)
    \bigg) \nonumber\\
& &\mbox{ }\mbox{ }\mbox{ }\mbox{ }\mbox{ }\mbox{ }\mbox{ }
 + 27\bigg( 6{\left( {p_K}\cdot {p_1} \right) }^2 -
     7({p_1}\cdot {p_2})m_{K}^2 + m_{K}^4 + m_{\pi}^4 -
     ({p_K}\cdot {p_2})\left( 7m_{K}^2 + 2m_{\pi}^2 \right) \nonumber\\
& &\mbox{ }\mbox{ }\mbox{ }\mbox{ }\mbox{ }\mbox{ }\mbox{ }\mbox{
}\mbox{ }
   \mbox{ }\mbox{ }\mbox{ }\mbox{ }\mbox{ }\mbox{ }\mbox{ }\mbox{ } +
     ({p_K}\cdot {p_1})\left( -{p_1}\cdot {p_2} - 7{p_K}\cdot {p_2} +
        3m_{K}^2 + 3m_{\pi}^2 \right)  \bigg)
%%% end of SimpC1
\nonumber\\ & &\mbox{ }\mbox{ }\mbox{ }\mbox{ }
    +9 d(-p_1 + p_K)
\nonumber\\ & &\mbox{ }\mbox{ }\mbox{ }\mbox{ }\mbox{ }\mbox{
}\mbox{ }\mbox{ } \times\big[
%%% SimpC2
-\left( ({p_1}\cdot {p_2})m_{K}^4 \right)  + 21m_{K}^6 -
  16({p_1}\cdot {p_2})m_{K}^2m_{\pi}^2 +
  3m_{K}^4m_{\pi}^2 - ({p_1}\cdot {p_2})m_{\pi}^4 -
  3m_{K}^2m_{\pi}^4\nonumber\\
& &\mbox{ }\mbox{ }\mbox{ }\mbox{ }\mbox{ }\mbox{ }\mbox{ }\mbox{
}
   \mbox{ }\mbox{ }\mbox{ }\mbox{ } - 21m_{\pi}^6 +
  ({p_K}\cdot {p_2})\left( -23m_{K}^4 - 8m_{K}^2m_{\pi}^2 +
     49m_{\pi}^4 \right)\nonumber\\
& &\mbox{ }\mbox{ }\mbox{ }\mbox{ }\mbox{ }\mbox{ }\mbox{ }\mbox{
}
   \mbox{ }\mbox{ }\mbox{ }\mbox{ }  +
  18({p_K}\cdot {p_1})\bigg( m_{K}^4 -
     m_{\pi}^2\left( 2{p_1}\cdot {p_2} - 2{p_K}\cdot {p_2} +
        m_{\pi}^2 \right)\bigg)
%%% end of SimpC2
\mbox{ }\big ]\nonumber\\ & &\mbox{ }\mbox{ }\mbox{ }\mbox{ }
    -3 d(-p_1 + p_K)^{2}
\nonumber\\ & &\mbox{ }\mbox{ }\mbox{ }\mbox{ }\mbox{ }\mbox{
}\mbox{ }\mbox{ } \times\big[
%%% SimpC3
-133({p_1}\cdot {p_2})m_{K}^6 + 171m_{K}^8 +
  138({p_1}\cdot {p_2})m_{K}^4m_{\pi}^2 -
  111({p_1}\cdot {p_2})m_{K}^2m_{\pi}^4\nonumber\\
& &\mbox{ }\mbox{ }\mbox{ }\mbox{ }\mbox{ }\mbox{ }\mbox{ }\mbox{
}
   \mbox{ }\mbox{ }\mbox{ }\mbox{ }
 -
  110({p_1}\cdot {p_2})m_{\pi}^6 - 171m_{\pi}^8 -
  90{\left( {p_K}\cdot {p_1} \right) }^2
   \left( m_{K}^4 - m_{\pi}^4 \right) \nonumber\\
& &\mbox{ }\mbox{ }\mbox{ }\mbox{ }\mbox{ }\mbox{ }\mbox{ }\mbox{
}
   \mbox{ }\mbox{ }\mbox{ }\mbox{ } +
  ({p_K}\cdot {p_2})\left( 31m_{K}^6 - 12m_{K}^4m_{\pi}^2 -
     51m_{K}^2m_{\pi}^4 + 248m_{\pi}^6 \right) \nonumber\\
& &\mbox{ }\mbox{ }\mbox{ }\mbox{ }\mbox{ }\mbox{ }\mbox{ }\mbox{
}
   \mbox{ }\mbox{ }\mbox{ }\mbox{ } +
  3({p_K}\cdot {p_1})\bigg( -51m_{K}^6 +
     11({p_1}\cdot {p_2})m_{\pi}^4 + 51m_{\pi}^6 -
     m_{K}^4\left( 19{p_1}\cdot {p_2} + 15m_{\pi}^2 \right) \nonumber\\
& &\mbox{ }\mbox{ }\mbox{ }\mbox{ }\mbox{ }\mbox{ }\mbox{ }\mbox{
}\mbox{ }
   \mbox{ }\mbox{ }\mbox{ }\mbox{ }\mbox{ }\mbox{ }\mbox{ }\mbox{ }
   \mbox{ }\mbox{ }\mbox{ }\mbox{ }\mbox{ }\mbox{ }\mbox{ }\mbox{ }
   \mbox{ }\mbox{ }\mbox{ }\mbox{ }
 +
     ({p_K}\cdot {p_2})\left( 43m_{K}^4 -
        56m_{K}^2m_{\pi}^2 - 59m_{\pi}^4 \right)\nonumber\\
& &\mbox{ }\mbox{ }\mbox{ }\mbox{ }\mbox{ }\mbox{ }\mbox{ }\mbox{
}\mbox{ }
   \mbox{ }\mbox{ }\mbox{ }\mbox{ }\mbox{ }\mbox{ }\mbox{ }\mbox{ }
   \mbox{ }\mbox{ }\mbox{ }\mbox{ }\mbox{ }\mbox{ }\mbox{ }\mbox{ }
   \mbox{ }\mbox{ }\mbox{ }\mbox{ }
 +
     5m_{K}^2\left( 16({p_1}\cdot {p_2})m_{\pi}^2 +
        3m_{\pi}^4 \right)  \bigg)
%%% end of SimpC3
\mbox{ }\big]\nonumber\\ & &\mbox{ }\mbox{ }\mbox{ }\mbox{ }
    + d(-p_1 + p_K)^{3}\left( m_{K}^2 - m_{\pi}^2 \right)\nonumber\\
& &\mbox{ }\mbox{ }\mbox{ }\mbox{ }\mbox{ }\mbox{ }\mbox{ }\mbox{
} \times \big[
%%% SimpC4
-593({p_1}\cdot {p_2})m_{K}^6 + 675m_{K}^8 -
  123({p_1}\cdot {p_2})m_{K}^4m_{\pi}^2 +
  324m_{K}^6m_{\pi}^2\nonumber\\
& &\mbox{ }\mbox{ }\mbox{ }\mbox{ }\mbox{ }\mbox{ }\mbox{ }\mbox{
}
   \mbox{ }\mbox{ }\mbox{ }\mbox{ } +
  153({p_1}\cdot {p_2})m_{K}^2m_{\pi}^4 +
  162m_{K}^4m_{\pi}^4 + 563({p_1}\cdot {p_2})m_{\pi}^6 +
  324m_{K}^2m_{\pi}^6 + 675m_{\pi}^8\nonumber\\
& &\mbox{ }\mbox{ }\mbox{ }\mbox{ }\mbox{ }\mbox{ }\mbox{ }\mbox{
}
   \mbox{ }\mbox{ }\mbox{ }\mbox{ } +
  540{\left( {p_K}\cdot {p_1} \right) }^2
   {\left( m_{K}^2 + m_{\pi}^2 \right) }^2\nonumber\\
& &\mbox{ }\mbox{ }\mbox{ }\mbox{ }\mbox{ }\mbox{ }\mbox{ }\mbox{
}
   \mbox{ }\mbox{ }\mbox{ }\mbox{ } +
  ({p_K}\cdot {p_2})\left( 461m_{K}^6 +
     255m_{K}^4m_{\pi}^2 - 21m_{K}^2m_{\pi}^4 -
     695m_{\pi}^6 \right) \nonumber\\
& &\mbox{ }\mbox{ }\mbox{ }\mbox{ }\mbox{ }\mbox{ }\mbox{ }\mbox{
}
   \mbox{ }\mbox{ }\mbox{ }\mbox{ } -
  (6{p_K}\cdot {p_1})
 \bigg( 252m_{K}^6 +
     151({p_1}\cdot {p_2})m_{\pi}^4 + 252m_{\pi}^6 +
     m_{K}^4\left( -161{p_1}\cdot {p_2} + 108m_{\pi}^2 \right)\nonumber\\
& &\mbox{ }\mbox{ }\mbox{ }\mbox{ }\mbox{ }\mbox{ }\mbox{ }\mbox{
}\mbox{ }
   \mbox{ }\mbox{ }\mbox{ }\mbox{ }\mbox{ }\mbox{ }\mbox{ }\mbox{ }
   \mbox{ }\mbox{ }\mbox{ }\mbox{ }\mbox{ }\mbox{ }\mbox{ }\mbox{ }
   \mbox{ }\mbox{ }\mbox{ }\mbox{ }
     + 39({p_K}\cdot {p_2})
      \left( 3m_{K}^4 + 2m_{K}^2m_{\pi}^2 -
        5m_{\pi}^4 \right)\nonumber\\
& &\mbox{ }\mbox{ }\mbox{ }\mbox{ }\mbox{ }\mbox{ }\mbox{ }\mbox{
}\mbox{ }
   \mbox{ }\mbox{ }\mbox{ }\mbox{ }\mbox{ }\mbox{ }\mbox{ }\mbox{ }
   \mbox{ }\mbox{ }\mbox{ }\mbox{ }\mbox{ }\mbox{ }\mbox{ }\mbox{ }
   \mbox{ }\mbox{ }\mbox{ }\mbox{ }  +
     2m_{K}^2\left( 5{p_1}\cdot {p_2}m_{\pi}^2 +
        54m_{\pi}^4 \right)  \bigg)
%%% end of SimpC4
\mbox{ }\big] \mbox{ }\bigg]\nonumber\\ & &+\log\left(d(-p_1 +
p_K)m_{\eta}^2\right)\nonumber\\ & &\mbox{ }\mbox{ }\times \bigg[
  - ({p_K}\cdot {p_2}) \nonumber\\
& &\mbox{ }\mbox{ }\mbox{ }\mbox{ }\mbox{ }\mbox{ }\mbox{ }\mbox{
} \times\big[
%%% SimpD1
27\left( 3{p_K}\cdot {p_1} + 10m_{K}^2 - 4m_{\pi}^2 \ \right)
\nonumber\\ & &\mbox{ }\mbox{ }\mbox{ }\mbox{ }\mbox{ }\mbox{
}\mbox{ }\mbox{ }
   \mbox{ }\mbox{ }\mbox{ }\mbox{ }
+ 2{d(-{p_1} + {p_K})}^4
   {\left( m_{K}^2 - m_{\pi}^2 \right) }^4
   \left( -2{p_K}\cdot {p_1} + m_{K}^2 + m_{\pi}^2 \right)\nonumber\\
& &\mbox{ }\mbox{ }\mbox{ }\mbox{ }\mbox{ }\mbox{ }\mbox{ }\mbox{
}
   \mbox{ }\mbox{ }\mbox{ }\mbox{ }   -
  9d(-{p_1} + {p_K})
   \bigg( 25m_{K}^4 + 16m_{K}^2m_{\pi}^2 -
     5m_{\pi}^4 + 6({p_K}\cdot {p_1})
      \left( 7m_{K}^2 - m_{\pi}^2 \right)  \bigg) \nonumber\\
& &\mbox{ }\mbox{ }\mbox{ }\mbox{ }\mbox{ }\mbox{ }\mbox{ }\mbox{
}
   \mbox{ }\mbox{ }\mbox{ }\mbox{ }  -
  {d(-{p_1} + {p_K})}^3
   {\left( m_{K}^2 - m_{\pi}^2 \right) }^2
   \bigg( 77m_{K}^4 + 80m_{K}^2m_{\pi}^2 -
     13m_{\pi}^4 \nonumber\\
& &\mbox{ }\mbox{ }\mbox{ }\mbox{ }\mbox{ }\mbox{ }\mbox{ }\mbox{
}\mbox{ }
   \mbox{ }\mbox{ }\mbox{ }\mbox{ }\mbox{ }\mbox{ }\mbox{ }\mbox{ }
   \mbox{ }\mbox{ }\mbox{ }\mbox{ }\mbox{ }\mbox{ }\mbox{ }\mbox{ }
   \mbox{ }\mbox{ }\mbox{ }\mbox{ }\mbox{ }\mbox{ }\mbox{ }\mbox{ }
   \mbox{ }\mbox{ }\mbox{ }\mbox{ }\mbox{ }\mbox{ }\mbox{ }\mbox{ }
   \mbox{ }\mbox{ }\mbox{ }\mbox{ }\mbox{ }\mbox{ }\mbox{ }\mbox{ }
   \mbox{ }\mbox{ }\mbox{ }\mbox{ }
- 18({p_K}\cdot {p_1})
      \left( 9m_{K}^2 - m_{\pi}^2 \right)  \bigg) \nonumber\\
& &\mbox{ }\mbox{ }\mbox{ }\mbox{ }\mbox{ }\mbox{ }\mbox{ }\mbox{
}
   \mbox{ }\mbox{ }\mbox{ }\mbox{ }  +
  3{d(-{p_1} + {p_K})}^2
   \bigg( 82m_{K}^6 + 186m_{K}^4m_{\pi}^2 -
     60m_{K}^2m_{\pi}^4 + 8m_{\pi}^6 \nonumber\\
& &\mbox{ }\mbox{ }\mbox{ }\mbox{ }\mbox{ }\mbox{ }\mbox{ }\mbox{
}\mbox{ }
   \mbox{ }\mbox{ }\mbox{ }\mbox{ }\mbox{ }\mbox{ }\mbox{ }\mbox{ }
   \mbox{ }\mbox{ }\mbox{ }\mbox{ }\mbox{ }\mbox{ }\mbox{ }\mbox{ }
   \mbox{ }\mbox{ }\mbox{ }\mbox{ }\mbox{ }\mbox{ }\mbox{ }\mbox{ }
  \mbox{ }\mbox{ }\mbox{ }\mbox{ }+
     3({p_K}\cdot {p_1})\left( -77m_{K}^4 +
        4m_{K}^2m_{\pi}^2 + m_{\pi}^4 \right)  \bigg)
%%% end of SimpD1
\mbox{ }\big]\nonumber\\ & &\mbox{ }\mbox{ }\mbox{ }\mbox{ }
  + ({p_1}\cdot {p_2}) \nonumber\\
& &\mbox{ }\mbox{ }\mbox{ }\mbox{ }\mbox{ }\mbox{ }\mbox{ }\mbox{
} \times\big[
%%% SimpD2
-27\left( 3{p_K}\cdot {p_1} + 8m_{K}^2 - 2m_{\pi}^2 \ \right)
\nonumber\\ & &\mbox{ }\mbox{ }\mbox{ }\mbox{ }\mbox{ }\mbox{
}\mbox{ }\mbox{ }
   \mbox{ }\mbox{ }\mbox{ }\mbox{ }  + 2{d(-{p_1} + {p_K})}^4
   {\left( m_{K}^2 - m_{\pi}^2 \right) }^4
   \left( -2{p_K}\cdot {p_1} + m_{K}^2 + m_{\pi}^2 \right)\nonumber\\
& &\mbox{ }\mbox{ }\mbox{ }\mbox{ }\mbox{ }\mbox{ }\mbox{ }\mbox{
}
   \mbox{ }\mbox{ }\mbox{ }\mbox{ }   -
  {d(-{p_1} + {p_K})}^3
   {\left( m_{K}^2 - m_{\pi}^2 \right) }^2
   \bigg( 83m_{K}^4 + 80m_{K}^2m_{\pi}^2 -
     19m_{\pi}^4  \nonumber\\
& &\mbox{ }\mbox{ }\mbox{ }\mbox{ }\mbox{ }\mbox{ }\mbox{ }\mbox{
}\mbox{ }
   \mbox{ }\mbox{ }\mbox{ }\mbox{ }\mbox{ }\mbox{ }\mbox{ }\mbox{ }
   \mbox{ }\mbox{ }\mbox{ }\mbox{ }\mbox{ }\mbox{ }\mbox{ }\mbox{ }
   \mbox{ }\mbox{ }\mbox{ }\mbox{ }\mbox{ }\mbox{ }\mbox{ }\mbox{ }
   \mbox{ }\mbox{ }\mbox{ }\mbox{ }\mbox{ }\mbox{ }\mbox{ }\mbox{ }
   \mbox{ }\mbox{ }\mbox{ }\mbox{ }\mbox{ }\mbox{ }\mbox{ }\mbox{ }
   \mbox{ }\mbox{ }\mbox{ }\mbox{ }
   -6({p_K}\cdot {p_1})
      \left( 29m_{K}^2 - 5m_{\pi}^2 \right)  \bigg) \nonumber\\
& &\mbox{ }\mbox{ }\mbox{ }\mbox{ }\mbox{ }\mbox{ }\mbox{ }\mbox{
}
   \mbox{ }\mbox{ }\mbox{ }\mbox{ }  -
  9d(-{p_1} + {p_K})
   \bigg( 67m_{K}^4 - 32m_{K}^2m_{\pi}^2 + m_{\pi}^4 +
  6({p_K}\cdot {p_1})\left( 7m_{K}^2 - m_{\pi}^2 \right) \bigg) \nonumber\\
& &\mbox{ }\mbox{ }\mbox{ }\mbox{ }\mbox{ }\mbox{ }\mbox{ }\mbox{
}
   \mbox{ }\mbox{ }\mbox{ }\mbox{ }
 + 3{d(-{p_1} + {p_K})}^2
   \bigg( -3({p_K}\cdot {p_1})
      \left( 119m_{K}^4 - 52m_{K}^2m_{\pi}^2 +
        5m_{\pi}^4 \right) \nonumber\\
& &\mbox{ }\mbox{ }\mbox{ }\mbox{ }\mbox{ }\mbox{ }\mbox{ }\mbox{
}\mbox{ }
   \mbox{ }\mbox{ }\mbox{ }\mbox{ }\mbox{ }\mbox{ }\mbox{ }\mbox{ }
   \mbox{ }\mbox{ }\mbox{ }\mbox{ }\mbox{ }\mbox{ }\mbox{ }\mbox{ }
   \mbox{ }\mbox{ }\mbox{ }\mbox{ }\mbox{ }\mbox{ }\mbox{ }\mbox{ }
  \mbox{ }\mbox{ }\mbox{ }\mbox{ } +
     2\left( 74m_{K}^6 + 84m_{K}^4m_{\pi}^2 -
        57m_{K}^2m_{\pi}^4 + 7m_{\pi}^6 \right)  \bigg)
%%% end of SimpD2
\mbox{ }\big] \mbox{ }\bigg]\nonumber\\ & &-27 \log (d(-p_1 +
p_K)m_{\pi}^2)\nonumber\\ & & \mbox{ }\mbox{ }\times \bigg[
%%% SimpE1
-12{\left( {p_K}\cdot {p_1} \right) }^2 +
  11({p_K}\cdot {p_1})({p_K}\cdot {p_2}) - 6({p_K}\cdot {p_1})m_{K}^2 +
  4({p_K}\cdot {p_2})m_{K}^2 - 2m_{K}^4
\nonumber\\ & &\mbox{ }\mbox{ }\mbox{ }\mbox{ }\mbox{ }\mbox{
}\mbox{ }
 -
  6({p_K}\cdot {p_1})m_{\pi}^2 +
  8({p_K}\cdot {p_2})m_{\pi}^2 - 2m_{\pi}^4
\nonumber\\ & &\mbox{ }\mbox{ }\mbox{ }\mbox{ }\mbox{ }\mbox{
}\mbox{ } +
  8{d(-{p_1} + {p_K})}^5
   {\left( m_{K}^2 - m_{\pi}^2 \right) }^5
   {\left( -2{p_K}\cdot {p_1} + m_{K}^2 + m_{\pi}^2 \right) }^2\nonumber\\
& &\mbox{ }\mbox{ }\mbox{ }\mbox{ }\mbox{ }\mbox{ }\mbox{ }
 +
  ({p_1}\cdot {p_2})\left( -{p_K}\cdot {p_1} + 6m_{K}^2 +
     2m_{\pi}^2 \right)
%%% end of SimpE1
\nonumber\\ & &\mbox{ }\mbox{ }\mbox{ }\mbox{ }
    -2 d(-p_1 + p_K)^{4}\times
    \left( 2{p_K}\cdot {p_1} - m_{K}^2 - m_{\pi}^2 \right)\times
    \left( m_{K}^2 - m_{\pi}^2 \right)^{3}\nonumber\\
& &\mbox{ }\mbox{ }\mbox{ }\mbox{ }\mbox{ }\mbox{ }\mbox{ }\mbox{
} \times\big[
%%% SimpE2
9({p_1}\cdot {p_2})m_{K}^2 - 16m_{K}^4 -
  9({p_1}\cdot {p_2})m_{\pi}^2 - 8m_{K}^2m_{\pi}^2 -
  16m_{\pi}^4 \nonumber\\
& &\mbox{ }\mbox{ }\mbox{ }\mbox{ }\mbox{ }\mbox{ }\mbox{ }\mbox{
}
   \mbox{ }\mbox{ }\mbox{ }\mbox{ }
   - 9({p_K}\cdot {p_2})
   \left( m_{K}^2 - m_{\pi}^2 \right)  +
  20({p_K}\cdot {p_1})\left( m_{K}^2 + m_{\pi}^2 \right)
%%% end of SimpE2
\mbox{ }\big] \nonumber\\ & &\mbox{ }\mbox{ }\mbox{ }\mbox{ }
    +d(-p_1 + p_K)\nonumber\\
& &\mbox{ }\mbox{ }\mbox{ }\mbox{ }\mbox{ }\mbox{ }\mbox{ }\mbox{
} \times\big[
%%% SimpE3
-23({p_1}\cdot {p_2})m_{K}^4 + 14m_{K}^6 +
  2m_{K}^4m_{\pi}^2 - ({p_1}\cdot {p_2})m_{\pi}^4 -
  2m_{K}^2m_{\pi}^4 \nonumber\\
& &\mbox{ }\mbox{ }\mbox{ }\mbox{ }\mbox{ }\mbox{ }\mbox{ }\mbox{
}
   \mbox{ }\mbox{ }\mbox{ }\mbox{ }
- 14m_{\pi}^6 +
  ({p_K}\cdot {p_2})\left( -7m_{K}^4 + 31m_{\pi}^4 \right)
\nonumber\\ & &\mbox{ }\mbox{ }\mbox{ }\mbox{ }\mbox{ }\mbox{
}\mbox{ }\mbox{ }
   \mbox{ }\mbox{ }\mbox{ }\mbox{ }
+
  2({p_K}\cdot {p_1})\bigg( -7({p_1}\cdot {p_2})m_{K}^2 +
     6m_{K}^4 - 11({p_1}\cdot {p_2})m_{\pi}^2 -
     6m_{\pi}^4
\nonumber\\ & &\mbox{ }\mbox{ }\mbox{ }\mbox{ }\mbox{ }\mbox{
}\mbox{ }\mbox{ }
   \mbox{ }\mbox{ }\mbox{ }\mbox{ }\mbox{ }\mbox{ }\mbox{ }\mbox{ }
   \mbox{ }\mbox{ }\mbox{ }\mbox{ }\mbox{ }\mbox{ }\mbox{ }\mbox{ }
   \mbox{ }\mbox{ }\mbox{ }\mbox{ }\mbox{ }\mbox{ }
+ ({p_K}\cdot {p_2})
      \left( 7m_{K}^2 + 11m_{\pi}^2 \right)  \bigg)
%%% end of SimpE3
\mbox{ }\big] \nonumber\\ & &\mbox{ }\mbox{ }\mbox{ }\mbox{ }
    +d(-p_1 + p_K)^{3} \left( m_{K}^2 - m_{\pi}^2 \right)\nonumber\\
& &\mbox{ }\mbox{ }\mbox{ }\mbox{ }\mbox{ }\mbox{ }\mbox{ }\mbox{
} \times\big[
%%% SimpE4
-47({p_1}\cdot {p_2})m_{K}^6 + 50m_{K}^8 -
  9({p_1}\cdot {p_2})m_{K}^4m_{\pi}^2 +
  24m_{K}^6m_{\pi}^2 +
  15({p_1}\cdot {p_2})m_{K}^2m_{\pi}^4
\nonumber\\ & &\mbox{ }\mbox{ }\mbox{ }\mbox{ }\mbox{ }\mbox{
}\mbox{ }\mbox{ }
   \mbox{ }\mbox{ }\mbox{ }\mbox{ }
 +
  12m_{K}^4m_{\pi}^4 + 41({p_1}\cdot {p_2})m_{\pi}^6
  +24m_{K}^2m_{\pi}^6 + 50m_{\pi}^8 +
  40{\left( {p_K}\cdot {p_1} \right) }^2
   {\left( m_{K}^2 + m_{\pi}^2 \right) }^2
\nonumber\\ & &\mbox{ }\mbox{ }\mbox{ }\mbox{ }\mbox{ }\mbox{
}\mbox{ }\mbox{ }
   \mbox{ }\mbox{ }\mbox{ }\mbox{ }
+
  ({p_K}\cdot {p_2})\left( 37m_{K}^6 + 19m_{K}^4m_{\pi}^2 -
     5m_{K}^2m_{\pi}^4 - 51m_{\pi}^6 \right)
\nonumber\\ & &\mbox{ }\mbox{ }\mbox{ }\mbox{ }\mbox{ }\mbox{
}\mbox{ }\mbox{ }
   \mbox{ }\mbox{ }\mbox{ }\mbox{ }
  -
  2({p_K}\cdot {p_1})\bigg( 56m_{K}^6 +
     33{p_1}\cdot {p_2}m_{\pi}^4 + 56m_{\pi}^6 +
     3m_{K}^4\left( -13{p_1}\cdot {p_2} + 8m_{\pi}^2\right)
\nonumber\\ & &\mbox{ }\mbox{ }\mbox{ }\mbox{ }\mbox{ }\mbox{
}\mbox{ }\mbox{ }
   \mbox{ }\mbox{ }\mbox{ }\mbox{ }\mbox{ }\mbox{ }\mbox{ }\mbox{ }
   \mbox{ }\mbox{ }\mbox{ }\mbox{ }\mbox{ }\mbox{ }\mbox{ }\mbox{ }
   \mbox{ }\mbox{ }\mbox{ }\mbox{ }\mbox{ }\mbox{ }
  + ({p_K}\cdot {p_2})\left( 29m_{K}^4 +
        14m_{K}^2m_{\pi}^2 - 43m_{\pi}^4 \right)
\nonumber\\ & &\mbox{ }\mbox{ }\mbox{ }\mbox{ }\mbox{ }\mbox{
}\mbox{ }\mbox{ }
   \mbox{ }\mbox{ }\mbox{ }\mbox{ }\mbox{ }\mbox{ }\mbox{ }\mbox{ }
   \mbox{ }\mbox{ }\mbox{ }\mbox{ }\mbox{ }\mbox{ }\mbox{ }\mbox{ }
   \mbox{ }\mbox{ }\mbox{ }\mbox{ }\mbox{ }\mbox{ }
 +
     6m_{K}^2\left( ({p_1}\cdot {p_2})m_{\pi}^2 +
        4m_{\pi}^4 \right)  \bigg)
%%% end of SimpE4
\mbox{ }\big] \nonumber\\ & &\mbox{ }\mbox{ }\mbox{ }\mbox{ }
    -d(-p_1 + p_K)^{2}\nonumber\\
& &\mbox{ }\mbox{ }\mbox{ }\mbox{ }\mbox{ }\mbox{ }\mbox{ }\mbox{
} \times\big[
%%% SimpE5
-20{\left( {p_K}\cdot {p_1} \right) }^2
   \left( m_{K}^4 - m_{\pi}^4 \right)
\nonumber\\ & &\mbox{ }\mbox{ }\mbox{ }\mbox{ }\mbox{ }\mbox{
}\mbox{ }\mbox{ }
   \mbox{ }\mbox{ }\mbox{ }\mbox{ }
+
  ({p_K}\cdot {p_1})\bigg( -34m_{K}^6 +
     9{p_1}\cdot {p_2}m_{\pi}^4 + 34m_{\pi}^6 +
     m_{K}^4\left( 27{p_1}\cdot {p_2} - 10m_{\pi}^2 \right)
\nonumber\\ & &\mbox{ }\mbox{ }\mbox{ }\mbox{ }\mbox{ }\mbox{
}\mbox{ }\mbox{ }
   \mbox{ }\mbox{ }\mbox{ }\mbox{ }\mbox{ }\mbox{ }\mbox{ }\mbox{ }
   \mbox{ }\mbox{ }\mbox{ }\mbox{ }\mbox{ }\mbox{ }\mbox{ }\mbox{ }
   \mbox{ }\mbox{ }\mbox{ }\mbox{ }\mbox{ }\mbox{ }
+
     3({p_K}\cdot {p_2})\left( m_{K}^4 -
        12m_{K}^2m_{\pi}^2 - 13m_{\pi}^4 \right)
\nonumber\\ & &\mbox{ }\mbox{ }\mbox{ }\mbox{ }\mbox{ }\mbox{
}\mbox{ }\mbox{ }
   \mbox{ }\mbox{ }\mbox{ }\mbox{ }\mbox{ }\mbox{ }\mbox{ }\mbox{ }
   \mbox{ }\mbox{ }\mbox{ }\mbox{ }\mbox{ }\mbox{ }\mbox{ }\mbox{ }
   \mbox{ }\mbox{ }\mbox{ }\mbox{ }\mbox{ }\mbox{ }
 +
     2m_{K}^2\left( (18{p_1}\cdot {p_2})m_{\pi}^2 +
        5m_{\pi}^4 \right)  \bigg)
%%% end of SimpE5
\nonumber\\ & &\mbox{ }\mbox{ }\mbox{ }\mbox{ }\mbox{ }\mbox{
}\mbox{ }\mbox{ }
   \mbox{ }\mbox{ }\mbox{ }\mbox{ }
+ 2\times \bigg(
%%% SimpE6
-23({p_1}\cdot {p_2})m_{K}^6 + 19m_{K}^8 +
  6({p_1}\cdot {p_2})m_{K}^4m_{\pi}^2 -
  6({p_1}\cdot {p_2})m_{K}^2m_{\pi}^4
\nonumber\\ & &\mbox{ }\mbox{ }\mbox{ }\mbox{ }\mbox{ }\mbox{
}\mbox{ }\mbox{ }
   \mbox{ }\mbox{ }\mbox{ }\mbox{ }\mbox{ }\mbox{ }\mbox{ }\mbox{ }
   \mbox{ }\mbox{ }\mbox{ }\mbox{ }\mbox{ }\mbox{ }\mbox{ }
 -
  13({p_1}\cdot {p_2})m_{\pi}^6 - 19m_{\pi}^8
\nonumber\\ & &\mbox{ }\mbox{ }\mbox{ }\mbox{ }\mbox{ }\mbox{
}\mbox{ }\mbox{ }
   \mbox{ }\mbox{ }\mbox{ }\mbox{ }\mbox{ }\mbox{ }\mbox{ }\mbox{ }
   \mbox{ }\mbox{ }\mbox{ }\mbox{ }\mbox{ }\mbox{ }\mbox{ }
+
  ({p_K}\cdot {p_2})\left( 8m_{K}^6 + 9m_{K}^4m_{\pi}^2 -
     9m_{K}^2m_{\pi}^4 + 28m_{\pi}^6 \right)
%%% end of SimpE6
\bigg) \mbox{ }\big] \mbox{ }\bigg]\nonumber\\ & &+ 27 B(-p_1 +
p_K,{m_{\pi }},{m_K})\times LL(-p_1 + p_K,{m_{\pi }},{m_K})
\nonumber\\ & & \mbox{ }\mbox{ }\times \bigg[
%%% SimpF1
12{\left( {p_K}\cdot {p_1} \right) }^2 -
  11({p_K}\cdot {p_1})({p_K}\cdot {p_2}) + 6({p_K}\cdot {p_1})m_{K}^2 -
  4({p_K}\cdot {p_2})m_{K}^2 + 2m_{K}^4
\nonumber\\ & &\mbox{ }\mbox{ }\mbox{ }\mbox{ }\mbox{ }\mbox{
}\mbox{ } +
  6({p_K}\cdot {p_1})m_{\pi}^2 -
  8({p_K}\cdot {p_2})m_{\pi}^2 + 2m_{\pi}^4
\nonumber\\ & &\mbox{ }\mbox{ }\mbox{ }\mbox{ }\mbox{ }\mbox{
}\mbox{ } +
  8{d(-{p_1} + {p_K})}^4
   {\left( m_{K}^2 - m_{\pi}^2 \right) }^4
   {\left( -2{p_K}\cdot {p_1} + m_{K}^2 + m_{\pi}^2 \right) }^2
\nonumber\\ & &\mbox{ }\mbox{ }\mbox{ }\mbox{ }\mbox{ }\mbox{
}\mbox{ }
-
  6{d(-{p_1} + {p_K})}^3
   \left( 2{p_K}\cdot {p_1} - m_{K}^2 - m_{\pi}^2 \right)
   {\left( m_{K}^2 - m_{\pi}^2 \right) }^2
\nonumber\\ & &\mbox{ }\mbox{ }\mbox{ }\mbox{ }\mbox{ }\mbox{
}\mbox{ }
   \mbox{ }\mbox{ }\mbox{ }\mbox{ }\mbox{ }
\times
   \bigg( 3({p_1}\cdot {p_2})m_{K}^2 - 4m_{K}^4 -
     3({p_1}\cdot {p_2})m_{\pi}^2 - 4m_{\pi}^4
\nonumber\\ & &\mbox{ }\mbox{ }\mbox{ }\mbox{ }\mbox{ }\mbox{
}\mbox{ }\mbox{ }
   \mbox{ }\mbox{ }\mbox{ }\mbox{ }\mbox{ }\mbox{ }\mbox{ }\mbox{ }
-
     3({p_K}\cdot {p_2})\left( m_{K}^2 - m_{\pi}^2 \right)  +
     4({p_K}\cdot {p_1})\left( m_{K}^2 + m_{\pi}^2 \right)  \bigg)
\nonumber\\ & &\mbox{ }\mbox{ }\mbox{ }\mbox{ }\mbox{ }\mbox{
}\mbox{ } + ({p_1}\cdot {p_2})\bigg( {p_K}\cdot {p_1} -
     2\left( 3m_{K}^2 + m_{\pi}^2 \right)  \bigg)
%%% end of SimpF1
\nonumber\\ & &\mbox{ }\mbox{ }\mbox{ }\mbox{ }
 + d(-p_1 + p_K)^{2}\nonumber\\
& &\mbox{ }\mbox{ }\mbox{ }\mbox{ }\mbox{ }\mbox{ }\mbox{ }\mbox{
} \times\big[
%%% SimpF2
-29({p_1}\cdot {p_2})m_{K}^6 + 26m_{K}^8 +
  9({p_1}\cdot {p_2})m_{K}^4m_{\pi}^2 -
  8m_{K}^6m_{\pi}^2 -
  3({p_1}\cdot {p_2})m_{K}^2m_{\pi}^4 \nonumber\\
& &\mbox{ }\mbox{ }\mbox{ }\mbox{ }\mbox{ }\mbox{ }\mbox{ }\mbox{
}
   \mbox{ }\mbox{ }\mbox{ }\mbox{ }
-
  4m_{K}^4m_{\pi}^4 + 23({p_1}\cdot {p_2})m_{\pi}^6 -
  8m_{K}^2m_{\pi}^6 + 26m_{\pi}^8 \nonumber\\
& &\mbox{ }\mbox{ }\mbox{ }\mbox{ }\mbox{ }\mbox{ }\mbox{ }\mbox{
}
   \mbox{ }\mbox{ }\mbox{ }\mbox{ }
-
  8{\left( {p_K}\cdot {p_1} \right) }^2
   \left( m_{K}^4 - 6m_{K}^2m_{\pi}^2 + m_{\pi}^4 \right)  \nonumber\\
& &\mbox{ }\mbox{ }\mbox{ }\mbox{ }\mbox{ }\mbox{ }\mbox{ }\mbox{
}
   \mbox{ }\mbox{ }\mbox{ }\mbox{ }
+ ({p_K}\cdot {p_2})\left( 19m_{K}^6 + m_{K}^4m_{\pi}^2 +
     13m_{K}^2m_{\pi}^4 - 33m_{\pi}^6 \right)  \nonumber\\
& &\mbox{ }\mbox{ }\mbox{ }\mbox{ }\mbox{ }\mbox{ }\mbox{ }\mbox{
}
   \mbox{ }\mbox{ }\mbox{ }\mbox{ }
-
  2({p_K}\cdot {p_1})\bigg( 20m_{K}^6 +
     5m_{\pi}^4\left( 3{p_1}\cdot {p_2} + 4m_{\pi}^2 \
\right)  - m_{K}^4\left( 21{p_1}\cdot {p_2} + 4m_{\pi}^2 \ \right)
\nonumber\\ & &\mbox{ }\mbox{ }\mbox{ }\mbox{ }\mbox{ }\mbox{
}\mbox{ }\mbox{ }
   \mbox{ }\mbox{ }\mbox{ }\mbox{ }\mbox{ }\mbox{ }\mbox{ }\mbox{ }
   \mbox{ }\mbox{ }\mbox{ }\mbox{ }\mbox{ }\mbox{ }\mbox{ }\mbox{ }
   \mbox{ }\mbox{ }\mbox{ }\mbox{ }
+ ({p_K}\cdot {p_2})\left( 11m_{K}^4 +
        14m_{K}^2m_{\pi}^2 - 25m_{\pi}^4 \right) \nonumber\\
& &\mbox{ }\mbox{ }\mbox{ }\mbox{ }\mbox{ }\mbox{ }\mbox{ }\mbox{
}
   \mbox{ }\mbox{ }\mbox{ }\mbox{ }\mbox{ }\mbox{ }\mbox{ }\mbox{ }
   \mbox{ }\mbox{ }\mbox{ }\mbox{ }\mbox{ }\mbox{ }\mbox{ }\mbox{ }
   \mbox{ }\mbox{ }\mbox{ }\mbox{ }
 +
     m_{K}^2\left( 6({p_1}\cdot {p_2})m_{\pi}^2 -
        4m_{\pi}^4 \right)  \bigg)
%%% end of SimpF2
\mbox{ }\big] \nonumber\\ & &\mbox{ }\mbox{ }\mbox{ }\mbox{ }
 + d(-p_1 + p_K)\nonumber\\
& &\mbox{ }\mbox{ }\mbox{ }\mbox{ }\mbox{ }\mbox{ }\mbox{ }\mbox{
} \times\big[
%%% SimpF3
17({p_1}\cdot {p_2})m_{K}^4 - 12m_{K}^6 -
  8({p_1}\cdot {p_2})m_{K}^2m_{\pi}^2 -
  3({p_1}\cdot {p_2})m_{\pi}^4 - 12m_{\pi}^6
 \nonumber\\
& &\mbox{ }\mbox{ }\mbox{ }\mbox{ }\mbox{ }\mbox{ }\mbox{ }\mbox{
}
   \mbox{ }\mbox{ }\mbox{ }\mbox{ }
 +
  12{\left( {p_K}\cdot {p_1} \right) }^2
   \left( m_{K}^2 + m_{\pi}^2 \right)  +
  ({p_K}\cdot {p_2})\left( 3m_{K}^4 - 12m_{K}^2m_{\pi}^2 +
     23m_{\pi}^4 \right)
\nonumber\\ & &\mbox{ }\mbox{ }\mbox{ }\mbox{ }\mbox{ }\mbox{
}\mbox{ }\mbox{ }
   \mbox{ }\mbox{ }\mbox{ }\mbox{ }
+
  ({p_K}\cdot {p_1})\bigg( ({p_K}\cdot {p_2})
      \left( -25m_{K}^2 + 11m_{\pi}^2 \right)
  \nonumber\\
& &\mbox{ }\mbox{ }\mbox{ }\mbox{ }\mbox{ }\mbox{ }\mbox{ }\mbox{
}
   \mbox{ }\mbox{ }\mbox{ }\mbox{ }\mbox{ }\mbox{ }\mbox{ }\mbox{ }
   \mbox{ }\mbox{ }\mbox{ }\mbox{ }\mbox{ }\mbox{ }\mbox{ }\mbox{ }
   \mbox{ }\mbox{ }\mbox{ }
     -3\left( 2m_{K}^4 + 7({p_1}\cdot {p_2})m_{\pi}^2 +
        2m_{\pi}^4 - m_{K}^2
         \left( 5{p_1}\cdot {p_2} + 4m_{\pi}^2 \right)  \right) \bigg)
%%% end of SimpF3
\mbox{ }\big] \mbox{ }\bigg]\mbox{ } \nonumber\\
&+&(p_{1}\leftrightarrow p_{2})\bigg\} \nonumber\\ &
&+\frac{\alpha^{(27,1)}}{144\sqrt{2}f^{2}\,f_{\pi}^{2}f_{K}\pi^{2}}
\times \bigg[
%%% SimpB1
-72{d(-{p_1} + {p_K})}^4
   {\left( m_{K}^2 - m_{\pi}^2 \right) }^4
   {\left( -2{p_K}\cdot {p_1} + m_{K}^2 + m_{\pi}^2 \right) }^2\nonumber\\
& &\mbox{ }\mbox{ }\mbox{ }\mbox{ }\mbox{ }\mbox{ }\mbox{ }\mbox{
}\mbox{ } -72{d(-{p_2} + {p_K})}^4
   {\left( m_{K}^2 - m_{\pi}^2 \right) }^4
   {\left( -2{p_K}\cdot {p_2} + m_{K}^2 + m_{\pi}^2 \right) }^2\nonumber\\
& &\mbox{ }\mbox{ }\mbox{ }\mbox{ }\mbox{ }\mbox{ }\mbox{ }\mbox{
}\mbox{ } +4{d(-{p_1} + {p_K})}^3
   \left( 2{p_K}\cdot {p_1} - m_{K}^2 - m_{\pi}^2 \right)
   {\left( m_{K}^2 - m_{\pi}^2 \right) }^2\nonumber\\
& &\mbox{ }\mbox{ }\mbox{ }\mbox{ }\mbox{ }\mbox{ }\mbox{ }\mbox{
}\mbox{ } \mbox{ }\mbox{ }\mbox{ }\mbox{ }
 \times\bigg( 41({p_1}\cdot {p_2})m_{K}^2 - 63m_{K}^4 -
     41({p_1}\cdot {p_2})m_{\pi}^2 - 18m_{K}^2m_{\pi}^2\nonumber\\
& &\mbox{ }\mbox{ }\mbox{ }\mbox{ }\mbox{ }\mbox{ }\mbox{ }\mbox{
}\mbox{ } \mbox{ }\mbox{ }\mbox{ }\mbox{ }\mbox{ }\mbox{ }\mbox{ }
   -63m_{\pi}^4 - 41({p_K}\cdot {p_2})
      \left( m_{K}^2 - m_{\pi}^2 \right)  +
     72({p_K}\cdot {p_1})\left( m_{K}^2 + m_{\pi}^2 \right)
\bigg) \nonumber\\ & &\mbox{ }\mbox{ }\mbox{ }\mbox{ }\mbox{
}\mbox{ }\mbox{ }\mbox{ }\mbox{ } + 4{d(-{p_2} + {p_K})}^3
   \left( 2{p_K}\cdot {p_2} - m_{K}^2 - m_{\pi}^2 \right)
   {\left( m_{K}^2 - m_{\pi}^2 \right) }^2\nonumber\\
& &\mbox{ }\mbox{ }\mbox{ }\mbox{ }\mbox{ }\mbox{ }\mbox{ }\mbox{
}\mbox{ } \mbox{ }\mbox{ }\mbox{ }\mbox{ }
 \times\bigg( 41({p_1}\cdot {p_2})m_{K}^2 - 63m_{K}^4 -
     41({p_1}\cdot {p_2})m_{\pi}^2 - 18m_{K}^2m_{\pi}^2 \nonumber\\
& &\mbox{ }\mbox{ }\mbox{ }\mbox{ }\mbox{ }\mbox{ }\mbox{ }\mbox{
}\mbox{ } \mbox{ }\mbox{ }\mbox{ }\mbox{ }\mbox{ }\mbox{ }\mbox{ }
- 63m_{\pi}^4 - 41({p_K}\cdot {p_1})
      \left( m_{K}^2 - m_{\pi}^2 \right)  +
     72({p_K}\cdot {p_2})\left( m_{K}^2 + m_{\pi}^2 \right)  \
\bigg)\nonumber\\ & &\mbox{ }\mbox{ }\mbox{ }\mbox{ }\mbox{
}\mbox{ }\mbox{ }\mbox{ }\mbox{ } - 18{\textrm{div}}(-{p_2} +
{p_K})
   \bigg( 6{\left( {p_K}\cdot {p_2} \right) }^2 -
     7{p_1}\cdot {p_2}m_{K}^2 + m_{K}^4 + m_{\pi}^4\nonumber\\
& &\mbox{ }\mbox{ }\mbox{ }\mbox{ }\mbox{ }\mbox{ }\mbox{ }\mbox{
}\mbox{ } \mbox{ }\mbox{ }\mbox{ }\mbox{ }\mbox{ }\mbox{ }\mbox{ }
-
     {p_K}\cdot {p_1}\left( 7{p_K}\cdot {p_2} + 7m_{K}^2 +
        2m_{\pi}^2 \right)  +
     {p_K}\cdot {p_2}\left( -{p_1}\cdot {p_2} + 3m_{K}^2 +
        3m_{\pi}^2 \right)  \bigg)
%%% end od SimpB1
\nonumber\\ & &\mbox{ }\mbox{ }\mbox{ }\mbox{ }
     - 3\times\big[
%%% SimpB2
\mbox{ }36\left( 2 + {\textrm{div}}(-{p_1} + {p_K}) \ \right)
{\left( {p_K}\cdot {p_1} \right) }^2 +
  72{\left( {p_K}\cdot {p_2} \right) }^2 -
  278({p_1}\cdot {p_2})m_{K}^2\nonumber\\
& &\mbox{ }\mbox{ }\mbox{ }\mbox{ }\mbox{ }\mbox{ }\mbox{ }\mbox{
}
      \mbox{ }\mbox{ }\mbox{ }
 -42{\textrm{div}}(-{p_1} + {p_K})
   ({p_1}\cdot {p_2})m_{K}^2 + 42m_{K}^4 +
  6{\textrm{div}}(-{p_1} + {p_K})m_{K}^4\nonumber\\
& &\mbox{ }\mbox{ }\mbox{ }\mbox{ }\mbox{ }\mbox{ }\mbox{ }\mbox{
}
      \mbox{ }\mbox{ }\mbox{ } +
  (10{p_1}\cdot {p_2})m_{\pi}^2 + 4m_{K}^2m_{\pi}^2 +
  42m_{\pi}^4 + 6{\textrm{div}}(-{p_1} +
     {p_K})m_{\pi}^4 \nonumber\\
& &\mbox{ }\mbox{ }\mbox{ }\mbox{ }\mbox{ }\mbox{ }\mbox{ }\mbox{
}
      \mbox{ }\mbox{ }\mbox{ }+
  ({p_K}\cdot {p_2})
 \bigg( -28{p_1}\cdot {p_2} -
     \left( 71 + 42{\textrm{div}}(-{p_1} +
           {p_K}) \right) m_{K}^2\nonumber\\
& &\mbox{ }\mbox{ }\mbox{ }\mbox{ }\mbox{ }\mbox{ }\mbox{ }\mbox{
}
   \mbox{ }\mbox{ }\mbox{ }\mbox{ }\mbox{ }\mbox{ }\mbox{ }\mbox{ }
   \mbox{ }\mbox{ }\mbox{ }\mbox{ }\mbox{ }\mbox{ }\mbox{ }\mbox{ }
   \mbox{ }\mbox{ }\mbox{ }\mbox{ }
 + \left( 1 - 12{\textrm{div}}(-{p_1} +
           {p_K}) \right) m_{\pi}^2 \bigg)
\nonumber\\ & &\mbox{ }\mbox{ }\mbox{ }\mbox{ }\mbox{ }\mbox{
}\mbox{ }\mbox{ }
      \mbox{ }\mbox{ }\mbox{ }
 + ({p_K}\cdot {p_1})
\bigg( -28{p_1}\cdot {p_2} -
     6{\textrm{div}}(-{p_1} + {p_K})
      ({p_1}\cdot {p_2})\nonumber\\
& &\mbox{ }\mbox{ }\mbox{ }\mbox{ }\mbox{ }\mbox{ }\mbox{ }\mbox{
}
   \mbox{ }\mbox{ }\mbox{ }\mbox{ }\mbox{ }\mbox{ }\mbox{ }\mbox{ }
   \mbox{ }\mbox{ }\mbox{ }\mbox{ }\mbox{ }\mbox{ }\mbox{ }\mbox{ }
   \mbox{ }\mbox{ }\mbox{ }\mbox{ }
 - 2\left( 100 +
        21{\textrm{div}}(-{p_1} + {p_K}) \right)
      ({p_K}\cdot {p_2})\nonumber\\
& &\mbox{ }\mbox{ }\mbox{ }\mbox{ }\mbox{ }\mbox{ }\mbox{ }\mbox{
}
   \mbox{ }\mbox{ }\mbox{ }\mbox{ }\mbox{ }\mbox{ }\mbox{ }\mbox{ }
   \mbox{ }\mbox{ }\mbox{ }\mbox{ }\mbox{ }\mbox{ }\mbox{ }\mbox{ }
   \mbox{ }\mbox{ }\mbox{ }\mbox{ }
    + \left( -71 +
        18{\textrm{div}}(-{p_1} + {p_K}) \right)
      m_{K}^2 + m_{\pi}^2 +
     18{\textrm{div}}(-{p_1} + {p_K})
      m_{\pi}^2 \bigg)
\mbox{ }\big]
%%% end of SimpB2
\nonumber\\ & &\mbox{ }\mbox{ }\mbox{ }\mbox{ }
 +3 d(-p_2 + p_K)\nonumber\\
& &\mbox{ }\mbox{ }\mbox{ }\mbox{ }\mbox{ }\mbox{ }\mbox{ }\mbox{
} \times\big[
%%% SimpB3
-89({p_1}\cdot {p_2})m_{K}^4 + 65m_{K}^6 -
  24({p_1}\cdot {p_2})m_{K}^2m_{\pi}^2 +
  15m_{K}^4m_{\pi}^2 + 41({p_1}\cdot {p_2})m_{\pi}^4\nonumber\\
& &\mbox{ }\mbox{ }\mbox{ }\mbox{ }\mbox{ }\mbox{ }\mbox{ }\mbox{
}
   \mbox{ }\mbox{ }\mbox{ }\mbox{ }
 +15m_{K}^2m_{\pi}^4 + 65m_{\pi}^6 -
  64{\left( {p_K}\cdot {p_2} \right) }^2
   \left( m_{K}^2 + m_{\pi}^2 \right)\nonumber\\
& &\mbox{ }\mbox{ }\mbox{ }\mbox{ }\mbox{ }\mbox{ }\mbox{ }\mbox{
}
   \mbox{ }\mbox{ }\mbox{ }\mbox{ }
 -2({p_K}\cdot {p_2})\left( 7m_{K}^4 -
     16{p_1}\cdot {p_2}m_{\pi}^2 + 7m_{\pi}^4 +
     2m_{K}^2\left( -10{p_1}\cdot {p_2} + m_{\pi}^2 \right)  \
\right) \nonumber\\ & &\mbox{ }\mbox{ }\mbox{ }\mbox{ }\mbox{
}\mbox{ }\mbox{ }\mbox{ }
   \mbox{ }\mbox{ }\mbox{ }\mbox{ }
  + ({p_K}\cdot {p_1})\left( -31m_{K}^4 +
     24m_{K}^2m_{\pi}^2 - 113m_{\pi}^4 +
     8{p_K}\cdot {p_2}\left( 13m_{K}^2 + 2m_{\pi}^2 \right)  \
\right)
%%% end of SimpB3
\mbox{ }\big]\nonumber\\ & &\mbox{ }\mbox{ }\mbox{ }\mbox{ }
    +3 d(-p_1 + p_K)\nonumber\\
& &\mbox{ }\mbox{ }\mbox{ }\mbox{ }\mbox{ }\mbox{ }\mbox{ }\mbox{
} \times\big[
%%% SimpB4
-89({p_1}\cdot {p_2})m_{K}^4 + 65m_{K}^6 -
  24({p_1}\cdot {p_2})m_{K}^2m_{\pi}^2 +
  15m_{K}^4m_{\pi}^2\nonumber\\
& &\mbox{ }\mbox{ }\mbox{ }\mbox{ }\mbox{ }\mbox{ }\mbox{ }\mbox{
}
   \mbox{ }\mbox{ }\mbox{ }\mbox{ }
 + 41({p_1}\cdot {p_2})m_{\pi}^4 +
  15m_{K}^2m_{\pi}^4 + 65m_{\pi}^6 -
  64{\left( {p_K}\cdot {p_1} \right) }^2
   \left( m_{K}^2 + m_{\pi}^2 \right) \nonumber\\
& &\mbox{ }\mbox{ }\mbox{ }\mbox{ }\mbox{ }\mbox{ }\mbox{ }\mbox{
}
   \mbox{ }\mbox{ }\mbox{ }\mbox{ }
 +({p_K}\cdot {p_2})\left( -31m_{K}^4 + 24m_{K}^2m_{\pi}^2 -
     113m_{\pi}^4 \right) \nonumber\\
& &\mbox{ }\mbox{ }\mbox{ }\mbox{ }\mbox{ }\mbox{ }\mbox{ }\mbox{
}
   \mbox{ }\mbox{ }\mbox{ }\mbox{ }
 +(2{p_K}\cdot {p_1})
\bigg( -7m_{K}^4 +
     16{p_1}\cdot {p_2}m_{\pi}^2 - 7m_{\pi}^4 -
     2m_{K}^2\left( -10{p_1}\cdot {p_2} + m_{\pi}^2 \right)  \nonumber\\
& &\mbox{ }\mbox{ }\mbox{ }\mbox{ }\mbox{ }\mbox{ }\mbox{ }\mbox{
}
   \mbox{ }\mbox{ }\mbox{ }\mbox{ }\mbox{ }\mbox{ }\mbox{ }\mbox{ }
   \mbox{ }\mbox{ }\mbox{ }\mbox{ }\mbox{ }\mbox{ }\mbox{ }\mbox{ }
   \mbox{ }\mbox{ }\mbox{ }\mbox{ }\mbox{ }\mbox{ }
    +({p_K}\cdot {p_2})\left( 52m_{K}^2 + 8m_{\pi}^2 \right)  \
\bigg)
%%% end of SimpB4
\mbox{ }\big]\nonumber\\ & &\mbox{ }\mbox{ }\mbox{ }\mbox{ }
    -2 d(-p_2 + p_K)^{2}\nonumber\\
& &\mbox{ }\mbox{ }\mbox{ }\mbox{ }\mbox{ }\mbox{ }\mbox{ }\mbox{
} \times\big[
%%% SimpB5
-202({p_1}\cdot {p_2})m_{K}^6 + 165m_{K}^8 +
  66({p_1}\cdot {p_2})m_{K}^2m_{\pi}^4 -
  42m_{K}^4m_{\pi}^4 \nonumber\\
& &\mbox{ }\mbox{ }\mbox{ }\mbox{ }\mbox{ }\mbox{ }\mbox{ }\mbox{
}
   \mbox{ }\mbox{ }\mbox{ }\mbox{ }
+ 136({p_1}\cdot {p_2})m_{\pi}^6 +
  165m_{\pi}^8 + 48{\left( {p_K}\cdot {p_2} \right) }^2
   \left( m_{K}^4 + 4m_{K}^2m_{\pi}^2 + m_{\pi}^4 \right) \nonumber\\
& &\hspace{-0in} + 2({p_K}\cdot {p_1})\left( m_{K}^2 - m_{\pi}^2
\right)
   \bigg( 77m_{K}^4 + 101m_{K}^2m_{\pi}^2 +
     92m_{\pi}^4 - 30{p_K}\cdot {p_2}
      \left( 4m_{K}^2 + 5m_{\pi}^2 \right)  \bigg)\nonumber\\
& &\mbox{ }\mbox{ }\mbox{ }\mbox{ }\mbox{ }\mbox{ }\mbox{ }\mbox{
}
   \mbox{ }\mbox{ }\mbox{ }\mbox{ }  -
  6({p_K}\cdot {p_2})
  \bigg( 53m_{K}^6 +
     34({p_1}\cdot {p_2})m_{\pi}^4 + 53m_{\pi}^6 -
     m_{K}^4\left( 56{p_1}\cdot {p_2} + 5m_{\pi}^2 \right) \nonumber\\
& &\mbox{ }\mbox{ }\mbox{ }\mbox{ }\mbox{ }\mbox{ }\mbox{ }\mbox{
}
   \mbox{ }\mbox{ }\mbox{ }\mbox{ }\mbox{ }\mbox{ }\mbox{ }\mbox{ }
   \mbox{ }\mbox{ }\mbox{ }\mbox{ }\mbox{ }\mbox{ }\mbox{ }\mbox{ }
   \mbox{ }\mbox{ }\mbox{ }\mbox{ }\mbox{ }\mbox{ }
    +m_{K}^2\left( 22{p_1}\cdot {p_2}m_{\pi}^2 -
        5m_{\pi}^4 \right)  \bigg)
%%% end of SimpB5
\mbox{ }\big]\nonumber\\ & &\mbox{ }\mbox{ }\mbox{ }\mbox{ }
    -2 d(-p_1 + p_K)^{2}\nonumber\\
& &\mbox{ }\mbox{ }\mbox{ }\mbox{ }\mbox{ }\mbox{ }\mbox{ }\mbox{
} \times\big[
%%% SimpB6
-202({p_1}\cdot {p_2})m_{K}^6 + 165m_{K}^8 +
  66({p_1}\cdot {p_2})m_{K}^2m_{\pi}^4 -
  42m_{K}^4m_{\pi}^4\nonumber\\
& &\mbox{ }\mbox{ }\mbox{ }\mbox{ }\mbox{ }\mbox{ }\mbox{ }\mbox{
}
   \mbox{ }\mbox{ }\mbox{ }\mbox{ }
 + 136({p_1}\cdot {p_2})m_{\pi}^6 +
  165m_{\pi}^8 + 48{\left( {p_K}\cdot {p_1} \right) }^2
   \left( m_{K}^4 + 4m_{K}^2m_{\pi}^2 + m_{\pi}^4 \right)\nonumber\\
& &\mbox{ }\mbox{ }\mbox{ }\mbox{ }\mbox{ }\mbox{ }\mbox{ }\mbox{
}
   \mbox{ }\mbox{ }\mbox{ }\mbox{ }
+ 2({p_K}\cdot {p_2})\left( 77m_{K}^6 +
     24m_{K}^4m_{\pi}^2 - 9m_{K}^2m_{\pi}^4 -
     92m_{\pi}^6 \right) \nonumber\\
& &\mbox{ }\mbox{ }\mbox{ }\mbox{ }\mbox{ }\mbox{ }\mbox{ }\mbox{
}
   \mbox{ }\mbox{ }\mbox{ }\mbox{ }
-6({p_K}\cdot {p_1}) \bigg( 53m_{K}^6 +
     34{p_1}\cdot {p_2}m_{\pi}^4 + 53m_{\pi}^6 -
     m_{K}^4\left( 56{p_1}\cdot {p_2} + 5m_{\pi}^2 \right)\nonumber\\
& &\mbox{ }\mbox{ }\mbox{ }\mbox{ }\mbox{ }\mbox{ }\mbox{ }\mbox{
}
   \mbox{ }\mbox{ }\mbox{ }\mbox{ }\mbox{ }\mbox{ }\mbox{ }\mbox{ }
   \mbox{ }\mbox{ }\mbox{ }\mbox{ }\mbox{ }\mbox{ }\mbox{ }\mbox{ }
   \mbox{ }\mbox{ }\mbox{ }\mbox{ }\mbox{ }\mbox{ }
    +m_{K}^2\left( (22{p_1}\cdot {p_2})m_{\pi}^2 -
        5m_{\pi}^4 \right)\nonumber\\
& &\mbox{ }\mbox{ }\mbox{ }\mbox{ }\mbox{ }\mbox{ }\mbox{ }\mbox{
}
    +10{(p_K}\cdot {p_2})\left( 4m_{K}^4 + m_{K}^2m_{\pi}^2 -
        5m_{\pi}^4 \right)
\bigg)
%%% end of SimpB6
\mbox{ }\big]\,\bigg]\,.\eea

\subsubsection{Result for the matrix elements of $\op_{7,8}$ in Full
QCD with General Kinematics:}

For the EWP operators the corresponding expressions are:
\begin{equation} \la \pi^+
\pi^0|\op_{7,8}|K^+\ra=O_{7,8}^{tree}\,[1+\frac{m_K^2}{16 \pi^2
f^2}I_{zf}]+O_{7,8}^a+O_{7,8}^b+O_{7,8}^{c+d}+\textrm{counterterms}
\, , \eeq where the counterterms are given in \eq{eq:o4fullgen},
$I_{zf}$ is given in \eq{eq:izffull},
\begin{equation}
%
% lowest order
%
O_{7,8}^{\mathrm{tree}} =
\frac{2\sqrt{2}\gamma^{(8,8)}}{f_{\pi}^{2}f_{K}} \, ,
\end{equation}
\begin{eqnarray}
%
% tadpole diagram
%
O_{7,8}^{a} &=&
-\frac{3\gamma^{(8,8)}\log\left(\frac{m^{2}_{K}}{\mu^{2}}\right)m^{2}_{K}}
{4\sqrt{2}f^{2}\,f_{\pi}^{2}f_{K}\pi^{2}} -
\frac{7\gamma^{(8,8)}\log\left(
\frac{m^{2}_{\pi}}{\mu^{2}}\right)m^{2}_{\pi}}{8\sqrt{2}f^{2}\,f_{\pi}^{2}f_{K}\pi^{2}}
+\frac{\gamma^{(8,8)}\log\left(\frac{m^{2}_{\eta}}{\mu^{2}}\right)
(-4m^{2}_{K}+m^{2}_{\pi})}{24\sqrt{2}f^{2}\,f_{\pi}^{2}f_{K}\pi^{2}}
\, , \nn \\
\end{eqnarray}
\begin{eqnarray}
% scattering diagram
%
O_{7,8}^{b} &=& -\frac{\gamma^{(8,8)}
\left(\textrm{div}(p_1+p_2)-\log(d(p_1+p_2)m^{2}_{\pi})\right)(3
p_{1}\cdot p_{2} -m^{2}_{\pi})}
{6\sqrt{2}f^{2}\,f_{\pi}^{2}f_{K}\pi^{2}}\nonumber\\ & & - \frac
{\gamma^{(8,8)}\left[m^{2}_{\pi}\left(-3+4d(p_1+p_2)m^{2}_{\pi}\right)
 + (p_{1}\cdot p_{2})\left(6+4d(p_1+p_2)m^{2}_{\pi}\right)\right]}
{6\sqrt{2}f^{2}\,f_{\pi}^{2}f_{K}\pi^{2}}\nonumber\\ & & +
\frac{\gamma^{(8,8)}L1(p_1+p_2,m_{\pi})
\sqrt{2-8d(p_1+p_2)m^{2}_{\pi}}}{12f^{2}\,f_{\pi}^{2}f_{K}\pi^{2}}\nonumber\\
& &\hspace{-0.3in}\times \left[
m^{2}_{\pi}\left(-1+2d(p_1+p_2)m^{2}_{\pi}\right) + (p_{1}\cdot
p_{2}) \left(3+2d(p_1+p_2)m^{2}_{\pi}\right) \right] \, ,
\end{eqnarray}
and
\begin{eqnarray}
%
% hook diagram
%
O_{7,8}^{c+d} &=&
\frac{\gamma^{(8,8)}}{864\sqrt{2}f^{2}\,f_{\pi}^{2}f_{K}\pi^{2}}\nonumber\\
&\times&\bigg \{ -3\,B(-p_1 + p_K,{m_{\eta }},{m_K})\,
 \times LL(-p_1 + p_K,{m_{\eta}},{m_K})
\nonumber\\ & &\mbox{ }\times\bigg [ -27\,{p_K}\cdot {p_1} -
6\,{m_{K}^{2}} + 6\,{m_{\pi}^2} \nonumber\\ & &\mbox{ }\mbox{
}\mbox{ } +
  {{d(-p_1 + p_K)}^2}\,
   {{\left( {m_{K}^{2}} - {m_{\pi}^2} \right) }^2}\,
   \left( -2\,{p_K}\cdot {p_1} + {m_{K}^{2}} + {m_{\pi}^2}
\right) \nonumber\\ & &\mbox{ }\mbox{ }\mbox{ }  +
  d(-p_1 + p_K)\,
   \left( -7\,{{{m_K}}^4} - 16\,{m_{K}^{2}}\,{m_{\pi}^2} +
     5\,{m_{\pi}^4} + 3\,{p_K}\cdot {p_1}\,
      \left( 7\,{m_{K}^{2}} - {m_{\pi}^2} \right)  \right)
\bigg ] \nonumber\\ & &+3\,B(-p_2 + p_K,{m_{\eta }},{m_K})\,
 \times LL(-p_2 + p_K,{m_{\eta}},{m_K})
\nonumber\\ & &\mbox{ }\times\bigg [ -27\,{p_K}\cdot {p_2} -
6\,{m_{K}^{2}} + 6\,{m_{\pi}^2} \nonumber\\ & &\mbox{ }\mbox{
}\mbox{ } +
  {{d(-p_2 + p_K)}^2}\,
   {{\left( {m_{K}^{2}} - {m_{\pi}^2} \right) }^2}\,
   \left( -2\,{p_K}\cdot {p_2} + {m_{K}^{2}} + {m_{\pi}^2}
\right) \nonumber\\ & &\mbox{ }\mbox{ }\mbox{ }  +
  d(-p_2 + p_K)\,
   \left( -7\,{{{m_K}}^4} - 16\,{m_{K}^{2}}\,{m_{\pi}^2} +
     5\,{m_{\pi}^4} + 3\,{p_K}\cdot {p_2}\,
      \left( 7\,{m_{K}^{2}} - {m_{\pi}^2} \right)  \right)
\bigg ]\nonumber\\ & &+\log\left (d(-p_1+p_K) m^{2}_{\eta}\right
)\nonumber\\ & &\mbox{ }\times\bigg [ 9\,\left( 9\,{p_K}\cdot
{p_1} + 2\,{m_{K}^2} - 2\,{m_{\pi}^2} \right) \nonumber\\ &
&\mbox{ }\mbox{ }\mbox{ }
      - 9\,d(-p_1 +p_K)\,
   \left( {m_{K}^2} - {m_{\pi}^2} \right) \,
   \left( -2\,{p_K}\cdot {p_1} + 7\,{m_{K}^2} - {m_{\pi}^2} \right)
\nonumber\\ & &\mbox{ }\mbox{ }\mbox{ }
 +
  {{d(-p_1 + p_K)}^3}\,
   {{\left( {m_{K}^2} - {m_{\pi}^2} \right) }^3}\,
   \left( -2\,{p_K}\cdot {p_1} + {m_{K}^2} + {m_{\pi}^2} \right)
\nonumber\\ & &\mbox{ }\mbox{ }\mbox{ }
 -
  {{d(-p_1 + p_K)}^2}\,
   \left( {m_{K}^2} - {m_{\pi}^2} \right) \,
   \left( 28\,{m_{K}^4} + 34\,{m_{K}^2}\,{m_{\pi}^2} -
     8\,{m_{\pi}^4} + 9\,{p_K}\cdot {p_1}\,
      \left( -7\,{m_{K}^2} + {m_{\pi}^2} \right)  \right)
\bigg ]\nonumber\\ & &+\log\left (d(-p_2+p_K) m^{2}_{\eta}\right
)\nonumber\\ & &\mbox{ }\times\bigg [ 9\,\left( 9\,{p_K}\cdot
{p_2} + 2\,{m_{K}^2} - 2\,{m_{\pi}^2} \right) \nonumber\\ &
&\mbox{ }\mbox{ }\mbox{ }
      - 9\,d(-p_2 + p_K)\,
   \left( {m_{K}^2} - {m_{\pi}^2} \right) \,
   \left( -2\,{p_K}\cdot {p_2} + 7\,{m_{K}^2} - {m_{\pi}^2} \right)
\nonumber\\ & &\mbox{ }\mbox{ }\mbox{ }
 +
  {{d(-p_2 + p_K)}^3}\,
   {{\left( {m_{K}^2} - {m_{\pi}^2} \right) }^3}\,
   \left( -2\,{p_K}\cdot {p_2} + {m_{K}^2} + {m_{\pi}^2} \right)
\nonumber\\ & &\mbox{ }\mbox{ }\mbox{ }
 -
  {{d(-p_2 + p_K)}^2}\,
   \left( {m_{K}^2} - {m_{\pi}^2} \right) \,
   \left( 28\,{m_{K}^4} + 34\,{m_{K}^2}\,{m_{\pi}^2} -
     8\,{m_{\pi}^4} + 9\,{p_K}\cdot {p_2}\,
      \left( -7\,{m_{K}^2} + {m_{\pi}^2} \right)  \right)
\bigg ]\nonumber\\ & &+9 \,B(-p_1 + p_K,{m_{\eta }},{m_K})\,
 \times LL(-p_1 + p_K,{m_{\eta}},{m_K})
\nonumber\\ & &\mbox{ }\times\bigg [ 21\,{p_K}\cdot {p_1} -
4\,\left( {m_{K}^2} + {m_{\pi}^2} \right)\nonumber\\ & &\mbox{
}\mbox{ }\mbox{ }  +
  7\,{{d(-p_1 + p_K)}^2}\,
   {{\left( {m_{K}^2} - {m_{\pi}^2} \right) }^2}\,
   \left( -2\,{p_K}\cdot {p_1} + {m_{K}^2} + {m_{\pi}^2} \right)\nonumber\\
& &\mbox{ }\mbox{ }\mbox{ }  +
  d(-p_1 + p_K)\,
   \left( -3\,{m_{K}^4} - 8\,{m_{K}^2}\,{m_{\pi}^2} -
     3\,{m_{\pi}^4} + 7\,{p_K}\cdot {p_1}\,
      \left( {m_{K}^2} + {m_{\pi}^2} \right)  \right)
\bigg ]\nonumber\\ & &+9 \,B(-p_2 + p_K,{m_{\eta }},{m_K})\,
 \times LL(-p_2 + p_K,{m_{\eta}},{m_K})
\nonumber\\ & &\mbox{ }\times\bigg [ 21\,{p_K}\cdot {p_2} -
4\,\left( {m_{K}^2} + {m_{\pi}^2} \right)\nonumber\\ & &\mbox{
}\mbox{ }\mbox{ }  +
  7\,{{d(-p_2 + p_K)}^2}\,
   {{\left( {m_{K}^2} - {m_{\pi}^2} \right) }^2}\,
   \left( -2\,{p_K}\cdot {p_2} + {m_{K}^2} + {m_{\pi}^2} \right)\nonumber\\
& &\mbox{ }\mbox{ }\mbox{ }  +
  d(-p_2 + p_K)\,
   \left( -3\,{m_{K}^4} - 8\,{m_{K}^2}\,{m_{\pi}^2} -
     3\,{m_{\pi}^4} + 7\,{p_K}\cdot {p_2}\,
      \left( {m_{K}^2} + {m_{\pi}^2} \right)  \right)
\bigg ]\nonumber\\ & &+6\times\bigg[ -22\,{{d(-p_1 + p_K)}^2}\,
   {{\left( {m_{K}^2} - {m_{\pi}^2} \right) }^2}\,
   \left( -2\,{p_K}\cdot {p_1} + {m_{K}^2} + {m_{\pi}^2} \right)\nonumber\\
& &\mbox{ }\mbox{ }\mbox{ }  -
  22\,{{d(-p_2 + p_K)}^2}\,
   {{\left( {m_{K}^2} - {m_{\pi}^2} \right) }^2}\,
   \left( -2\,{p_K}\cdot {p_2} + {m_{K}^2} + {m_{\pi}^2} \right)\nonumber\\
& &\mbox{ }\mbox{ }\mbox{ }    +
  6\,[ -15\,\left( 2 + \textrm{div}(-p_1 +
          p_K) \right) \,{p_K}\cdot {p_1}   -
     15\,\left( 2 + \textrm{div}(-p_2 + p_K)
         \right) \,{p_K}\cdot {p_2}
\nonumber\\ & &\mbox{ }\mbox{ }\mbox{ }\mbox{ }\mbox{ }\mbox{
}\mbox{ }\mbox{ }\mbox{ }
 - 3\,{m_{K}^2} +
     \textrm{div}(-p_1 + p_K)\,{m_{K}^2} +
     \textrm{div}(-p_2 + p_K)\,{m_{K}^2} +
     9\,{m_{\pi}^2}
\nonumber\\ & &\mbox{ }\mbox{ }\mbox{ }\mbox{ }\mbox{ }\mbox{
}\mbox{ }\mbox{ }\mbox{ } + 3\,\textrm{div}(-p_1 +
        p_K)\,{m_{\pi}^2} +
     3\,\textrm{div}(-p_2 + p_K)\,
      {m_{\pi}^2} ]
\nonumber\\ & &\mbox{ }\mbox{ }\mbox{ } +
  d(-p_1 + p_K)\,
   \left( 37\,{m_{K}^4} + 70\,{m_{K}^2}\,{m_{\pi}^2} +
     13\,{m_{\pi}^4} - 12\,{p_K}\cdot {p_1}\,
      \left( 7\,{m_{K}^2} + 3\,{m_{\pi}^2} \right)  \right)
\nonumber\\ & &\mbox{ }\mbox{ }\mbox{ } +
  d(-p_2 + p_K)\,
   \left( 37\,{m_{K}^4} + 70\,{m_{K}^2}\,{m_{\pi}^2} +
     13\,{m_{\pi}^4} - 12\,{p_K}\cdot {p_2}\,
      \left( 7\,{m_{K}^2} + 3\,{m_{\pi}^2} \right)  \right)
\bigg]\nonumber\\
% last piece
& &+2\log\left(d(-p_1+p_K)m^{2}_{K})\right)\nonumber\\ & &\mbox{
}\times\bigg[ 9\,d(-p_1 + p_K)\,
   \left( {m_{K}^2} - {m_{\pi}^2} \right) \,
   \left( 6\,{p_K}\cdot {p_1} + 3\,{m_{K}^2} - {m_{\pi}^2} \right)\nonumber\\
& &\mbox{ }\mbox{ }\mbox{ }    +
  31\,{{d(-p_1 + p_K)}^3}\,
   {{\left( {m_{K}^2} - {m_{\pi}^2} \right) }^3}\,
   \left( -2\,{p_K}\cdot {p_1} + {m_{K}^2} + {m_{\pi}^2} \right)\nonumber\\
& &\mbox{ }\mbox{ }\mbox{ }   -
  9\,\left( -15\,{p_K}\cdot {p_1} + {m_{K}^2} + 3\,{m_{\pi}^2} \right)
\nonumber\\ & &\hspace{-0.2in}
      - {{d(-p_1 + p_K)}^2}\,
   \left( {m_{K}^2} - {m_{\pi}^2} \right) \,
   \left( 31\,{m_{K}^4} + 82\,{m_{K}^2}\,{m_{\pi}^2} +
     49\,{m_{\pi}^4} - 9\,{p_K}\cdot {p_1}\,
      \left( 7\,{m_{K}^2} + 11\,{m_{\pi}^2} \right)  \right)
\bigg]\nonumber\\ &
&+2\log\left((d(-p_2+p_K)m^{2}_{K}\right)\nonumber\\ & &\mbox{
}\times\bigg[ 9\,d(-p_2 + p_K)\,
   \left( {m_{K}^2} - {m_{\pi}^2} \right) \,
   \left( 6\,{p_K}\cdot {p_2} + 3\,{m_{K}^2} - {m_{\pi}^2} \right)\nonumber\\
& &\mbox{ }\mbox{ }\mbox{ }   +
  31\,{{d(-p_2 + p_K)}^3}\,
   {{\left( {m_{K}^2} - {m_{\pi}^2} \right) }^3}\,
   \left( -2\,{p_K}\cdot {p_2} + {m_{K}^2} + {m_{\pi}^2} \right)\nonumber\\
& &\mbox{ }\mbox{ }\mbox{ }   -
  9\,\left( -15\,{p_K}\cdot {p_2} + {m_{K}^2} + 3\,{m_{\pi}^2} \right)
  \nonumber\\
& &\hspace{-0.2in}- {{d(-p_2 + p_K)}^2}\,
   \left( {m_{K}^2} - {m_{\pi}^2} \right) \,
   \left( 31\,{m_{K}^4} + 82\,{m_{K}^2}\,{m_{\pi}^2} +
     49\,{m_{\pi}^4} - 9\,{p_K}\cdot {p_2}\,
      \left( 7\,{m_{K}^2} + 11\,{m_{\pi}^2} \right)  \right)
\bigg]\nonumber\\ &
&-9\log\left(d(-p_1+p_K)m^{2}_{\pi}\right)\nonumber\\ & &\mbox{
}\times\bigg[ -21\,{p_K}\cdot {p_1} + 4\,\left( {m_{K}^2} +
{m_{\pi}^2} \right)\nonumber\\ & &\mbox{ }\mbox{ }\mbox{ }  -
  d(-p_1 + p_K)\,
   \left( {m_{K}^2} - {m_{\pi}^2} \right) \,
   \left( -14\,{p_K}\cdot {p_1} + {m_{K}^2} + {m_{\pi}^2} \right)\nonumber\\
& &\mbox{ }\mbox{ }\mbox{ }  +
  7\,{{d(-p_1 + p_K)}^3}\,
   {{\left( {m_{K}^2} - {m_{\pi}^2} \right) }^3}\,
   \left( -2\,{p_K}\cdot {p_1} + {m_{K}^2} + {m_{\pi}^2} \right)\nonumber\\
& &\hspace{-0.2in} +
  {{d(-p_1 + p_K)}^2}\,
   \left( 21\,{p_K}\cdot {p_1}\,
      \left( {m_{K}^4} - {m_{\pi}^4} \right)  -
     2\,\left( 5\,{m_{K}^6} + 6\,{m_{K}^4}\,{m_{\pi}^2} -
        6\,{m_{K}^2}\,{m_{\pi}^4} - 5\,{m_{\pi}^6} \right)
     \right)
\bigg]\nonumber\\ &
&-9\log\left(d(-p_2+p_K)m^{2}_{\pi}\right)\nonumber\\ & &\mbox{
}\times\bigg[ -21\,{p_K}\cdot {p_2} + 4\,\left( {m_{K}^2} +
{m_{\pi}^2} \right)\nonumber\\ & &\mbox{ }\mbox{ }\mbox{ }   -
  d(-p_2 + p_K)\,
   \left( {m_{K}^2} - {m_{\pi}^2} \right) \,
   \left( -14\,{p_K}\cdot {p_2} + {m_{K}^2} + {m_{\pi}^2} \right)\nonumber\\
& &\mbox{ }\mbox{ }\mbox{ }   +
  7\,{{d(-p_2 + p_K)}^3}\,
   {{\left( {m_{K}^2} - {m_{\pi}^2} \right) }^3}\,
   \left( -2\,{p_K}\cdot {p_2} + {m_{K}^2} + {m_{\pi}^2} \right)\nonumber\\
& &\hspace{-0.5in} +
  {{d(-p_2 + p_K)}^2}\,
   \left( 21\,{p_K}\cdot {p_2}\,
      \left( {m_{K}^4} - {m_{\pi}^4} \right)  -
     2\,\left( 5\,{m_{K}^6} + 6\,{m_{K}^4}\,{m_{\pi}^2} -
        6\,{m_{K}^2}\,{m_{\pi}^4} - 5\,{m_{\pi}^6} \right)
     \right)
\bigg] \bigg\} \, . \eea

\section{Results in Quenched QCD}\label{sec:quenched}

Finally we present the results for matrix elements in quenched
QCD. The results for general kinematics, i.e. for arbitrary quark
masses and momenta are too lengthy to be exhibited in this paper;
we present them on the web site~\cite{quenchedfull}. In the
following subsection we present the results for the matrix
elements with the SPQR kinematics.

In the quenched approximation, the integral
$I^q_{zf}$~\footnote{We introduce a superscript $q$ in this
appendix to denote the quenched approximation.}, which corresponds
to the one-loop contributions from the renormalization of the
mesonic wave-functions and from the replacement of the factor
$1/f^3$ which appears at lowest-order in the chiral expansion, by
$1/(f_Kf_\pi^2)$ is given by ($y = m^{2}_{\pi}/m^{2}_{K}$):
\begin{equation}
I_{zf}^q=-\frac{2m_0^2}{9m_K^2}\l[1+\frac{1}{2(1-y)}\ln{\frac{y}{2-y}}\r]+
\frac{2\alpha}{9}\l[1+\frac{y(2-y)}{2(1-y)}\ln{\frac{y}{2-y}} \r].
\label{eq:izfquenched}\end{equation}

As for full QCD, the results in this appendix are defined in terms
of the variables, $y = m^{2}_{\pi}/m^{2}_{K}$, $z=m_\pi/m_K$ and
$\omega=E_\pi/m_\pi$, where $E_\pi$ is the energy of the pion
whose momentum is (in general) not equal to zero.

\subsection{$K\to\pi^+\pi^0$ Decays with SPQR Kinematics in Quenched QCD}
\label{subsec:spqrquenched}

\subsubsection{Result for the matrix elements of $\op_4$ in
Quenched QCD with SPQR kinematics}

The matrix elements of $\op_4$ with the SPQR kinematics in the
quenched approximation is given by
\begin{eqnarray}
\la\pi^{+}\pi^{0}|\op_4|K^{+}\ra &=&
\frac{-6\sqrt{2}\,m_K^2}{f_{K}f^{2}_{\pi}}\alpha^{(27,1)}\,\Bigg\{
 \frac{z(1+\omega+2z)}{2\omega}
 \left( 1 + \frac{m^{2}_{K}}{16\pi^{2}f^{2}}
 \left ( I_{zf}^q\right )\right ) \nonumber\\
 &&\hspace{1.2in}+ \frac{m^{2}_{K}}{16\pi^{2}f^{2}}\, \left(
  I_{a}^q + I_{b}^q + I_{c+d}^q\right )\Bigg\},
  \label{eq:spqrquenched271}\eea where $I^q_{zf}$ is given in \eq{eq:izfquenched} and the
$I^q_{a,b,c+d}$ are as follows:
\begin{eqnarray}
%
% tadpole diagram
I^q_a &=& \left (\frac{-1}
   {3\omega}\right)\Bigg \{
 \left ( -2z^{2}+\omega-2z(1+\omega)\right ){\mathrm{log}}(z^{2})\nonumber\\
&+&
 \left ( 2z^{2}-\omega+2z(1+\omega)+2z^{3}(1+\omega)+z^{4}(6+\omega)\right )
 {\mathrm{log}}\left ( \frac{m^{2}_{\pi}}{\mu^{2}}\right )
\Bigg\} \nonumber\\ &-& \frac{\alpha\,
 (3+2z+3\omega)\left ( 2-2z^{2}+z^{2}(-2+z^{2})
 {\mathrm{log}}\left ( -1+\frac{2}{z^{2}}\right )\right )}
 {18z^{3}(-1+z^{2})\omega}\nonumber\\
&- & \frac{m^{2}_{0}\,  (3+2z+3\omega)\left ( -2+2z^{2}+
 {\mathrm{log}}\left ( -1+\frac{2}{z^{2}}\right )\right )}
 {18z(-1+z^{2})\omega m^{2}_{\pi}}\mbox{ },
\end{eqnarray}
\begin{eqnarray}
%
% scattering diagram
I^q_b &=& \left (\frac{-z^3}
  {3\omega^{2}}\right )\Bigg \{ 3(1+2z+\omega) + 3 \sqrt{\frac{1-\omega}{1+\omega}}
(1+2z+\omega)
 {\mathrm{log}}\left (
  \frac{-1+\sqrt{\frac{1-\omega}{1+\omega}}}
  {1+\sqrt{\frac{1-\omega}{1+\omega}}}\right )\nonumber\\
& &\mbox{ }\mbox{ }\mbox{ }+
 \left [ -3-2\omega+\omega^{2}-2z(3-3\omega+\omega^{2})\right ]
   {\mathrm{log}}\left (\frac{m^{2}_{\pi}}{\mu^{2}}\right )
\Bigg\} \,,\end{eqnarray} and
\begin{eqnarray}
%
% hook diagrams
I_{c+d}^q &=& \left ( \frac{-1}
  {18(-1+z)\omega^{2}(-2z+\omega+z^{2}\omega)^{3}}
 \right )\times\nonumber\\
& &\hspace{-0.3in}\Bigg \{
 -3\,\bigg [ 10\,z\,{\omega }^4 - 2\,{\omega }^5 +
       2\,z^9\,{\omega }^2\,\left( 3 + {\omega }^2 \right)  -
       2\,z^2\,{\omega }^3\,\left( 9 - 6\,\omega  + 5\,{\omega }^2 \
\right)  \nonumber\\ & &
   - z^3\,{\omega }^2\,\left( -22 + 56\,\omega  -
          41\,{\omega }^2 + {\omega }^3 \right)  -
       z^8\,{\omega }^2\,\left( 12 + 42\,\omega  - 9\,{\omega }^2 +
          {\omega }^3 \right)\nonumber\\
& &\mbox{ }\mbox{ }\mbox{ }\mbox{ }\mbox{ }\mbox{ }\mbox{ }
   \mbox{ }\mbox{ }\mbox{ }\mbox{ } +
       z^4\,\omega \,\left( -32 + 90\,\omega  - 64\,{\omega }^2 +
          15\,{\omega }^3 - 17\,{\omega }^4 \right)\nonumber\\
& &\mbox{ }\mbox{ }\mbox{ }\mbox{ }\mbox{ }\mbox{ }\mbox{ }
   \mbox{ }\mbox{ }\mbox{ }\mbox{ } +
       z^7\,\left( -8 + 24\,\omega  + 56\,{\omega }^2 -
          18\,{\omega }^3 + 55\,{\omega }^4 - 3\,{\omega }^5
  \right)\nonumber\\
& &\mbox{ }\mbox{ }\mbox{ }\mbox{ }\mbox{ }\mbox{ }\mbox{ }
   \mbox{ }\mbox{ }\mbox{ }\mbox{ }+
       2\,z^5\,\left( 12 - 18\,\omega  + 28\,{\omega }^2 -
          33\,{\omega }^3 + 36\,{\omega }^4 + 6\,{\omega }^5
  \right)  \nonumber\\
& &\mbox{ }\mbox{ }\mbox{ }\mbox{ }\mbox{ }\mbox{ }\mbox{ }
   \mbox{ }\mbox{ }\mbox{ }\mbox{ }-
       2\,z^6\,\left( 8 + 14\,\omega  - 23\,{\omega }^2 +
          48\,{\omega }^3 + 13\,{\omega }^5 \right)  \bigg ] \,\log (z^2)
\nonumber\\ & & -6\,\left( -1 + z \right) \,z^3\,{\sqrt{1 -
{\omega }^2}}\,
     \bigg [ 3\,z^5\,{\omega }^2 +
       3\,{\omega }^2\,\left( -1 + 2\,\omega  \right)  -
       3\,z^4\,{\omega }^2\,\left( 1 + 2\,\omega  \right)\nonumber\\
& &\mbox{ }\mbox{ }\mbox{ }\mbox{ }\mbox{ }\mbox{ }\mbox{ }
   \mbox{ }\mbox{ }\mbox{ }\mbox{ }  -
       3\,z\,\omega \,\left( -4 + 6\,\omega  + {\omega }^3 \right)  +
       6\,z^2\,\left( -2 + 3\,\omega  - {\omega }^2 + {\omega }^3 \
\right)\nonumber\\ & &\mbox{ }\mbox{ }\mbox{ }\mbox{ }\mbox{
}\mbox{ }\mbox{ }
   \mbox{ }\mbox{ }\mbox{ }\mbox{ }  +
   z^3\,\left( -4 + 12\,\omega  - 7\,{\omega }^2 +
          5\,{\omega }^4 \right)  \bigg ]
     \log \left(\frac{1 - {\sqrt{1 - {\omega }^2}}}
       {1 + {\sqrt{1 - {\omega }^2}}}\right)\nonumber\\
& &\hspace{-0.5in}+\left( -1 + z \right) \,\left( -2\,z + \omega +
z^2\,\omega \right) \,
     \bigg [ 2\,z^2\,\bigg ( -9\,{\omega }^2\,
           \left( 1 - 4\,\omega  + {\omega }^2 \right)+
           z^5\,{\omega }^2\,\left( 7 + 9\,\omega  + 2\,{\omega }^2 \ \right)
\nonumber\\ & &\mbox{ }\mbox{ }\mbox{ }\mbox{ }\mbox{ }\mbox{
}\mbox{ } \mbox{ } +
 z\,\omega \,\left( 36 - 128\,\omega  + 45\,{\omega }^2 -
             7\,{\omega }^3 \right)  -
          z^4\,\omega \,\left( -2 + 45\,\omega  + 2\,{\omega }^2 +
             9\,{\omega }^3 \right)  \nonumber\\
& & \hspace{-0.6in}- 2\,z^2\,\left( 18 - 64\,\omega  + 45\,{\omega
}^2
-
             26\,{\omega }^3 + 9\,{\omega }^4 \right)
+ z^3\,\left( -8 + 72\,\omega  - 101\,{\omega }^2 +
             54\,{\omega }^3 + 19\,{\omega }^4 \right)  \bigg ) \nonumber\\
& &+ 3\,{\left( -2\,z + \omega  + z^2\,\omega  \right) }^2\,
        \bigg ( 2\,{\omega }^2 + 2\,z^4\,{\omega }^2 +
          2\,z\,\omega \,\left( 1 + \omega  \right)  +
          z^3\,\left( 2 + 3\,\omega  + {\omega }^2 \right)\nonumber\\
& & + 2\,z^2\,\left( 3 - 5\,\omega  + 3\,{\omega }^2 \right)
\bigg) \log \left(\frac{{{m_{\pi }}}^2}{{\mu }^2}\right) \,\bigg]
\Bigg\} \nonumber\\
%
% \alpha starts here
%
 &+&\alpha\left(\frac{-z^2}{18(-1+z)^{5}
 (1+z)^{2}\omega^{2}(-2z+\omega+z^{2}\omega)^{3}}\right )\times\nonumber\\
&&\left (\frac{1}{2z\omega-2z^{3}\omega-\omega^{2}+z^{4}\omega^{2}
 - z^{2}(-1+\omega^{2})}\right )\,\sqrt{(-1+z)^{3}(1+z)}\times\nonumber\\
&& \bigg \{ \frac{-1}{z}\sqrt{(-1+z)^{3}(1+z)}
 \bigg [ 2z\omega-2z^{3}\omega-\omega^{2}+
 z^{4}\omega^{2}-z^{2}(-1+\omega^{2}) \bigg ]\nonumber\\
& &\mbox{ }\mbox{ }\mbox{ }\times\Bigg[ z
%SimpA
[-12\,{\omega }^4 + 4\,z^{11}\,{\omega }^4 +
  z^{10}\,{\omega }^3\,( -6 + 9\,\omega  - 5\,{\omega }^2 )  -
  2\,z\,{\omega }^3\,( -23 + 8\,\omega  + {\omega }^2 )\nonumber\\
& & \mbox{ }\mbox{ }\mbox{ }\mbox{ }\mbox{ }\mbox{ }\mbox{ }\mbox{
}  +
  2\,z^2\,{\omega }^2\,\left( -45 + 61\,\omega  + 9\,{\omega }^2 +
     {\omega }^3 \right)  + z^9\,{\omega }^2\,
   \left( 18 - 72\,\omega  - 3\,{\omega }^2 + 5\,{\omega }^3
\right)\nonumber\\ & & \mbox{ }\mbox{ }\mbox{ }\mbox{ }\mbox{
}\mbox{ }\mbox{ }\mbox{ }-
  2\,z^3\,\omega \,\left( -27 + 123\,\omega  - 60\,{\omega }^2 +
     54\,{\omega }^3 + 14\,{\omega }^4 \right) \nonumber\\
& & \mbox{ }\mbox{ }\mbox{ }\mbox{ }\mbox{ }\mbox{ }\mbox{ }\mbox{
} -
  2\,z^7\,\omega \,\left( 42 - 92\,\omega  + 15\,{\omega }^2 -
     16\,{\omega }^3 + 17\,{\omega }^4 \right)\nonumber\\
& & \mbox{ }\mbox{ }\mbox{ }\mbox{ }\mbox{ }\mbox{ }\mbox{ }\mbox{
} +
  2\,z^8\,\omega \,\left( -10 + 69\,\omega  - 45\,{\omega }^2 +
     18\,{\omega }^3 + 17\,{\omega }^4 \right)\nonumber\\
& & \mbox{ }\mbox{ }\mbox{ }\mbox{ }\mbox{ }\mbox{ }\mbox{ }\mbox{
}-
  z^6\,\omega \,\left( 78 + 100\,\omega  + 42\,{\omega }^2 +
     39\,{\omega }^3 + 63\,{\omega }^4 \right)\nonumber\\
& & \mbox{ }\mbox{ }\mbox{ }\mbox{ }\mbox{ }\mbox{ }\mbox{ }\mbox{
}+
  2\,z^4\,\left( -2 + 85\,\omega  - 108\,{\omega }^2 + 126\,{\omega }^3 -
     54\,{\omega }^4 + 14\,{\omega }^5 \right)\nonumber\\
& & \mbox{ }\mbox{ }\mbox{ }\mbox{ }\mbox{ }\mbox{ }\mbox{ }\mbox{
}  +
  z^5\,\left( 4 + 126\,\omega  - 192\,{\omega }^2 + 204\,{\omega }^3 +
     19\,{\omega }^4 + 63\,{\omega }^5 \right)]
% end of SimpA
\, {\mathrm{log}}\left ( -1+\frac{2}{z^{2}}\right )\nonumber\\ %
&&\mbox{ }\mbox{ }\mbox{ }\mbox{ }\mbox{ }\mbox{ }\mbox{ }\mbox{ }
 -2 (1+z) \omega \bigg ( (-1+z)\times \nonumber\\
% SimpB
& &\mbox{ }\mbox{ }\mbox{ }\mbox{ }\mbox{ }\mbox{ } \bigg[-\left(
{\omega }^3\,\left( 1 + \omega  \right)  \right)  +
  3\,z^9\,{\omega }^2\,\left( -4 - 3\,\omega  + {\omega }^2 \right)  +
  z\,{\omega }^2\,\left( 6 + 27\,\omega  + {\omega }^2 \right)\nonumber\\
& & \mbox{ }\mbox{ }\mbox{ }\mbox{ }\mbox{ }\mbox{ }\mbox{ }\mbox{
}+
  z^8\,\omega \,\left( 24 + 66\,\omega  + 25\,{\omega }^2 -
     3\,{\omega }^3 \right)  -
  z^2\,\omega \,\left( 12 + 107\,\omega  + 17\,{\omega }^2 +
     22\,{\omega }^3 \right)\nonumber\\
& & \mbox{ }\mbox{ }\mbox{ }\mbox{ }\mbox{ }\mbox{ }\mbox{ }\mbox{
}-
  z^7\,\omega \,\left( 132 + 79\,\omega  + 39\,{\omega }^2 +
     22\,{\omega }^3 \right)  -
  4\,z^5\,\left( 3 - 8\,\omega  - 8\,{\omega }^2 + 14\,{\omega }^3 +
     {\omega }^4 \right) \nonumber\\
& & \mbox{ }\mbox{ }\mbox{ }\mbox{ }\mbox{ }\mbox{ }\mbox{ }\mbox{
}+
 2\,z^4\,
   \left( -46 - 8\,\omega  - 64\,{\omega }^2 - 35\,{\omega }^3 +
     2\,{\omega }^4 \right)  \nonumber\\
& & \mbox{ }\mbox{ }\mbox{ }\mbox{ }\mbox{ }\mbox{ }\mbox{ }\mbox{
}+
  z^6\,\left( 72 + 16\,\omega  + 109\,{\omega }^2 + 67\,{\omega }^3 +
     22\,{\omega }^4 \right) \nonumber\\
& & \mbox{ }\mbox{ }\mbox{ }\mbox{ }\mbox{ }\mbox{ }\mbox{ }\mbox{
} +
  z^3\,\left( 8 + 160\,\omega  + 41\,{\omega }^2 + 97\,{\omega }^3 +
     22\,{\omega }^4 \right)\bigg]\nonumber\\
% end of SimpB
% SimpC
& &\mbox{ }\mbox{ }\mbox{ }\mbox{ }\mbox{ }\mbox{ } +z(1+z)
\bigg[6\,{\omega }^3 + 3\,z^8\,{\omega }^2\,
   \left( -3 - 2\,\omega  + {\omega }^2 \right)  +
  z\,{\omega }^2\,\left( -23 - 3\,\omega  + 2\,{\omega }^2 \right)\nonumber\\
& & \mbox{ }\mbox{ }\mbox{ }\mbox{ }\mbox{ }\mbox{ }\mbox{ }\mbox{
}+
  3\,z^7\,\omega \,\left( 6 + 21\,\omega  + 4\,{\omega }^2 -
     3\,{\omega }^3 \right)  +
  z^6\,\omega \,\left( -126 - 89\,\omega  - 33\,{\omega }^2 +
     2\,{\omega }^3 \right) \nonumber\\
& &- 3\,z^2\,\omega \,\left( -14 + 7\,\omega + 3\,{\omega }^2 +
     2\,{\omega }^3 \right)  +
  3\,z^5\,\left( 16 + 40\,\omega  + 47\,{\omega }^2 + 19\,{\omega }^3 +
     6\,{\omega }^4 \right) \nonumber\\
& & \mbox{ }\mbox{ }\mbox{ }\mbox{ }\mbox{ }\mbox{ }\mbox{ }\mbox{
}+
  z^3\,\left( -24 + 54\,\omega  + 11\,{\omega }^2 + 78\,{\omega }^3 +
     13\,{\omega }^4 \right)  \nonumber\\
& & \mbox{ }\mbox{ }\mbox{ }\mbox{ }\mbox{ }\mbox{ }\mbox{ }\mbox{
}-
  z^4\,\left( 48 + 36\,\omega  + 145\,{\omega }^2 + 78\,{\omega }^3 +
     23\,{\omega }^4 \right)\bigg]\,{\mathrm{log}}(z^{2})\,\bigg)
 \,\Bigg]\nonumber\\
% end of SimpC
% SimpD
&+ & 3\,{\left( 1 + z \right) }^2\,\left( -4 + 10\,z - 3\,z^2 -
7\,z^3 +
    4\,z^4 \right)\omega \,{\left( -2\,z + \omega  + z^2\,\omega  \
\right) }^3\times\nonumber\\ & & \hspace{0.2in}\left [
2\,z\,\omega - 2\,z^3\,\omega - {\omega }^2 +
    z^4\,{\omega }^2 - z^2\,\left( -1 + {\omega }^2 \right)
\right ]\times\nonumber\\ & & \hspace{1in}
  \log \left (\frac{1 + z - z^2 - {\sqrt{{\left( -1 + z \right) }^3\,
          \left( 1 + z \right) }}}{1 + z - z^2 +
      {\sqrt{{\left( -1 + z \right) }^3\,\left( 1 + z \right) }}}\right )
\nonumber\\
% end of SimpD
% SimpE
&-&(-1+z)z^{2}\sqrt{(-1+z)^{3}(1+z)}\times \nonumber\\ %%
&&\Bigg[\sqrt{1-\omega^{2}}\bigg[-6\,{\omega }^4 -
6\,z^{10}\,{\omega }^4 + 27\,z^9\,{\omega }^5 +
  9\,{\omega }^6 - 9\,z\,{\omega }^3\,\left( -2 + 3\,{\omega }^2
  \right)\nonumber\\
& & \mbox{ }\mbox{ }\mbox{ }\mbox{ }\mbox{ }\mbox{ }\mbox{ }\mbox{
}-
  3\,z^7\,\omega \,\left( 4 - 3\,{\omega }^2 + 7\,{\omega }^4
\right) \nonumber\\ & & \mbox{ }\mbox{ }\mbox{ }\mbox{ }\mbox{
}\mbox{ }\mbox{ }\mbox{ }-
  3\,z^8\,{\omega }^2\,\left( -6 + 10\,{\omega }^2 + 7\,{\omega }^4 \
\right)  + z^6\,{\omega }^2\,\left( -38 + 67\,{\omega }^2 +
     55\,{\omega }^4 \right)\nonumber\\
& & \mbox{ }\mbox{ }\mbox{ }\mbox{ }\mbox{ }\mbox{ }\mbox{ }\mbox{
}+
  z^3\,\omega \,\left( 2 - 67\,{\omega }^2 + 113\,{\omega }^4 \right)  +
  z^5\,\left( 10\,\omega  + 40\,{\omega }^3 - 92\,{\omega }^5
\right) \nonumber\\ & & \mbox{ }\mbox{ }\mbox{ }\mbox{ }\mbox{
}\mbox{ }\mbox{ }\mbox{ }+
  z^2\,\left( -10\,{\omega }^2 + 29\,{\omega }^4 - 25\,{\omega }^6 \
\right)  \nonumber\\ & & \mbox{ }\mbox{ }\mbox{ }\mbox{ }\mbox{
}\mbox{ }\mbox{ }\mbox{ }+ z^4\,\left( 4 + 18\,{\omega }^2 -
42\,{\omega }^4 -
     22\,{\omega }^6 \right)\bigg]
% end of SimpE
\,{\mathrm{log}}\left (
\frac{1-\sqrt{1-\omega^{2}}}{1+\sqrt{1-\omega^{2}}}\right
)\nonumber\\ &+& \frac{1}{z^{2}}
\sqrt{2z\omega-2z^{3}\omega-\omega^{2}+z^{4}\omega^{2}-
z^{2}(-1+\omega^{2})}\times\nonumber\\
% SimpF
& & \bigg[-12\,{\omega }^5 + 6\,z^{11}\,{\omega }^4\,
   \left( -3 + {\omega }^2 \right)  +
  22\,z\,{\omega }^4\,\left( 2 + {\omega }^2 \right)  -
  z^2\,{\omega }^3\,\left( 38 + 73\,{\omega }^2 \right)\nonumber\\
& & \mbox{ }\mbox{ }\mbox{ }\mbox{ }\mbox{ }\mbox{ }\mbox{ }\mbox{
}-
  z^8\,\omega \,\left( -12 + 107\,{\omega }^2 + {\omega }^4 \right)  -
  2\,z^3\,{\omega }^2\,\left( -8 + 15\,{\omega }^2 + 5\,{\omega }^4 \
\right)  \nonumber\\ & & \mbox{ }\mbox{ }\mbox{ }\mbox{ }\mbox{
}\mbox{ }\mbox{ }\mbox{ } + 2\,z^7\,{\omega }^2\,
   \left( 48 + 55\,{\omega }^2 + 35\,{\omega }^4 \right)\nonumber\\
& & \mbox{ }\mbox{ }\mbox{ }\mbox{ }\mbox{ }\mbox{ }\mbox{ }\mbox{
}  -
  2\,z^6\,\omega \,\left( 7 + 50\,{\omega }^4 \right)  +
  z^4\,\omega \,\left( 2 + 97\,{\omega }^2 + 177\,{\omega }^4 \right)  +
  z^{10}\,\left( 48\,{\omega }^3 + 9\,{\omega }^5 \right)\nonumber\\
& & \mbox{ }\mbox{ }\mbox{ }\mbox{ }\mbox{ }\mbox{ }\mbox{ }\mbox{
}+
  z^9\,\left( -42\,{\omega }^2 + 11\,{\omega }^4 - 35\,{\omega }^6 \
\right)  - z^5\,\left( 4 + 58\,{\omega }^2 + 129\,{\omega }^4 +
     49\,{\omega }^6 \right)\bigg]\nonumber\\
% end of SimpF
%
& &\mbox{ }\mbox{ }\mbox{ }\mbox{ }\mbox{ }\mbox{ }\mbox{ }\mbox{
} \,{\mathrm{log}}\left (
\frac{z+\omega-z^{2}\omega-\sqrt{2z\omega-2z^{3}\omega-\omega^{2}
+z^{4}\omega^{2}-z^{2}(-1+\omega^{2})}}
{z+\omega-z^{2}\omega+\sqrt{2z\omega-2z^{3}\omega-\omega^{2}
+z^{4}\omega^{2}-z^{2}(-1+\omega^{2})}}\right )\mbox{ } \Bigg
]\mbox{ } \bigg \}\nonumber\\
%
%
% m_{0} starts here
%
+&m^{2}_{0}&\times\left (\frac{-z^4}{18(-1+z)^{5}
 (1+z)^{2}\omega^{2}(-2z+\omega+z^{2}\omega)^{3}m^{2}_{\pi}}
  \right ) \times\nonumber\\
&&\left (\frac{1}{2z\omega-2z^{3}\omega-\omega^{2}+z^{4}\omega^{2}
 - z^{2}(-1+\omega^{2})}\right )\,\sqrt{(-1+z)^{3}(1+z)}\times\nonumber\\
&& \Bigg\{ \frac{-1}{z}\sqrt{(-1+z)^{3}(1+z)}
 \bigg [ 2z\omega-2z^{3}\omega-\omega^{2}+
 z^{4}\omega^{2}-z^{2}(-1+\omega^{2}) \bigg ]\times\nonumber\\
& &\mbox{ }\mbox{ }\mbox{ }\bigg [
%SimpA
[-\left( {\omega }^4\,\left( 1 + \omega  \right)  \right)  -
  3\,z^9\,{\omega }^3\,\left( -3 - 2\,\omega  + {\omega }^2 \right)  +
  z\,{\omega }^3\,\left( 6 + 15\,\omega  + {\omega }^2 \right)\nonumber\\
& &+
  z^8\,{\omega }^2\,\left( -12 - 45\,\omega  + 4\,{\omega }^2 +
     3\,{\omega }^3 \right)  -
  z^2\,{\omega }^2\,\left( 12 + 49\,\omega  + 2\,{\omega }^2 +
     5\,{\omega }^3 \right)  \nonumber\\
& &\mbox{ }\mbox{ }\mbox{ }\mbox{ }\mbox{ }\mbox{ }\mbox{ }\mbox{
}+
  z^3\,\omega \,\left( 8 + 74\,\omega  - 35\,{\omega }^2 +
     18\,{\omega }^3 + 5\,{\omega }^4 \right) \nonumber\\
& &\mbox{ }\mbox{ }\mbox{ }\mbox{ }\mbox{ }\mbox{ }\mbox{ }\mbox{
}+
  z^4\,\omega \,\left( -42 + 106\,\omega  - 81\,{\omega }^2 +
     33\,{\omega }^3 + 12\,{\omega }^4 \right)  \nonumber\\
& &\mbox{ }\mbox{ }\mbox{ }\mbox{ }\mbox{ }\mbox{ }\mbox{ }\mbox{
}+
  z^7\,\omega \,\left( 6 + 84\,\omega  - 75\,{\omega }^2 +
     30\,{\omega }^3 + 13\,{\omega }^4 \right) \nonumber\\
& &\mbox{ }\mbox{ }\mbox{ }\mbox{ }\mbox{ }\mbox{ }\mbox{ }\mbox{
} +
  z^5\,\left( 4 - 86\,\omega  + 110\,{\omega }^2 - 141\,{\omega }^3 +
     27\,{\omega }^4 - 12\,{\omega }^5 \right) \nonumber\\
& &\hspace{-0.1in}- z^6\,\left( 4 + 54\,\omega  - 154\,{\omega }^2
+ 93\,{\omega }^3 -
     38\,{\omega }^4 + 13\,{\omega }^5 \right) ]
% end SimpA
\times {\mathrm{log}}\left ( -1 + \frac{2}{z^{2}}\right
)\nonumber\\ %%
% SimpB
&+& 2 (1+z) \omega \bigg[\left( -1 + z \right) \,{\left( -2\,z +
\omega + z^2\,\omega \right) }^2\times \nonumber\\ && %
\left(4\,z^4\,\omega - \omega \,\left( 1 + \omega \right)
-z^3\,\left( 3 + {\omega }^2 \right)  + z^2\,\left( -5 - 2\,\omega
+ {\omega }^2 \right)  + z\,\left( 2 + 5\,\omega  + {\omega }^2
\right)  \right)\nonumber\\ %%
& &-
  z\,\left( 1 + z \right) \,\bigg ( -6\,{\omega }^3 -
     3\,z^5\,{\omega }^2\,\left( -10 - 3\,\omega  + {\omega }^2 \right)  +
     z^6\,{\omega }^2\,\left( -4 - 3\,\omega  + {\omega }^2\right)\nonumber\\
& &+
     z\,{\omega }^2\,\left( 20 + 9\,\omega  + 7\,{\omega }^2 \right)  -
     2\,z^4\,\omega \,\left( 18 + 7\,\omega  + 18\,{\omega }^2 +
        2\,{\omega }^3 \right) \nonumber\\
& &-
     3\,z^2\,\omega \,\left( 12 + 2\,\omega  + 9\,{\omega }^2 +
        7\,{\omega }^3 \right)  +
     z^3\,\left( 24 + 46\,{\omega }^2 + 30\,{\omega }^3 +
        20\,{\omega }^4 \right)  \bigg ) \,\log (z^2)\bigg]
% end SimpB
\bigg] \nonumber\\
% SimpC
& +& 3\,{\left( 1 + z \right) }^2\,\left( 1 - 3\,z + 2\,z^2
\right) \,\omega \,
  {\left( -2\,z + \omega  + z^2\,\omega  \right) }^3\times\nonumber\\
& &\hspace{-.4in}
  \left( 2\,z\,\omega  - 2\,z^3\,\omega  - {\omega }^2 + z^4\,{\omega }^2 -
    z^2\,\left( -1 + {\omega }^2 \right)  \right)% \,\nonumber\\
%& &\mbox{ }\mbox{ }\mbox{ }\mbox{ }\mbox{ }\times
  \log \left(\frac{1 + z - z^2 - {\sqrt{{\left( -1 + z \right) }^3\,
          \left( 1 + z \right) }}}{1 + z - z^2 +
      {\sqrt{{\left( -1 + z \right) }^3\,\left( 1 + z \right) }}}\right)
\nonumber\\
% end SimpC
% SimpD
&-&(-1+z)\sqrt{(-1+z)^{3}(1+z)}\times\nonumber\\ & & \Bigg[
\sqrt{1-\omega^{2}}\,[6\,z^{10}\,{\omega }^4 + 9\,z\,{\omega }^5 -
3\,{\omega }^6 +
  3\,z^2\,{\omega }^4\,\left( -3 + 5\,{\omega }^2 \right)  -
  3\,z^9\,{\omega }^3\,\left( 2 + 7\,{\omega }^2 \right) \nonumber\\
& &+ 2\,z^4\,{\omega }^2\,\left( -1 + 17\,{\omega }^2 + 5\,{\omega
}^4 \ \right)  + z^8\,{\omega }^2\,\left( -10 + 32\,{\omega }^2 +
     11\,{\omega }^4 \right) \nonumber\\
& &+ z^7\,\omega \,\left( 14 - 13\,{\omega }^2 + 23\,{\omega }^4
\right)  + 2\,z^5\,\omega \,\left( -7 + 2\,{\omega }^2 +
26\,{\omega }^4 \right)  +
  z^3\,\left( 15\,{\omega }^3 - 63\,{\omega }^5 \right)\nonumber\\
& &- z^6\,\left( 4 - 24\,{\omega }^2 + 75\,{\omega }^4 +
29\,{\omega }^6 \ \right)\,] \,{\mathrm{log}}\left(
\frac{1-\sqrt{1-\omega^{2}}}{1+\sqrt{1-\omega^{2}}}\right
)\nonumber\\
% end SimpD
 & & +
\sqrt{2z\omega-2z^{3}\omega-\omega^{2}
+z^{4}\omega^{2}-z^{2}(-1+\omega^{2})}\times\nonumber\\ & &
% SimpE
[3\,{\omega }^5 + 2\,z^9\,{\omega }^4\,\left( -4 + {\omega }^2
\right)  -
  2\,z\,{\omega }^4\,\left( 4 + 5\,{\omega }^2 \right)  +
  z^8\,{\omega }^3\,\left( 14 + 13\,{\omega }^2 \right)\nonumber\\
& &\mbox{ }\mbox{ }\mbox{ }\mbox{ }\mbox{ }\mbox{ }\mbox{ }\mbox{
}-
  z^7\,{\omega }^4\,\left( 23 + 13\,{\omega }^2 \right)  +
  z^2\,{\omega }^3\,\left( -1 + 37\,{\omega }^2 \right)  -
  z^3\,{\omega }^2\,\left( 4 + 3\,{\omega }^2 + 5\,{\omega }^4
  \right)\nonumber\\
& &\mbox{ }\mbox{ }\mbox{ }\mbox{ }\mbox{ }\mbox{ }\mbox{ }\mbox{
}-
  2\,z^4\,\omega \,\left( -5 + 8\,{\omega }^2 + 24\,{\omega }^4 \right)  +
  z^6\,\left( -10\,\omega  + 3\,{\omega }^3 - 5\,{\omega }^5
\right)\nonumber\\ & &\mbox{ }\mbox{ }\mbox{ }\mbox{ }\mbox{
}\mbox{ }\mbox{ }\mbox{ }+
  z^5\,\left( 4 - 8\,{\omega }^2 + 54\,{\omega }^4 + 22\,{\omega }^6 \right)]
% end SimpE
\nonumber\\ & &\hspace{-0.3in}\times {\mathrm{log}}\left (
\frac{z+\omega-z^{2}\omega-\sqrt{2z\omega-2z^{3}\omega-\omega^{2}
+z^{4}\omega^{2}-z^{2}(-1+\omega^{2})}}
{z+\omega-z^{2}\omega+\sqrt{2z\omega-2z^{3}\omega-\omega^{2}
+z^{4}\omega^{2}-z^{2}(-1+\omega^{2})}}\right ) \Bigg]\mbox{
}\Bigg \} .\label{eq:icdqgen}
\end{eqnarray}

\subsubsection{Results for the matrix elements of $\op_{7,8}$ in
quenched QCD with SPQR kinematics}

The matrix elements of the EWP operators are given by:
\begin{equation} \la\pi^{+}\pi^{0}|\op_{7,8}|K^{+}\ra =
 \frac{2\sqrt{2}}{f_{K}f^{2}_{\pi}}
 \gamma^{(8,8)}\,\left[1+\frac{m^{2}_{K}}{16\pi^{2}f^{2}}
  \left ( I^q_{zf} +
  J^q_{a} + J^q_{b} +
  J^q_{c+d}\right)\right]
\label{eq:spqrquenched88}\end{equation}
where $I_{zf}$, the one-loop contribution from the renormalization
of the mesons' wavefunctions and the replacement of $f^3$ by
$f_Kf_\pi^2$ is given in \eq{eq:izfquenched} and the $J_{a,b,c+d}$
are as follows:
\begin{eqnarray}
%
% tadpole diagram
 J^q_a &=&
 \frac{ \left ( {\mathrm{log}}(z^{2})+
  (-1+z^{2}){\mathrm{log}}(\frac{m^{2}_{\pi}}{\mu^{2}})\right )
  }{3}- \frac{\alpha
   \left ( 2 - 2z^{2}+z^{2}(-2+z^{2})
   {\mathrm{log}}(-1+\frac{2}{z^{2}})\right )}
   {9 (-1+z^{2})}\nonumber\\
 & &\mbox{ }-
 \frac{m^{2}_{0}z^2
  \left ( -2 +2z^{2}+{\mathrm{log}}(-1+\frac{2}{z^{2}})
   \right )}{9 (-1+z^{2}) m^{2}_{\pi}}\mbox{ } ,
\end{eqnarray}
\begin{equation}
%
% scattering diagram
  J^q_b=
 \frac{-2z^2\left ( 3+3\sqrt{\frac{1-\omega}{1+\omega}}
   {\mathrm{log}}
  \left (\frac{-1+\sqrt{\frac{1-\omega}{1+\omega}}}
   {1+\sqrt{\frac{1-\omega}{1+\omega}}} \right )
   + (-3+\omega){\mathrm{log}}(\frac{m^{2}_{\pi}}{\mu^{2}})
    \right ) }
     {3\omega}\mbox{ }\end{equation}
and
\begin{eqnarray}
%
% hook diagrams
 J^q_{c+d} &=& \left (
    \frac{1}{3\, ( -1 + z ) \,
    \omega \,( -2\,z + \omega  + z^2\,\omega)}\times \right )
 \nonumber \\
 && \bigg \{ \left[ {\omega }^2 +
       z\,\omega \,\left( 1 + 2\,\omega  \right)  +
       z^2\,\left( -3 - 7\,\omega  + {\omega }^2 \right)  +
       z^3\,\left( 3 + 2\,{\omega }^2 \right)  \right] \,
     {\mathrm{log}} \left(z^2\right)  \nonumber \\
   & & + 3\,\left( -1 + z \right) \,z^2\,
     {\sqrt{1 - {\omega }^2}}\,
     {\mathrm{log}} \left(\frac{1 - {\sqrt{1 - {\omega }^2}}}
       {1 + {\sqrt{1 - {\omega }^2}}}\right)\nonumber \\
    & &\hspace{-0.4in}+
    \left( -1 + z \right) \,
     \left( -2\,z + \omega  + z^2\,\omega  \right)\,
     \left[ -3\,z\,\left( 1 + \omega  \right)  +
       \left( \omega  + z^2\,\omega  +
          3\,z\,\left( 1 + \omega  \right)  \right) \,
        {\mathrm{log}} \left(\frac{{{m^{2}_{\pi}}}}{{\mu}^2}\right)
    \right]
    \bigg \}  \nonumber \\
%
% \alpha starts here
%
%
+&\alpha&\,\left ( \frac{z^2}{18(-1+z)^{5}(1+z)^{2}
\omega(-2z+\omega+z^{2}\omega)^{2}} \right )\times\nonumber\\
&&\hspace{-0.2in}\left(\frac{1}{2z\omega-2z^{3}\omega-
\omega^{2}+z^{4}\omega^{2}-z^{2}(-1+\omega^{2})} \right
)\,\sqrt{(-1+z)^{3}(1+z)}\times\nonumber\\ && \bigg \{
\frac{2}{z^{2}}\sqrt{(-1+z)^{3}(1+z)}\bigg [
2z\omega-2z^{3}\omega-\omega^{2}+z^{4}\omega^{2}-z^{2}(-1+\omega)^{2}\bigg]
\times\nonumber\\ & &\bigg [
% SimpA
z\,[ z^6\,\left( 5 - 7\,\omega  \right) \,\omega  +
     z^7\,\left( -5 + \omega  \right) \,{\omega }^2 +
     2\,z^8\,{\omega }^3 - 3\,{\omega }^2\,\left( 1 + \omega  \right)  +
     z\,{\omega }^2\,\left( 15 + \omega  \right)\nonumber\\
& &\mbox{ }\mbox{ }\mbox{ }\mbox{ }\mbox{ }\mbox{ }\mbox{ }\mbox{
} \mbox{ }\mbox{ }\mbox{ }+
     z^5\,\omega \,\left( 7 - 10\,\omega  + 3\,{\omega }^2 \right)  +
     2\,z^4\,\omega \,\left( 4 + 5\,\omega  + 3\,{\omega }^2
 \right)\nonumber\\
& &\hspace{-0.3in}+
     z^2\,\left( 3 - 12\,\omega  + 17\,{\omega }^2 - 10\,{\omega }^3 \
\right)  - z^3\,\left( 3 + 20\,\omega  - 7\,{\omega }^2 +
        12\,{\omega }^3 \right) ]\log \left(-1 + \frac{2}{z^{2}}\right)\nonumber\\ & &+
  \left( -1 + z^2 \right) \,\omega \,
   [ -\left( z\,\left( -5 + \omega  \right) \,\omega  \right)  -
     2\,{\omega }^2 - 4\,z^5\,{\omega }^2 +
     z^4\,\omega \,\left( 13 + \omega  \right)\nonumber\\
& &\mbox{ }\mbox{ }\mbox{ }\mbox{ }\mbox{ }\mbox{ }\mbox{ }\mbox{
} \mbox{ }\mbox{ }\mbox{ }-
     z^2\,\left( 2 - 7\,\omega  + {\omega }^2 \right)  -
     z^3\,\left( 10 + \omega  + 5\,{\omega }^2 \right)\nonumber\\
& &\mbox{ }\mbox{ }\mbox{ }\mbox{ }\mbox{ }\mbox{ }\mbox{ }\mbox{
} \mbox{ }\mbox{ }\mbox{ } +
     3\,z\,\left( 1 + z \right) \,
      \bigg ( \omega \,\left( 1 + \omega  \right)  +
        z^4\,\omega \,\left( 1 + \omega  \right)  -
        2\,z\,\omega \,\left( 3 + \omega  \right)\nonumber\\
& &\mbox{ }\mbox{ }\mbox{ }\mbox{ }\mbox{ }\mbox{ }\mbox{ }\mbox{
} \mbox{ }\mbox{ }\mbox{ }\mbox{ }\mbox{ }\mbox{ }\mbox{ }\mbox{ }
\mbox{ }\mbox{ }\mbox{ }\mbox{ }\mbox{ }\mbox{ }\mbox{ }\mbox{ }
\mbox{ }\mbox{ }\mbox{ }\mbox{ }\mbox{ }\mbox{ }-
        2\,z^3\,\omega \,\left( 3 + \omega  \right)  +
        2\,z^2\,\left( 2 + \omega  + 3\,{\omega }^2 \right) \bigg ) \,
      \log (z^2)\mbox{ } ]\mbox{ }
% end of SimpA
\bigg]\nonumber\\ &-&\left( 2 + z - 4\,z^2 - z^3 + 2\,z^4 \right)
\,\omega \,
  {\left( -2\,z + \omega  + z^2\,\omega  \right) }^2\times\nonumber\\
& & \left[ 2\,z\,\omega  - 2\,z^3\,\omega  - {\omega }^2 +
z^4\,{\omega }^2 -
    z^2\,\left( -1 + {\omega }^2 \right)  \right] \times\nonumber\\
& &\mbox{ }\mbox{ }\mbox{ }\mbox{ }\mbox{ }
  \log \left (\frac{1 + z - z^2 - {\sqrt{{\left( -1 + z \right) }^3\,
          \left( 1 + z \right) }}}{1 + z - z^2 +
      {\sqrt{{\left( -1 + z \right) }^3\,\left( 1 + z \right) }}}\right)
\nonumber\\ &-&3(-1+z)\sqrt{(-1+z)^{3}(1+z)}\times\nonumber\\ &
&\hspace{-0.95in}\bigg [
% SimpC
z\,{\sqrt{1 - {\omega }^2}}\,[ -2\,z^7\,{\omega }^3 +
     z^2\,\left( 2 - 5\,{\omega }^2 \right)  +
     {\omega }^2\,\left( -2 + 3\,{\omega }^2 \right)  +
     z^6\,{\omega }^2\,\left( 4 + 3\,{\omega }^2 \right)+ 2\,z^3\,\omega \,\left( -2 + 7\,{\omega }^2
\right) \nonumber\\ & &\hspace{-0.2in}+
     z\,\left( 4\,\omega  - 6\,{\omega }^3 \right)  -
     2\,z^5\,\left( \omega  + 2\,{\omega }^3 \right)  +
     z^4\,\left( {\omega }^2 - 6\,{\omega }^4 \right)  ]
     \log \left( \frac{1 - {\sqrt{1 - {\omega }^2}}}
       {1 + {\sqrt{1 - {\omega }^2}}} \right)\nonumber\\
& &\mbox{ }\mbox{ }\mbox{ }\mbox{ }\mbox{ }\mbox{ }\mbox{ }\mbox{
} \mbox{ }\mbox{ }\mbox{ }  +
  {\sqrt{2\,z\,\omega  - 2\,z^3\,\omega  - {\omega }^2 +
       z^4\,{\omega }^2 - z^2\,\left( -1 + {\omega }^2
\right) }}\,\nonumber\\ & &\mbox{ }\mbox{ }\mbox{ }\mbox{ }\mbox{
}\mbox{ }\mbox{ }\mbox{ } \mbox{ }\mbox{ }\mbox{ } \times
   [ 2\,z^7\,{\omega }^3 - 2\,{\omega }^4 +
     z^5\,\omega \,\left( 2 + 3\,{\omega }^2 \right)  -
     z^6\,{\omega }^2\,\left( 4 + 3\,{\omega }^2 \right)\nonumber\\
& &\mbox{ }\mbox{ }\mbox{ }\mbox{ }\mbox{ }\mbox{ }\mbox{ }\mbox{
} \mbox{ }\mbox{ }\mbox{ }+
     z\,\omega \,\left( -2 + 7\,{\omega }^2 \right)  +
     z^3\,\left( 2\,\omega  - 14\,{\omega }^3 \right)  +
     z^2\,\left( -2 + 5\,{\omega }^2 - 2\,{\omega }^4 \right)\nonumber\\
& &\mbox{ }\mbox{ }\mbox{ }\mbox{ }\mbox{ }\mbox{ }\mbox{ }\mbox{
} \mbox{ }\mbox{ }\mbox{ }  +
     z^4\,\left( {\omega }^2 + 7\,{\omega }^4 \right)]\times \nonumber\\
& &\log \left(\frac{z + \omega - z^2\,\omega -{\sqrt{2\,z\,\omega
- 2\,z^3\,\omega  - {\omega }^2 + z^4\,{\omega }^2 - z^2\,\left(
-1 + {\omega }^2 \right) }}}{z + \omega  - z^2\,\omega  +
{\sqrt{2\,z\,\omega  - 2\,z^3\,\omega  - {\omega }^2 +
z^4\,{\omega }^2 - z^2\,\left( -1 + {\omega }^2 \right)
}}}\right)\,
% end of SimpC
\bigg ] \mbox{ } \bigg \}\nonumber\\
%
% m_{0} starts here
%
%
+&m^{2}_{0}& \left ( \frac{z^2}{18(-1+z)^{5}(1+z)^{2}
\omega(-2z+\omega+z^{2}\omega)^{2} m^{2}_{\pi}} \right
)\times\nonumber\\ &&\left(\frac{1}{2z\omega-2z^{3}\omega-
\omega^{2}+z^{4}\omega^{2}-z^{2}(-1+\omega^{2})} \right )\, z \,
\sqrt{(-1+z)^{3}(1+z)}\times\nonumber\\ &&\bigg \{
\frac{1}{z}\sqrt{(-1+z)^{3}(1+z)}\bigg [
2z\omega-2z^{3}\omega-\omega^{2}+z^{4}\omega^{2}-z^{2}(-1+\omega)^{2}\bigg
] \nonumber\times\\ & &\bigg [
% SimpA
2( -1 + z^2 )\omega
   [ z ( -5 + \omega ) \omega  + 2\omega^{2} +
     4z^{5}\omega^{2} - z^{4}\omega ( 13 + \omega )\nonumber\\
& &\mbox{ }\mbox{ }\mbox{ }\mbox{ }\mbox{ }\mbox{ }\mbox{ }\mbox{
} \mbox{ }\mbox{ }\mbox{ }+
     z^{2} ( 2 - 7\omega  + \omega^{2} )  +
     z^{3} ( 10 + \omega  + 5\omega^{2} )  ] \nonumber\\
& &\mbox{ }\mbox{ }\mbox{ }\mbox{ }\mbox{ }\mbox{ } \mbox{ }\mbox{
}\mbox{ } +
  [ z\,\left( -5 + \omega  \right) \,{\omega }^2 + 2\,{\omega }^3 +
     3\,z^6\,{\omega }^2\,\left( 1 + \omega  \right)  +
     z^5\,\omega \,\left( -6 - 15\,\omega  + {\omega }^2 \right)\nonumber\\
& &\mbox{ }\mbox{ }\mbox{ }\mbox{ }\mbox{ }\mbox{ }\mbox{ }\mbox{
} \mbox{ }\mbox{ }\mbox{ }+
     z^2\,\omega \,\left( 8 - 7\,\omega  + {\omega }^2 \right)  +
     2\,z^3\,\left( -3 + 2\,\omega  - 7\,{\omega }^2 + 4\,{\omega }^3 \
\right) \nonumber\\ & &\mbox{ }\mbox{ }\mbox{ }\mbox{ }\mbox{
}\mbox{ }\mbox{ }\mbox{ } \mbox{ }\mbox{ }\mbox{ } +
2\,z^4\,\left( 3 + 9\,\omega  - 5\,{\omega }^2 +
        4\,{\omega }^3 \right)  ] \,
   \log \left(-1 + \frac{2}{z^{2}}\right)
% end SimpA
\bigg ]\nonumber\\ & &\mbox{ }\mbox{ }\mbox{ }\mbox{ }-
% SimpB
3( -1 + z^2 ) \omega
  {\left( -2\,z + \omega  + z^2\,\omega  \right) }^2\times \nonumber\\
& &
  \left[ 2\,z\,\omega  - 2\,z^3\,\omega  - {\omega }^2 + z^4\,{\omega }^2 -
    z^2\,\left( -1 + {\omega }^2 \right)  \right]\times \nonumber\\
& &
  \log \left(\frac{1 + z - z^2 - {\sqrt{{\left( -1 + z \right) }^3\,
          \left( 1 + z \right) }}}{1 + z - z^2 +
      {\sqrt{{\left( -1 + z \right) }^3\,\left( 1 + z \right) }}}\right)
% end SimpB
\nonumber\\ & &\mbox{ }\mbox{ }\mbox{ }\mbox{ }
-3(-1+z)\sqrt{(-1+z)^{3}(1+z)}\times\nonumber\\ & &\bigg[
% SimpC
{\sqrt{1 - {\omega }^2}}\,[ 6\,z^5\,\omega  + 3\,z^2\,{\omega }^2
+
     2\,z\,{\omega }^3 + 2\,z^7\,{\omega }^3 - {\omega }^4 -
     z^6\,{\omega }^2\,\left( 6 + {\omega }^2 \right)\nonumber\\
& &- 2\,z^3\,\omega \,\left( 2 + 3\,{\omega }^2 \right)  +
     z^4\,\left( -2 + 5\,{\omega }^2 + 2\,{\omega }^4 \right) ]
\log \left( \frac{1 - {\sqrt{1 - {\omega }^2}}}
       {1 + {\sqrt{1 - {\omega }^2}}} \right)\nonumber\\
& &\mbox{ }\mbox{ }\mbox{ }\mbox{ }\mbox{ }\mbox{ }\mbox{ }\mbox{
} \mbox{ }\mbox{ }\mbox{ }  +
  {\sqrt{2\,z\,\omega  - 2\,z^3\,\omega  - {\omega }^2 +
       z^4\,{\omega }^2 - z^2\,\left( -1 + {\omega }^2 \right) }}\times\nonumber\\
& &
   [ 2\,z^2\,\omega  - 3\,z\,{\omega }^2 + {\omega }^3 +
     z^4\,\omega \,\left( -4 + {\omega }^2 \right)\nonumber\\
& &\mbox{ }\mbox{ }\mbox{ }\mbox{ }\mbox{ }\mbox{ }\mbox{ }\mbox{
} \mbox{ }\mbox{ }\mbox{ }   -
     z^5\,{\omega }^2\,\left( -2 + {\omega }^2 \right)  +
     z^3\,\left( 2 - {\omega }^2 + {\omega }^4 \right)  ]\times\nonumber\\
& & \hspace{-0.3in}\log \left(\frac{z + \omega  - z^2\,\omega  -
       {\sqrt{2\,z\,\omega  - 2\,z^3\,\omega  - {\omega }^2 +
           z^4\,{\omega }^2 - z^2\,\left( -1 + {\omega }^2 \right) }}}
    {z + \omega  - z^2\,\omega  +
       {\sqrt{2\,z\,\omega  - 2\,z^3\,\omega  - {\omega }^2 +
           z^4\,{\omega }^2 - z^2\,\left( -1 + {\omega }^2
 \right) }}}\right)\,
% end SimpC
\bigg] \mbox{ }\bigg \}\,.\label{eq:jcdqgen}
\end{eqnarray}
%

%\end{fmffile}

\begin{thebibliography}{99}

\bibitem{epeexp} NA48 Collaboration, V.Fanti et al.,
\PL{B465}{1999}{335};\\  KTEV Collaboration, A.~Alavi-Harati et
al., \PRL{83}{1999}{22}

\bibitem{spqr} SPQ${\textrm{CD}}$R Collaboration, Ph.Boucaud et
al., \\ \NP{Proc.Suppl.106}{2002}{329} ({\tt hep-lat/0110206})

\bibitem{LL}
L.~Lellouch and M.~L\"uscher, \CMP{219}{2001}{31}

\bibitem{dlinetal} C.-J.D. Lin, G. Martinelli, C.T. Sachrajda, M. Testa,
\NP{B619}{2001}{467}

\bibitem{cairns} C.-J.D. Lin, G. Martinelli, C.T. Sachrajda, M. Testa,
\NP{Proc.Suppl. 109}{2002}{218}

\bibitem{BG} C.~Bernard and M.F.L.~Golterman \PRD{53}{1996}{476}

\bibitem{GP} M.F.L.~Golterman and E.~Pallante
\NP{Proc.Suppl. 83}{2000}{250}

\bibitem{dt12quenched} C.-J.D.~Lin, G.~Martinelli, E.~Pallante,
C.T.~Sachrajda and G.~Villadoro, in preparation

\bibitem{ciuchini} M.~Ciuchini, E.~Franco, G.~Martinelli, L.~Reina and L.~Silvestrini
\ZP{C68}{1995}{239}

\bibitem{buras} G.~Buchalla, A.J.~Buras and M.E.~Lautenbacher, \textit{Rev.Mod.Phys.} \textbf{68}, 1125
(1996)

\bibitem{shifman} M.A.~Shifman, A.I.~Vainstein and V.I.~Zakharov,
\NP{B120}{1977}{316}

\bibitem{BPRD} C.~Bernard, T.~Draper, A.~Soni, H.D.~Politzer and
M.B.~Wise, \PRD{32}{1985}{2343}

\bibitem{bmgg} R.C.~Brower, G.~Maturana, M.B.~Gavela and R.~Gupta,
\PRL{53}{1984}{1318}

\bibitem{cmp} N.~Cabibbo, G.~Martinelli and R.~Petronzio,
\NP{B244}{1984}{381}

\bibitem{aoki} JLQCD Collaboration, S.~Aoki et al.,
\PRD{58}{1998}{054503}

\bibitem{spqr2} SPQ${\textrm{CD}}$R Collaboration, Ph.Boucaud et
al., \\ \NP{Proc.Suppl.106}{2002}{323} ({\tt hep-lat/0110169})

\bibitem{laiho} J.~Laiho and A.~Soni {\tt hep-ph/0203106}

\bibitem{CPPACS} CP-PACS Collaboration, J.~Noaki et al., {\tt
hep-lat/0108013} (2001)

\bibitem{RBC} RBC Collaboration, T.~Blum et al., {\tt
hep-lat/0110075} (2001)

\bibitem{ga-le} J.~Gasser and H.~Leutwyler, \NP{B250} {1985} {465}

\bibitem{bg} C.~Bernard and M.~Golterman, \PRD{46}{1992}{853}

\bibitem{KMW}
J. Kambor, J. Missimer and D. Wyler, \NP{B346}{1990}{17}

\bibitem{quenchedfull} C.-J.D.~Lin et al., {\tt
http://www.hep.phys.soton.ac.uk/kpipi/}

\bibitem{CG} V.~Cirigliano and E.~Golowich, \PL{B475} {2000} {351}

\bibitem{einp} G. Ecker et al., \NP{B591}{2000}{419}

\bibitem{GL}
M.F.L. Golterman and K.-C. Leung, \PRD{56}{1997}{2950}

\bibitem{GL2}
M.F.L. Golterman and K.-C. Leung, \PRD{58}{1998}{097503}

\bibitem{Glpart} M.F.L.~Golterman and K.C.~Leung,
\PRD{57}{1998}{5703}

\bibitem{BDHS} C.~Bernard, T.~Draper, G.~Hockney and A.~Soni,
\NP{Proc.Suppl. 4}{1988}{483}

\bibitem{dawson} C.~Dawson et al., \NP{B514}{1998}{313}

\bibitem{EP} E.~Pallante, \JHEP{9901}{1999}{012}

\bibitem{MT} L. Maiani and M. Testa \PL{B245}{1990}{585}

\bibitem{L1} M.~L\"uscher, \CMP{104}{1986}{177};
\NP{B354}{1991}{531}; \NP{B364}{1991}{237}

\bibitem{L2} M.~L\"uscher, \CMP{105}{1986}{153}

\bibitem{ciuchinietal} M.~Ciuchini, E.~Franco, G.~Martinelli and
L.~Silvestrini, \PL{B380}{1996}{353}

\bibitem{RuGo} K.~Rummukainen and S.~Gottlieb, \NP{B450}{1995}{397}

\bibitem{BPP}
J. Bijnens, E. Pallante and J. Prades, \NP{B521}{1998}{305}

\bibitem{PPS}
E. Pallante, A. Pich and I. Scimemi, \NP{B617}{2001}{441}

\bibitem{Mgprivate} M.F.L.~Golterman, Private Communication

\end{thebibliography}
\end{document}